\documentclass[12pt]{article}
\usepackage{fullpage,hyperref,graphicx,amssymb,amsmath}

\def\a{\alpha}\def\b{\beta}\def\g{\gamma}\def\d{\delta}\def\e{\epsilon}
\def\l{\lambda}\def\m{\mu}\def\n{\nu}\def\o{\omega}
\def\s{\sigma}\def\t{\tau}\def\z{\zeta}\def\th{\theta}
\def\varphi{\varphi}
\def\G{\Gamma}\def\D{\Delta}\def\O{\Omega}

\def\CH{\mathcal{H}}\def\CI{\mathcal{I}}
\def\CJ{\mathcal{J}}\def\CK{\mathcal{K}}\def\CL{\mathcal{L}}
\def\CM{\mathcal{M}}\def\CN{\mathcal{N}}

\def\IZ{\mathbb{Z}}
\def\IR{\mathbb{R}}\def\IC{\mathbb{C}}\def\IH{\mathbb{H}}
\def\IF{\mathbb{F}}\def\IP{\mathbb{P}}

\def\ibar{{\bar\imath}}\def\jbar{{\bar\jmath}}
\def\kbar{{\bar k}}

\def\pd{\partial}

\def\bi{\mathbf{i}}\def\bj{\mathbf{j}}\def\bk{\mathbf{k}}

\def\bx{\mathbf{x}}

\def\bomega{{\boldsymbol{\omega}}}
\def\grad{\boldsymbol{\nabla}}

\DeclareMathOperator{\Tr}{Tr}

\DeclareMathOperator{\re}{Re}  
\DeclareMathOperator{\im}{Im}  
\DeclareMathOperator{\Spin}{Spin}

\DeclareMathOperator{\Vol}{Vol}

\DeclareMathOperator{\Hol}{Hol}
\DeclareMathOperator{\GL}{GL}
\DeclareMathOperator{\SL}{SL}
\DeclareMathOperator{\SO}{SO}
\DeclareMathOperator{\U}{U}
\DeclareMathOperator{\SU}{SU}
\DeclareMathOperator{\Sp}{Sp}
\DeclareMathOperator{\USp}{USp}

\def\tilde{\widetilde}

\def\w{\wedge}

\def\ha{\dfrac12}\def\half{\tfrac12}

\long\def\symbolfootnote[#1]#2{\begingroup%
\def\thefootnote{\fnsymbol{footnote}}\footnote[#1]{#2}\endgroup}

\def\K3{\text{K3}}

\def\tx{{\tilde x}}
\def\ty{{\tilde y}}

\def\tV{{\tilde V}}
\def\Gbar{{\bar G}}
\def\mhat{{\hat m}}
\def\nhat{{\hat n}}
\def\phat{{\hat p}}
\def\1hat{{\hat 1}}
\def\2hat{{\hat 2}}
\def\3hat{{\hat 3}}
\def\4hat{{\hat 4}}
\def\ahat{{\hat\a}}
\def\bhat{{\hat\b}}
\def\ghat{{\hat\g}}

\begin{document}
\begin{titlepage}
\setcounter{page}{0}
\begin{flushright}
  arXiv:1206.1070\\ NSF-KITP-12-100\\ UPR 1238-T
\end{flushright}
\vspace*{\stretch{1}}
\begin{center}
  \huge M-theory/type IIA duality and\\ 
K3 in the Gibbons-Hawking approximation
\end{center}
\vspace*{\stretch{0.75}}
\begin{center}\hskip5.5pt  
  \large Michael B. Schulz\symbolfootnote[1]{mbschulz at brynmawr.edu}
  and
  Elliott F. Tammaro\symbolfootnote[2]{etammaro at brynmawr.edu}
\end{center}
\begin{center}
  \textit{Department of Physics, Bryn Mawr College\\
    Bryn Mawr, PA 19010, USA\\}
    \end{center}
\vspace*{\stretch{1}}
\begin{abstract}
  \normalsize We review the geometry of K3 surfaces and then describe
  this geometry from the point of view of an approximate metric of
  Gibbons-Hawking form.  This metric arises from the M-theory lift of
  the tree-level supergravity description of type~IIA string theory on
  the $T^3/\IZ_2$ orientifold, the D6/O6 orientifold T-dual to type~I
  on $T^3$.  At large base, it provides a good approximation to the
  exact K3 metric everywhere except in regions that can be made
  arbitrarily small.  The metric is hyperk\"ahler, and we give
  explicit expressions for the hyperk\"ahler forms as well as harmonic
  representatives of all cohomology classes in this soluble model.
  Finally, in the Gibbons-Hawking approximation, we compute the metric
  on the moduli space of metrics in two ways, first by projecting to
  transverse traceless deformations (using compensators), and then by
  computing the naive moduli space metric from dimensional reduction.
  In either case, we find agreement with the exact coset moduli space
  of K3 metrics.  The $T^3/\IZ_2$ orientifold provides a simple
  example of a warped compactification, and in a separate paper, this
  work will be applied to study warped Kaluza-Klein reduction on
  $T^3/\IZ_2$.
   \end{abstract}
  \vspace*{\stretch{5}}
  \flushleft{5 June 2012; revised 10 August 2016}
\end{titlepage}


\tableofcontents
\newpage


\section{Introduction}
\label{sec:Intro}

K3 surfaces are ubiquitous in string theory
compactifications~\cite{Aspinwall:1996mn}.  As the unique Calabi-Yau
manifolds of complex dimension 2, they are the simplest compact
special holonomy manifolds, and the simplest manifolds other than tori
admitting the covariantly constant spinors required for low energy
supersymmetry~\cite{Joyce:2000}.  K3 surfaces also feature prominently
in string duality~\cite{Witten:1995ex,Morrison:1996na,Morrison:1996pp,
  Aspinwall:1996mn,Aspinwall:2000fd}.  For example, in a chain of
dualities related by circle reduction, F-theory on K3 is dual to the
$E_8\times E_8$ heterotic string on $T^2$, M-theory on K3 is dual to
the heterotic or type I string on $T^3$, and type IIA string theory on
K3 is dual to the heterotic or type I string on $T^4$.  Fiberwise
application of this duality relates M-theory on manifolds of $G_2$
holonomy to the heterotic or type I string on Calabi-Yau
3-folds~\cite{Acharya:2004qe}.

The paper adds three contributions to the literature on K3: (1)~We
study the Gibbons-Hawking approximation to the global K3 metric and
provide the most complete description to date of all aspects of K3
geometry and topology visible through the lens of this intuitive and
soluble simplified model.  This model is motivated by the duality
between M-theory on K3 and type IIA string theory on the $T^3/\IZ_2$
orientifold (as opposed to \mbox{F-theory}/heterotic duality), and is
equivalent to the classical supergravity description of the
latter. (2)~We give a review of K3 for string theorists that differs
in character from previous reviews (for example, the excellent
reviews~\cite{Morrison:1988,Aspinwall:1996mn,Aspinwall:2000fd}), in
that it is less algebraic geometry oriented and more differential
geometry oriented compared to previous treatments, and we provide
correspondingly more differential geometry background (for example, on
complex, hypercomplex, and hyperk\"ahler structure; the Lichnerowicz
and Laplace-de Rham operators, and the relation between their
zero-modes; metric deformations and moduli space metrics).  (3)~We
give a complete and accessible description of the exact K3 homology
and cohomology lattices and their splittings, in terms of the natural
orbifold bases.  It was suprising to us that such an account did not
already exist in the literature, so we have sought to fill the void.

The model allows us to see the following at a level of explicitness
not generally available:

\begin{enumerate}
\item The hyperk\"ahler metric on K3 as a function of moduli.
\item The harmonic forms in this metric (including the hyperk\"ahler
  triple).
\item The metric deformations.  Up to a compensating diffeomorphism,
  the metric deformations obtained the explicit hyperk\"hler metric
  indeed agree with the required combinations of invariant tensors
  (hypercomplex structure) and harmonic forms.
\item Yau's Theorem.  The choice of hyperk\"ahler structure indeed
  determines the K3 metric in this simplified model, since both depend
  explicitly on the same coset moduli with the same identifications.
\item The moduli space metric.  We compute the
  diffeomorphism-invariant metric on the moduli space of metrics by
  projecting to transverse traceless deformations (using
  compensators), and find that it agrees with both the naive moduli
  space metric from dimensional reduction and the exact coset moduli
  space metric of K3.
\end{enumerate}

\subsection{Motivations and scope}

The motivation for the present work, and the duality of interest here,
is the lift from type IIA string theory on the $T^3/\IZ_2$ orientifold
to M-theory on K3.  Here $T^3/\IZ_2$ is the orientifold obtained from
type I on $T^3$ after T-duality with respect to all three torus
isometries.  This duality relates the conventional compactification of
M-theory on K3 to a simple IIA warped compactification, where the
warping is due to the 16 D6-branes and 8 O6-planes of $T^3/\IZ_2$.

This paper is part of a larger
investigation~\cite{SchulzTammaroA,SchulzTammaroB}, whose overall goal
is to elucidate the procedure for Kaluza-Klein reduction of warped
compactifications via duality to standard compactifications. (A
secondary goal along the way, is to learn about compactification on
manifolds of $\SU(2)$ structure~\cite{Schulz:2012uj}.)  Virtually all
phenomenologically relevant string theory compactifications are of
warped type, in which the overall scale factor of 4D spacetime varies
over the internal dimensions~\cite{Giddings:2001yu,Chan:2000ms}.  This
feature modifies 4D mass scales and couplings, and naturally realizes
the Randall-Sundrum approach to the hierarchy
problem~\cite{Randall:1999ee}.  A long standing obstacle to
quantitative prediction has been our incomplete understanding of the
analog of standard Kaluza-Klein reduction for warped compactification.
Warped Kaluza-Klein reduction was first studied in
Refs.~\cite{DeWolfe:2002nn,Giddings:2005ff} and our understanding was
greatly enhanced in Refs.~\cite{Shiu:2008ry,Douglas:2008jx,
  Frey:2008xw,Frey:2009qb,Underwood:2010pm}.  Our investigation takes
a complementary route, and will be useful in probing the formalism of
Refs.~\cite{Shiu:2008ry,Douglas:2008jx,Frey:2008xw}.

With this goal in mind, this paper provides a review of the geometry
of K3 surfaces emphasizing the tools that will be useful for the
duality, of which a novel feature is the role of an approximate K3
metric exactly dual to the classical supergravity description of
$T^3/\IZ_2$.  At the level of the tree-level type IIA supergravity
description of $T^3/\IZ_2$, the lift to M-theory gives a
``first-order'' metric on K3 of Gibbons-Hawking~\cite{Gibbons:1979zt}
form, through which it is possible to study the differential geometry
of K3 more explicitly than is generally the case for Calabi-Yau
manifolds.\footnote{An approximate metric of this form has a history
  of \emph{local} application to the geometry of K3 in the math
  literature.  See, for example, Ref.~\cite{Gross:2000}.}  For
physicists, differential geometry, the language of general relativity,
is more intuitive than algebraic geometry, however, exact Calabi-Yau
metrics are not known except in orbifold limits.  Therefore, one is
usually led to an algebro-geometric description, in which the
quantities appearing in the local supergravity equations---frame,
metric, deformations, harmonic forms, K\"ahler form---are analyzed
abstractly rather than explicitly.  In contrast, K3 in the
Gibbons-Hawking approximation provides a model in which one can
manipulate all of the moving parts of K3 at the level of explicitness
that one would really like.  There is an explicit metric whose
components are simply defined functions of quantities
$G^{\a\b},\b^{\a\b},x^{I\a}$ parametrizing an $\bigl(\SO(3)\times
\SO(19)\bigr)\backslash \SO(3,3+16)$ coset, in addition to overall
scaling from the volume modulus $V_{\K3}$.  One can write down the
harmonic forms in this metric, and see explicitly how they vary with
the moduli.  One can write down a frame and a triple of hyperk\"ahler
2-forms and see that all three 2-forms are closed, thus showing that
there is no torsion, and that the metric connection has $\SU(2)$
holonomy.  Finally, one can study the metric on the metric moduli
space, and here too, there is an explicit example of a novel
phenomenon.  The diffeomorphism invariant moduli space metric obtained
using the formalism of compensators agrees with the naive moduli space
metric from direct dimensional reduction.

We restrict the scope of this paper to the geometry of K3 and the
duality between a family of M-theory vacua on K3 and family of IIA
vacua on $T^3/\IZ_2$.  We steer clear of 7D effective field theory
questions, in which the moduli are promoted from \emph{parameters}
labeling the families to \emph{fields} depending on the noncompact
dimensions of spacetime.  The effective field theory analysis will
appear in Ref.~\cite{SchulzTammaroA}, in which the warped
compactification ansatz for type IIA on $T^3/\IZ_2$ is derived from
the standard compactification ansatz for M-theory on K3, at the level
of the tree level IIA supergravity description.  All of this is a
warm-up for Ref.~\cite{SchulzTammaroB}, in which we treat a similar
duality~\cite{Schulz:2004tt,Donagi:2008ht} relating the $T^6/\IZ_2$
type IIB orientifold with $\CN=2$ flux to a class of purely geometry
type IIA Calabi-Yau compactifications.  The latter will allow us to
deduce the warped dimensional reduction procedure for $T^6/\IZ_2$ from
that for the conventional Calabi-Yau dual.  (See
Ref.~\cite{Grimm:2012rg}, for recent related work which studies the
effective field theory and warped Kaluza-Klein reduction ansatz for an
F-theory compactification that yields gauge flux on 7-branes as
opposed to NS and RR 3-form flux.\footnote{The M-theory description of
  the local geometry of Ref.~\cite{Grimm:2012rg} builds on the
  Ooguri-Vafa metric Ref.~\cite{Ooguri:1996me} and involves a similar
  Gibbons-Hawking description and cohomology analysis to that of the
  present paper.  Among the results of Ref.~\cite{Grimm:2012rg}, it is
  shown that the M-theory warp factor modifies the real part of the
  gauge coupling function, and the modified M-theory 3-form potential
  corrects the imaginary part.  We thank the authors of
  Ref.~\cite{Grimm:2012rg} for alerting us to this work.})

As an interesting vista along the way to this goal, we observe in
Ref.~\cite{Schulz:2012uj} that the class of Calabi-Yau manifolds
arising in the duality of the previous paragraph are not only
manifolds of $\SU(3)$ holonony, but also manifolds of $\SU(2)$
structure.  That is, there exists a connection with torsion whose
holonomy group is further restricted from $\SU(3)$ to $\SU(2)$.  Just
as the dual choice of flux in the type IIB $T^6/\IZ_2$ dual
spontaneously breaks an $\CN=4$ low energy field theory to
$\CN=2$~\cite{Kachru:2002he,Schulz:2002eh,Schulz:2004ub,Schulz:2004tt},
the $\SU(2)$ structure gives the IIA Calabi-Yau compactification an
$\CN=4$ effective field theory in which the topology spontaneously
breaks the supersymmetry to $\CN=2$.  This will be discussed in
Ref.~\cite{Schulz:2012uj} based on a first order description very
similar to that of the present paper, together with a few exact
results.  For related work on the $\SU(2)$ structure of the Enriques
Calabi-Yau 3-fold~\cite{Ferrara:1995yx} and the resulting effective
field theory, see Refs.~\cite{TriendlStringsTalk,KashaniPoor:2013en}.
For earlier work on $\SU(2)$ structure compactifications, see
Refs.~\cite{Gauntlett:2003cy,Bovy:2005qq,ReidEdwards:2008rd,
  Triendl:2009ap,Louis:2009dq,Danckaert:2009hr,Spanjaard:2008zz,
  Danckaert:2011ju}.  \medskip

\subsection{Outline}

An outline of the paper is as follows:
\medskip

In Sec.~\ref{sec:ReviewK3}, we review the geometry of K3 surfaces,
beginning with their definition as compact complex surfaces of trivial
canonical bundle, and going on to review their holonomy, K\"ahler and
hyperk\"ahler structure, and (co)homology.  On general grounds, the
integer (co)homology lattice $H_2(\K3,\IZ)$ splits as
$(-E_8)\oplus(-E_8)\oplus (U_{1,1})^{\oplus3}$ and
$(-\Spin(32)/\IZ_2)\oplus (U_{1,1})^{\oplus3}$ where $(-E_8)$ denotes
the weight lattice of $E_8$ with the sign of the inner product
reversed, and similarly for $\Spin(32)/\IZ_2$.  These splittings are
not obvious in the homology basis from the Kummer construction of
$\K3$ as the resolution of $T^4/\IZ_2$.  Therefore, with future
applications in mind, we derive the explicit relations between the
Kummer, $(-E_8)\oplus(-E_8)\oplus (U_{1,1})^{\oplus3}$ and
$(-\Spin(32)/\IZ_2)\oplus (U_{1,1})^{\oplus3}$ homology bases.  These
relations, while certainly not new, do not seem to be written down in
the literature, and we expect that a clear discussion will be useful
to others.  Sec.~\ref{sec:ReviewK3} concludes with a discussion of the
moduli space of hyperk\"ahler structure and its relation to the moduli
space of K3 metrics.

In Sec.~\ref{sec:K3MetricGH}, we describe the approximate K3 metric of
Gibbons-Hawking form, obtained from the lift of the tree-level type
IIA supergravity description of $T^3/\IZ_2$ to M-theory.  After an
overview of the results of this section, we review the Gibbons-Hawking
multicenter metrics and the standard supergravity identifications
between type IIA string theory and M-theory.  Then, we consider in
succession the M-theory lift of a collection of $N$ D6-branes on
$\IR^3$, of a collection of $N$ D6-branes near an O6-plane on $\IR^3$,
and finally of the $T^3/\IZ_2$ orientifold with 16 D6 branes and 8
O6-planes.  In Sec.~\ref{sec:HarmApprox} we give the frame,
hyperk\"ahler forms, and a basis of harmonic forms in the approximate
metric, identifying the cohomology classes of the latter with the
those in the exact treatment of Sec.~\ref{sec:ModHkahler}.  Finally,
Sec.~\ref{sec:MetricModGH} is devoted to the moduli space of the
approximate K3 metric, treated from two points of view.  First,
focusing on the case of the $16\times3$ exceptional deformations, we
show that metric deformations due to the explicit dependence of the
approximate metric on moduli $G^{\a\b}$, $\b^{\a\b}$, and $x^{I\a}$
agree with the transverse traceless deformations generated by by
harmonic forms, up to compensating diffeomorphisms. The exact coset
moduli space follows.  Next we consider the naive moduli space metric
from dimensional reduction.  This differs from the previous moduli
space metric in two ways: the metric deformations are not projected to
transverse traceless components (i.e., no compensators), and instead,
there exists an additional term in the metric.  We find that naive
moduli space metric precisely agrees with the previous one.  Finally,
in Sec.~\ref{sec:Conclusions}, we conclude.

The Appendices contain additional background and technical details.
App.~\ref{app:HkahlerT} describes hyperk\"ahler structures on $T^4$.
App.~\ref{app:HomologyLattice} treats the homology lattice of K3,
deriving intersection numbers and the details of
$(-E_8)\oplus(-E_8)\oplus (U_{1,1})^{\oplus3}$ and
$(-\Spin(32)/\IZ_2)\oplus (U_{1,1})^{\oplus3}$ splittings.
App.~\ref{app:Lichnerowicz} reviews the Lichnerowicz operator, which
relates deformations of the Ricci tensor to metric deformations.
App.~\ref{app:MetDefHarm} provides background on the relation between
metric deformations and harmonic forms.  Finally,
App.~\ref{app:GHdefs} treats the metric deformations and compensating
vector fields of the Gibbons-Hawking multicenter metric, and evaluates
a class of integrals used elsewhere in the paper.


\section{Review of K3 surfaces}
\label{sec:ReviewK3}


\subsection{Definition}
\label{sec:Definition}

A K3 surface is a compact complex surface of trivial canonical bundle
and $h^{1,0} = 0$~\cite{Barth:2004}.  The latter condition is
necessary only to distinguish a K3 surface from $T^4$ (an abelian
surface).  Every K3 surface is K\"ahler~\cite{Siu:1983}, so K3
surfaces are the unique compact Calabi-Yau manifolds of complex
dimension 2.  As such, they are ubiquitous in compactifications of
string theory.  In contrast to Calabi-Yau 3-folds, all K3 surfaces are
deformations of one another, so they are all
diffeomorphic~\cite{Barth:2004}.  The name K3 was coined in 1958 by
Andr\'e Weil to honor the achievements of geometers Kummer, K\"ahler,
and Kodaira~\cite{Weil:1958}, a short time after mountain climbers
first ascended K2 in northern Kashmir, the world's second highest
peak~\cite{Aspinwall:1996mn}.  The two simplest constructions of a K3
surface are the quartic hypersurface in $\IP^4$,\footnote{The quartic
  hypersurface is an example of an algebraic K3 surface---the
  vanishing locus of a set of polynomial equations in a complex
  projective space.  Not all K3 surfaces are algebraic. In particular,
  the generic Kummer surface is not algebraic.} and the Kummer
surface, defined as the surface obtained by resolving the sixteen
orbifold singularities of $T^4/\IZ_2$.

In the next three sections we review the holonomy, hyperk\"ahler
structure, and homology of K3 surfaces.  Harmonic forms and the moduli
space of hyperk\"ahler structure are discussed next, in
Sec.~\ref{sec:ModHkahler}.  We conclude our overview of K3 surfaces
with a discussion of the metric on K3 metric moduli space in
Sec.~\ref{sec:K3MetricMod}.  For a more extensive review of the
geometry of K3 surfaces, we refer the reader to
Refs.~\cite{Morrison:1988,Aspinwall:1996mn,Joyce:2000,Barth:2004}.


\subsection{Holonomy}
\label{sec:Holonomy}

K3 surfaces are the unique Calabi-Yau 2-folds.  Recall that a
Calabi-Yau $n$-fold $X$ is defined to be a compact K\"ahler manifold
of complex dimension $n$ and trivial canonical bundle.  This
definition, together with Yau's theorem, implies an equivalent
definition in terms of global $\SU(n)$ holonomy.  For simply connected
$X$, the reasoning is as follows.\footnote{When $X$ is not simply
  connected, the existence of a metric of global $\SU(n)$ holonomy is
  still an equivalent definition of a Calabi-Yau manifold, even though
  the reasoning given here must be modified.  However, in this case
  it is a weaker condition to say that $X$ is Ricci-flat K{\"a}hler
  manifold than to say it is Calabi-Yau.  For example, an Enriques
  surface $\K3/\IZ_2$ is a Ricci-flat K{\"a}hler manifold, but it is not
  Calabi-Yau: The canonical bundle is nontrivial and $c_1$ vanishes in
  $H^2(X,\IR)$ but gives a $\IZ_2$ torsion class in $H^2(X,\IZ)$.  The
  restricted holonomy group $\Hol_0(g)$ is $\SU(2)$, but the global
  holonomy group $\Hol(g)$ is disconnected.}  (i) The trivial
canonical bundle implies vanishing first Chern class $c_1 \in
H^2(X,\IZ)$, which in turn implies the weaker condition that $c_1$
vanishes in $H^2(X,\IR)$. (ii) Since $X$ is K\"ahler, we can then
apply Yau's theorem: \emph{Let $X$ be a compact complex manifold, of
  dimension at least 2, with vanishing real first Chern class.
  Consider a fixed complex structure on $X$.  Given a real class in
  $H^{1,1}(X,\IC)$ of positive norm, there is a unique Ricci flat
  metric on $X$ with K\"ahler form in this cohomology class.}  (iii)
Finally, the Riemannian holonomy group $\Hol(g)$ of the Levi-Civita
connection of a Ricci flat metric $g$ follows from the classification
of Berger~\cite{Berger:1955}.

Berger proved that for a simply connected Riemannian manifold $M$ that
is not a reducible or symmetric space, the Riemannian holonomy must be
$\SO(m)$, $\U(m)$, $\SU(m)$, $\Sp(m)$, $\bigl(\Sp(m)\times
\Sp(1)\bigr)/\IZ_2^\text{center}$, $G_2$, or $\Spin(7)$.  (See, for
example, Ref.~\cite{Besse:1987,Joyce:2000}.)  Let us focus on the
subset of K\"ahler special holonomy groups $U(m)$, $\SU(m)$, and
$\Sp(m)$, and let $M$ be real $d$-dimensional.  As cited in
Ref.~\cite{Aspinwall:1996mn}, Berger's classification states that
%
%
\begin{enumerate}
\item $\Hol(M)\subset \U(\tfrac{d}2)$ if and only if $X$ is K\"ahler;
\item $\Hol(M)\subset \SU(\tfrac{d}2)$ if and only if $X$ is K\"ahler
  and Ricci-flat;
\item $\Hol(M)\subset \Sp(\tfrac{d}4)$ if and only if $X$ is
  hyperk\"ahler.
\end{enumerate}
Each line contains those that follow, since $U(\tfrac{d}2)\supset
\SU(\tfrac{d}2)\supset \Sp(\frac{d}4)$.  Thus, a Calabi-Yau $n$-fold is
a manifold of $\SU(n)$ holonomy.  In the special case $n=2$, we have
$\SU(2)\cong \Sp(1)$, and the last two categories collapse to one.
Therefore, a K3 surface with a Ricci-flat metric has $\SU(2)\cong
\Sp(1)$ holonomy and is hyperk\"ahler.

Ricci flat metrics on hyperk\"ahler 4-manifolds are solutions to 4D
Euclidean Einstein equations that can be shown to have (anti)selfdual
curvature 2-form.  For this reason, they are often referred to as
gravitational instantons, by analogy to (anti)selfdual Yang-Mills
instantons in 4D.  The only compact hyperk\"ahler 4-manifolds are
$\K3$ and $T^4$.  Noncompact hyperk\"ahler \mbox{4-manifolds}
asymptotic to $\IH/\G$, where $\G$ is a finite subgroup of $\Sp(1)$,
are known as asymptotically locally Euclidean (ALE) spaces.  As long
as $\G$ is not too large, they can be viewed as local models for the
resolution of orbifold singularities of K3 surfaces.  The prime
examples for which explicit metrics can be written down are Taub-NUT
spaces~\cite{Taub:1950ez,Newman:1963yy}, multi-Taub-NUT
spaces~\cite{Hawking:1976jb}, Eguchi-Hanson
spaces~\cite{Eguchi:1978gw}, and finally Gibbons-Hawking multicenter
spaces~\cite{Eguchi:1978gw,Gibbons:1979zt}, which include the previous
spaces as special cases.  The Gibbons-Hawking ansatz is discussed in
Sec.~\ref{sec:GHmulticenter}.  An important example in which the
metric is known only more implicitly is the Atiyah-Hitchin
space~\cite{Atiyah:1985dv,Atiyah:1985fd,Atiyah:1988jp}, the moduli
space of two $\SU(2)$ 't Hooft-Polyakov monopoles in four dimensions.
Additional constructions beyond the Gibbons-Hawking ansatz include
twistor theory~\cite{Hitchin:1900zr} and the hyperk\"ahler quotient
construction of
Kronheimer~\cite{Kronheimer:1989zs,Kronheimer:1989pu,Gibbons:1996nt}.


\subsection{Hyperk\"ahler structure}
\label{sec:Hyperkahler}

The definition of a hyperk\"ahler manifold as a $4m$-dimensional
Riemannian manifold of $\Sp(m)$ holonomy implies the existence of a
triple of integrable almost complex structures satisfying the
quaternionic algebra.  We now review the structure of K\"ahler and
hyperk\"ahler manifolds and following Chapter 7 of
Ref.~\cite{Joyce:2000} closely.


\subsubsection{K\"ahler manifolds}
\label{sec:KahlerMan}

It is useful to begin with a review of K\"ahler structure.  First,
recall that the complex number field $\IC$ is defined as an extension
of the reals by writing $z\in\IC$ as $z = a + ib$, where $a,b\in\IR$
and $i^2 = -1$, and comes with the involution of complex conjugation
$\bar z = z^* = a - ib$.  A K\"ahler manifold has the local structure
of $\IC^n$ together with its standard complex structure, Hermitian
metric, and K\"ahler 2-form.  For $\IC^n$ with holomorphic coordinates
$z^j = x^{2j-1}+ix^{2j}$ and antiholomorphic coordinates $z^\jbar =
x^{2j-1}-ix^{2j}$, these tensors are
\begin{subequations}
  \begin{align}
    ds^2 &= \sum_{m=1}^{2n} dx^{m}\otimes dx^m\\
    \CJ &= \sum_{j=1}^n dx^{2j-1}\otimes\pd_{2j} - dx^{2j}\otimes\pd_{2j-1},\\
    J &= \sum_{j=1}^n dx^{2j-1}\w dx^{2j},
  \end{align}
\end{subequations}
or, equivalently,
\begin{subequations}
  \begin{align}
    ds^2 &= \half\sum_{j=1}^n\bigl(dz^j\otimes dz^\jbar + dz^\jbar\otimes dz^j\bigr),\\
    \CJ &= \sum_{j=1}^n\bigl(dz^j\otimes\pd_j - dz^\jbar\otimes\pd_\jbar\bigr),\\
    J &= \half \sum_{j=1}^n dz^j\w dz^\jbar.
  \end{align}
\end{subequations}

On a general K\"ahler manifold $X$, analogous expressions hold in
which coordinate vectors and \mbox{1-forms} are replaced by the frame
and coframe (i.e., vielbein basis).  A $2n$ dimensional K\"ahler
manifold has three successive layers of structure:

\begin{enumerate}
\item Complex structure.  Recall that an almost complex structure
  (ACS) $\CJ$ on an even-dimensional oriented manifold $X$ is a smooth
  tensor field of rank (1,1) that maps the (co)tangent bundle to
  itself and that squares to one.  In physics conventions, it has
  index structure $\CJ_a{}^b$ and naturally acts on the cotangent
  bundle, mapping a 1-form $\o_p$ to $\CJ_p{}^q\o_q$.  Its transpose
  naturally acts on the tangent bundle.  An ACS gives a canonical
  isomorphism between each tangent space $T_pM\cong\IR^{2n}$ and
  $\IC^n$.  When the Nijenhuis tensor vanishes, the almost complex
  structure is integrable, and there exist complex coordinates on $X$
  with holomorphic transition functions $z'^j = f^j(z)$, independent
  of $z^\kbar$.  In this case, we drop the word \emph{almost}, and
  refer to $\CJ$ as the complex structure.  In the language of
  $G$-structures and intrinsic torsion, a complex manifold is a
  manifold of torsion free $\GL(n,\IC)$ structure: the structure group
  of the frame bundle is reduced from $\GL(2n,\IR)$ to $\GL(n,\IC)$,
  and the intrinsic torsion vanishes precisely when the Nijenhuis
  tensor vanishes.
\item Hermitian metric.  $X$ is endowed with a metric $g$ such that $g
  = \CJ g\CJ^T$, or in components, $g_{j\kbar} = (g_{\jbar k})^*$ and
  $g_{jk}=g_{\jbar\kbar} = 0$.  Define the fundamental form $J$ by
  $J_{pq} = \CJ_p{}^rg_{rq}$, or equivalently $J = ig_{j\kbar}dz^j\w
  dz^\kbar.$ The volume form on a Hermitian manifold is given by
  \begin{equation}
    \Vol_M = \frac1{n!}J^n.
  \end{equation}
\item K\"ahler structure.  $X$ is said to be K\"ahler, and $J$ called
  the K\"ahler form, when $dJ =0$.  Given an almost complex structure
  and Hermitian metric, the condition $dJ=0$ is equivalent to $\nabla
  \CJ=0$, to $\nabla J=0$, and to the condition that $X$ have a
  torsion free $\U(n)$ structure.  The group $\U(n)$ is the subgroup
  of $\GL(n,\IC)$ preserving $J$, and the intrinsic torsion vanishes
  precisely when $J$ is covariantly constant and the structure group
  of the frame bundle is realized as the Riemannian holonomy group.
  Note that $J_{j\kbar} = \half\d_{j\kbar}$ in the holomorphic coframe
  basis $\th^j$ (and conjugate basis $\th^\jbar)$, and $U\in\U(n)$
  acts on this basis as $\th^j\mapsto U^j{}_k\th^k$ (and
  $\th^\jbar\mapsto (\bar U)^\jbar{}_\kbar\th^\kbar$).
\end{enumerate}


\subsubsection{Hyperk\"ahler manifolds}
\label{sec:HkahlerMan}

Hyperk\"ahler manifolds are the natural generalizations of K\"ahler
manifolds when the complex numbers are generalized to the
quaternions.  Recall that the quaternionic number field $\IH$ is
defined by writing $q\in\IH$ as $q = a + b\bi + c\bj+ d\bk$, where
$a,b,c,d\in\IR$ and $Q=\{\pm1,\pm\bi,\pm\bj,\pm\bk\}$ is the
quaternionic group, with multiplication defined by\footnote{The last
  three relations can also be written more concisely as
  $\bi\bj\bk=-1$.}
\begin{equation}
  \bi^2 = \bj^2 = \bk^2 = -1,\quad 
  \bi\bj = \bk,\quad \bj\bk = \bi,\quad \bk\bi = \bj.
\end{equation}
As in the complex case, we define $\bar q = q^* = a - b\bi - c\bj
-d\bk$, $\re q = a$, and $\im q = b\bi + c\bj + d\bk$.  A
hyperk\"ahler manifold has the local structure of $\IH^n$ together
with its standard hypercomplex structure, Hermitian metric, and
hyperk\"ahler 2-forms.  For $\IH^n$ with quaternionic coordinates $q^j
= - x^{j4} + \bi x^{j1} + \bj x^{j2} +\bk x^{j3}$, these tensors
are\footnote{Here, to simplify notation, we write $a\w b = a\otimes b
  - b\otimes a$, even when $b$ is a vector rather than a 1-form.}
\begin{subequations}
  \begin{equation}
    \begin{split}
      \CJ^1 &= dx^{j2}\w\pd_{j3} + dx^{j1}\w\pd_{j4},\\
      \CJ^2 &= dx^{j3}\w\pd_{j1} + dx^{j2}\w\pd_{j4},\\
      \CJ^3 &= dx^{j1}\w\pd_{j2} + dx^{j3}\w\pd_{j4},
    \end{split}
  \end{equation}
  \begin{equation}
    \begin{split}
      J^1 &= dx^{j2}\w dx^{j3} + dx^{j1}\w dx^{j4},\\
      J^2 &= dx^{j3}\w dx^{j1} + dx^{j2}\w dx^{j4},\\
      J^3 &= dx^{j1}\w dx^{j2} + dx^{j3}\w dx^{j4},
    \end{split}
  \end{equation}
  and
  \begin{equation}
    ds^2 = \sum_{j=1}^m\sum_{\a=0}^3 dx^{j\a}\otimes dx^{j\a}
    = \re\Bigl(\sum_{j=1}^m dq^j\otimes dq^\jbar\Bigr).
  \end{equation}
\end{subequations}
Here, the action of $\CJ^1,\CJ^2,\CJ^3$ on the $dx^{j\a}$ induces the
same action on $dq^i$ as multiplication by $\bi,\bj,\bk$.  The
expressions for $J^\a$ and the metric can be written concisely as
\begin{equation}
  \sum_{j=1}^m dq^j \otimes dq^\jbar = g - J^1\bi - J^2\bj -J^3\bk.
\end{equation}
Note that the metric is K\"ahler with respect to an $S^2$ worth of
complex structures: if $(n_1)^2 + (n_2)^2 + (n_3)^2 = 1$, then $n_\a
\CJ^\a$ is a complex structure with corresponding K\"ahler form
$n_\a J^\a$.  With respect to the complex structure $\CJ^1$, the 2-form
\begin{equation}
  J= -J^2+iJ^3 = \sum_{j=1}^{2m}dz^{2j-1}\w dz^{2j}
\end{equation}
is a complex symplectic form, with similar definitions for $J$ in the
complex structure $n_\a\CJ^\a$.  On a general hyperk\"ahler manifold
$X$, these expressions hold when coordinate vectors and \mbox{1-forms}
are replaced by the frame and coframe (i.e., vielbein basis).  A
hyperk\"ahler manifold has three successive layers of structure:

\begin{enumerate}
\item Hypercomplex structure.  For an oriented 4$m$-dimensional
  manifold $X$, we define an almost hypercomplex structure (AHS) to be
  a triple of almost complex structures $\CJ^\a$, $\a=1,2,3$, whose
  action on the tangent bundle satisfies the quaternionic algebra.  In
  physics conventions, this means that $(\CJ^1)^T,(\CJ^2)^T,(\CJ^3)^T$
  satisfy the algebra of $\bi,\bj,\bk$, and $\CJ^1,\CJ^2,\CJ^3$
  satisfy the algebra of $-\bi,-\bj,-\bk$.  An AHS gives a canonical
  isomorphism between each tangent space $T_p M\cong\IR^{4m}$ and
  $\IH^m$.  When all three complex structures are integrable, we drop
  the word almost and refer to the triple as a hypercomplex structure.
  As above, this means that there is an $S^2$ worth of complex
  structures on $X$.  It does \emph{not}, however, mean that there is
  a coordinate integrability analogous to that in the complex case.  A
  hypercomplex structure does not imply the existence quaternionic
  coordinates with transition functions $q'^j = f^j(q)$, independent
  of $q^\jbar$.
\item Hyperhermitian metric.  $X$ is endowed with a metric $g$ that is
  hermitian with respect to each complex structure $\CJ^\a$, $\a=1,2,3$.
  Define the corresponding fundamental forms $J^\a$ by lowering the
  vector index of $\CJ^\a$.  Then, the volume form can be written
  \begin{equation}
    \Vol_M = \frac1{m!}(n_\a J^\a)^{2m}\quad\text{for any $n_\a\in S^2$.}
  \end{equation}
  and the complex volume form can be written
  \begin{equation}
    \O_M = \frac1{m!}J^{2m},
  \end{equation}
  with
  \begin{equation}
    \Vol_M = \frac1{4^m}\O_M\w\bar\O_M.
  \end{equation}
  Here, $J =(-n'_\a + in''_\a)J^\a$, where $n,n',n''$ is a right
  handed orthonormal basis for $\IR^3$.
\item Hyperk\"ahler structure.  $X$ is said to be hyperk\"ahler, with
  hyperk\"ahler 2-forms $J^\a$, $\a=1,2,3$, when all three 2-forms are
  closed, $dJ^a=0$.  Given a hypercomplex structure and a
  hyperhermitian metric, this is equivalent to the condition $\nabla
  J^\a$ for $\a=1,2,3$, and to the condition that $X$ have torsion
  free $\Sp(n)$ structure.  The group $\Sp(m)$ is the subgroup of
  $\GL(4m,\IR)$ preserving $g, J^1, J^2, J^3$.  It can also be viewed
  as the subgroup of $\GL(m,\IH)$ such that $\bar U^T U =1$, where
  $\bar U$ refers to conjugation in $\IH$.\footnote{There are two
    groups commonly referred to as $\Sp$.  The group $\Sp(m)$ relevant
    here is compact and has fundamental representation of real
    dimension $4m$; it is the straightforward generalization of the
    special unitary groups when $\IC$ is replaced by $\IH$, and is
    sometimes denoted $\USp(m)$.  The other group, $\Sp(2m,\IR)$
    preserves a skew symmetric form with $\pm I_n$ off diagonal; it is
    noncompact and has fundamental representation of real dimension
    $2m$.  The complexification of either is $\Sp(2m,\IC)$, and we
    have $\USp(2n) = \U(2m)\cap\SL(2m,\IC)$.}  It acts as on the
  quaternionic coframe $\th^j$ as $\th^j\mapsto U^j{}_k \th^k$ (and
  the conjugate basis as $\th^\jbar\mapsto (\bar U)^\jbar{}_\kbar
  \th^\kbar$).  The reader can easily check that this preserves the
  left hand side of
  \begin{equation}
    \sum_{j=1}^m\th^j\otimes\th^\jbar = g -J^1\bi-J^2\bj-J^3\bk.
  \end{equation}
  The intrinsic torsion vanishes precisely when the $J^\a$ are
  covariantly constant and the structure group of the frame bundle is
  realized as the Riemannian holonomy group.
\end{enumerate}

Since $\Sp(m)\subset\SU(2m)$, hyperk\"ahler manifolds are necessarily
Calabi-Yau, and their metrics are necessarily Ricci flat.  We will
discuss the hyperk\"ahler moduli space of $\K3$ and its identification
with the metric moduli space in Secs.~\ref{sec:ModHkahler} and
~\ref{sec:K3MetricMod}.


\subsection{Homology and cohomology}
\label{sec:Homology}

From the definition of K3 as a manifold of trivial canonical bundle
and $h^{1,0}=0$, the interesting part of the homology ring of K3 is
the second homology lattice $H_2(\K3,\IZ)$.  As we explain below, this
lattice is 22 dimensional, and is a selfdual, even unimodular lattice,
of signature $(3,19)$.  For a proof of this result on general grounds,
see the discussion in Sec.~2.3 of Ref.~\cite{Aspinwall:1996mn}.

Even selfdual lattices exist only in signature $(p,q)$ such that $p-q$
is divisible by 8.  For example, the signature $(16,0)$ case arises in
the construction of the heterotic string.  In this case, there are
exactly two such lattices: the root lattice of $E_8\times E_8$ and the
weight lattice of
$\Spin(32)/\IZ_2$.\footnote{\label{footnote:RootsWeights}For $E_8$,
  the root and weight lattices are the same.  $\Spin(32)$ is the
  universal cover of $SO(32)$.  Its weight lattice is the root lattice
  of $\SO(32)$ together with the weights of the vector representation
  and the spinor representations of each chirality.  The center of
  $\Spin(32)$ is $\IZ_2\times\IZ_2$, of which there are three
  nontrivial elements, each generating a $\IZ_2$.  Quotienting by one
  $\IZ_2$ eliminates the spinor weights and gives $\SO(32)$.
  Quotienting by the second $\IZ_2$ eliminates the vector and negative
  chirality spinor representations and leaves the group we have
  denoted $\Spin(32)/\IZ_2$.  Quotienting by the third $\IZ_2$
  eliminates the vector and positive chirality spinor representation.}
For $p,q>0$ the solution is unique up to lattice automorphism, and for
signature $(3,19)$ the lattice is
\begin{align}\label{eq:K3latticeSplittings}
  H_2(\K3,\IZ)
  &\cong(-E_8)\oplus(-E_8)\oplus(U_{1,1})^{\oplus3}\notag\\
  &\cong(-\Spin(32)/\IZ_2)\oplus(U_{1,1})^{\oplus3}.
\end{align}
Here, $U_{1,1}$ denotes the unique even selfdual lattice of signature
(1,1), with inner product $\bigl(\begin{smallmatrix} 0 & 1\\ 1&
  0\end{smallmatrix}\bigr)$.

The goal of this section is to see the
result~\eqref{eq:K3latticeSplittings} explicitly, starting from the
simplest possible description of K3.  In the discussion that follows,
we first describe the homology of K3 from the point of view of the
resolution of $T^4/\IZ_2$ in Sec.~\ref{sec:Abasis}.  This gives a
natural basis of $H_2$ in terms of the 2-tori inherited from $T^4$ and
the 16 exceptional cycles obtained by resolving the $2^4$ $A_1$
orbifold singularities.  This basis, however, does not generate
$H_2(\K3,\IZ)$ with integer coefficients.  We refer to this basis as
the $(A_1)^{16}$ basis, and compute the intersection pairing in this
basis in Sec.~\ref{sec:Intersection} and App.~\ref{app:Resolution}.
The lattice splittings~\eqref{eq:K3latticeSplittings} arise from the
resolution of limits in which K3 develops two $E_8$ singularities or a
$D_{16}$ singularity.  The remaining sections and
App.~\ref{app:Splittings} relate the $(A_1)^{16}$ basis to $(E_8)^2$
and $D_{16}$ bases, realizing the splittings of
Eq.~\eqref{eq:K3latticeSplittings}.


\subsubsection{$(A_1)^{16}$ basis}
\label{sec:Abasis}

We now specialize to K3 realized as a Kummer surface.  Let $S$ denote
the smooth resolution $S$ of the orbifold $T^4/\IZ_2$ obtained by
blowing up its sixteen $A_1$ orbifold singularities, and write $\pi:
S\to T^4/\IZ_2$.  Choose coordinates $x^\m$, $m=1,2,3,4$ on $T^4$ with
periodicities $x^m\cong x^m+1$, and consider the quotient by the
$\IZ_2$ involution
\begin{equation}
  \s\colon\quad (x^1,x^2,x^3,x^4) \mapsto (-x^1,-x^2,-x^3,-x^4).
\end{equation}
The involution has $2^4=16$ fixed points labeled by the elements of
$(\IF_2)^4$, i.e., by coordinates such that $2x^m\in\IF_2 = \{0,1\}$
for $m=1,2,3,4$.

A part of the (co)homology of $S$ is inherited directly from the
$T^4$.  Let us focus on $H_2(S,\IZ)$.  The subgroup inherited from
$H_2(T^4,\IZ)$ is generated by ``sliding cycles'' made up of $\IZ_2$
invariant pairs of 4-tori on the $T^4$.  We label these as follows:
\begin{subequations}
  \begin{equation}
    \begin{split}
      f^1 &= T^2_{x^2x^3}\times\{p\cup p'\subset T^2_{x^1x^4}\},\\
      f^2 &= T^2_{x^3x^1}\times\{p\cup p'\subset T^2_{x^2x^4}\},\\
      f^3 &= T^2_{x^1x^2}\times\{p\cup p'\subset T^2_{x^3x^4}\},
    \end{split}
  \end{equation}
  and
  \begin{equation}
    \begin{split}
      f_1 &= T^2_{x^1x^4}\times\{p\cup p'\subset T^2_{x^2x^3}\},\\
      f_2 &= T^2_{x^2x^4}\times\{p\cup p'\subset T^2_{x^3x^1}\},\\
      f_3 &= T^2_{x^3x^4}\times\{p\cup p'\subset T^2_{x^1x^2}\}.
    \end{split}
  \end{equation}
\end{subequations}
Here, $f^1$ is the union of a $T^2$ at fixed $x^1,x^4$ (spanning all
possible $x^2,x^3$ values), together with its $\IZ_2$ image.  The
point $p$ is any non fixed point, and $p'=-p$.  Poincar\'e
duality\footnote{On a $d$-dimensional manifold $X$, Poincar\'e duality
  identifies $[B]\in H_p(X,\IZ)$ with $[\b]\in H^{d-p}(X,\IZ)$, where
  $\int_B\g = \int\g\w\b$.}  identifies $H_2(T^4,\IZ)$ with
$H^2(T^4,\IZ)$.  In cohomology, the same classes $[f^\a],[f_\a]$,
$\a=1,2,3$ can be represented by
\begin{equation}
  \begin{split}
    f^1 &= 2dx^1\w dx^4,\\
    f^2 &= 2dx^2\w dx^4,\\
    f^3 &= 2dx^3\w dx^4,
  \end{split}
  \qquad\quad
  \begin{split}
    f_1 &= 2dx^2\w dx^3,\\
    f_2 &= 2dx^3\w dx^1,\\
    f_3 &= 2dx^1\w dx^2.
  \end{split}
\end{equation}
Since $H_2(T^4,\IZ)\cong H^2(T^4,\IZ)$, we use the same notation in
either case.\footnote{For notational simplicity, we also leave the
  various pullbacks and images implicit and do not distinguish
  between, for example, $f^1$ as a cycle in $T^4$, $T^4/\IZ_2$ and
  $S$.}
In addition, there are 16 exceptional cycles from resolving each of
the 16 orbifold singularities of $T^2/\IZ_2$,
\begin{equation}
  E_I,\quad\text{where}\quad I=1\dots 16.
\end{equation}

As explained in App.~\ref{app:Resolution}, the basis
$[f^\a],[f_\a],[E_I]$ generates $H_2(S,\IR)$ over the reals, but only
an order 2 sublattice of $H_2(S,\IZ)$ over the integers.  It misses
elements of $H_2(S,\IR)$ inherited from $\IZ_2$ invariant 2-tori in
$T^4$, of which there are 24:
\begin{subequations}
  \begin{equation}
    \begin{split}
      D^1_s &= T^2_{x^2x^3}\times\{p_s\subset T^2_{x^1x^4}\},\\
      D^2_s &= T^2_{x^3x^1}\times\{p_s\subset T^2_{x^2x^4}\},\\
      D^3_s &= T^2_{x^1x^2}\times\{p_s\subset T^2_{x^3x^4}\},
    \end{split}
  \end{equation}
  and
  \begin{equation}
    \begin{split}
      D_{1s} &= T^2_{x^1x^4}\times\{p_s\subset T^2_{x^2x^3}\},\\
      D_{2s} &= T^2_{x^2x^4}\times\{p_s\subset T^2_{x^3x^1}\},\\
      D_{3s} &= T^2_{x^3x^4}\times\{p_s\subset T^2_{x^1x^2}\},
    \end{split}
  \end{equation}
\end{subequations}
where $p_s$, for $s=1,2,3,4$, runs over the four $\IZ_2$ fixed points
$(0,0)$, $(0,\half)$, $(\half,0)$, $(\half,\half)\in T^2$,
respectively.  Under $T^4\to T^4/\s$, each of these twenty four 2-tori
maps to a $T^2/\IZ_2\cong S^2$.  The four fixed points of the latter
coincide with four of the sixteen fixed points of $T^4/\IZ^2$.
Therefore, after resolution of the 16 orbifold singularities, each
intersects 4 exceptional cycles.  We show in App.~\ref{app:Resolution}
that the $D^\a_s,D_{\a s}$ have cohomology classes
\begin{subequations}
  \begin{equation}
    [D^\a_s] = \tfrac12[f^\a] -\tfrac12\sum_{\text{four $I$}} E_I
  \end{equation}
  and
  \begin{equation}
    [D_{\a s}]  = \tfrac12[f_\a] -\tfrac12\sum_{\text{four $I$}} E_I,
  \end{equation}
\end{subequations}
where the four $E_I$ that appear in each case can be found in
App.~\ref{app:Resolution}.  The complete integer homology lattice
$H_2(S,\IZ)$ is generated by the integer span of
$[f^\a],[f_\a],[E_I],[D^\a_s],[D_{\a s}]$.

In what follows, we will refer to
\begin{equation}
  \chi^{(A)}{}_a = (f^\a,f_\a,E_I)
\end{equation}
as the $(A_1)^{16}$ basis, since the $\chi^{(A)}{}_I$, for
$I=1,\dots,16$ have zero volume at the $(A_1)^{16}$ orbifold locus in
moduli space.  Equivalently, the triple of hyperk\"ahler classes
$[J^1],[J^2],[J^3]$ are linear combination of only the
$[\chi^{(A)\,\a}]$ and $[\chi^{(A)}{}_\a]$ at the $(A_1)^{16}$
orbifold point.  In Secs.~\ref{sec:Dbasis} and \ref{sec:Ebasis}, we
will define bases $\chi^{(D)}{}_a$ and $\chi^{(E)}{}_a$ having the
analogous property at the $D_{16}$ and $E_8\times E_8$ orbifold loci
of the moduli space of K3.

It will be helpful to define also a half-integer basis
\begin{equation}
  \xi^{(A)}{}_a = (f^a,f_a,e^{(A)}{}_I),
\end{equation}
where the \emph{roots} $\chi^{(A)}{}_I$ of $(A_1)^{16}$ are
related to the \emph{orthonormal} basis $e^{(A)}{}_I$ via
\begin{equation}
  \chi^{(A)}_{2i-1} = e^{(A)}_{2i} - e^{(A)}{}_{2i-1},\quad
  \chi^{(A)}_{2i} = e^{(A)}_{2i} + e^{(A)}{}_{2i-1},
\end{equation}
for $i=1,\dots 16$, as a consequence of which the cohomology classes
$[e^{(A)}_I]$ are $\half\IZ$-valued.  The terms root and orthonormal
will become meaningful once we define an inner product.


\subsubsection{Intersection inner product}
\label{sec:Intersection}

The inner product on the (co)homology lattice is the intersection
number in homology or, equivalently, the cup product in cohomology.
If $[B],[C]\in H_2(S,\IZ)$ are Poincar\'e dual to $[\b],[\g]\in
H^2(S,\IZ)$, respectively, then we have
\begin{equation}\label{eq:IntersectionPairing}
  \# A\cap B = [\a]\cup[\b] = \int_S\a\w\b.
\end{equation}
which we denote by $\a\cdot\b$.  Thus we have
\begin{equation}
  f^\a\cdot f_\b = \int_{T^4/\IZ_2}f^a\w f_\b 
  = \half \int_{T^4}f^\a\w f_\b = 2\d^\a_\b,
\end{equation}
while $f^\a\cdot f^\b = f_\a\cdot f_\b = 0$.\footnote{Alternatively
  two 2-tori on $T^4$ intersecting transversely have 1 point of
  intersection.  Each $f$ is a pair of 2-tori.  Therefore, two $f$'s
  intersect in 4 points on $T^4$, or equivalently, 2 points on
  $T^4/\IZ_2$.}  The $E_I$ give
\begin{equation}
  E_I\cdot E_J = -2\d_{IJ},
\end{equation}
since they come from blowing up independent points, and each is a
2-spheres.\footnote{As discussed in Ref.~\cite{Aspinwall:1996mn},
  using $c_1(S)=0$, the self intersection of a genus $g$ Riemann
  surface on $S$ can be shown to be $2g-2$.}  Finally, $E_I\cdot f^\a
= E_I\cdot f_\a = 0$, since the sliding cycles $f$ of can be moved
away from the fixed points of $T^4/\IZ_2$.  Thus, in the basis
$\chi^{(A)}{}_a = (f^\a,f_\a,E_\a)$, the lattice inner product is
\begin{equation}
  \chi^{(A)}{}_a\cdot \chi^{(A)}{}_b 
  = \eta^{(A)}{}_{ab} =
  \begin{pmatrix}
    0 & 2 & 0\\
    2 & 0 & 0\\
    0 & 0 & -(A_1)^{16}
  \end{pmatrix},
\end{equation}
where $\bigl(-(A_1)^{16}\bigr)_{IJ} = -2\d_{IJ}$ is minus the Cartan
matrix of $(A_1)^{16}$.  If we use the orthonormal basis instead of
the root basis, then this becomes
\begin{equation}\label{eq:EtaDef}
  \xi^{(A)}{}_a\cdot \xi^{(A)}{}_b 
  = \eta_{ab} =
  \begin{pmatrix}
    0 &2 & 0\\
    2 & 0 & 0\\
    0 & 0 & -1
  \end{pmatrix}.
\end{equation}
As already noted, the basis $\chi^{(A)}_a = (f^\a,f_\a,E_I)$ does not
generate the (co)homology lattice with integer coefficients, since the
$D^\a_s$ have half-integer coefficients in this basis.  Equivalently,
the integer homology lattice $H_2(\K3,\IZ)$ does not split as sum of
the integer sublattices $\langle f^\a,f_\a\rangle$ and $\langle
f^\a,f_\a\rangle^\perp = (-A_1)^{\oplus16}$.  Here angle brackets
denote ``span of'' and $\perp$ denotes the orthogonal complement with
respect to the inner product~\eqref{eq:IntersectionPairing}.
Secs.~\ref{sec:Dbasis} and \ref{sec:Ebasis}, we will see that the
$D_{16}$ and $(E_8)^2$ bases $\chi^{(D)}_a$ $\chi^{(E)}_a$ are better
behaved in this regard.  The integer homology lattice $H_2(\K3,\IZ)$
\emph{does} split as $(-\Spin(32)/\IZ_2)\oplus U_{3,3}$ and
$(-E_8)\oplus(-E_8)\oplus U_{3,3}{}'$.


\subsubsection{$D_{16}$ basis}
\label{sec:Dbasis}

To relate the $(A_1)^{16}$ (co)homology basis $[\chi^{(A)}{}_a]$ to a
$D_{16}$ basis $[\chi^{(D)}{}_a]$ we seek a linear transformation
\begin{equation}
    [\chi^{(A)}{}_a] = V_a{}^b[\chi^{(D)}{}_b]
\end{equation}
such that the $[\chi^{(D)}{}_a]$ are integer (co)homology classes with
inner product
\begin{equation}
  \chi^{(D)}{}_a\cdot \chi^{(D)}{}_b 
  = \eta^{(D)}{}_{ab} =
  \begin{pmatrix}
    0 &2 & 0\\
    2 & 0 & 0\\
    0 & 0 & -D_{16}
  \end{pmatrix},
\end{equation}
where $\bigl(-D_{16})_{IJ}$ is minus the Cartan matrix of $\SO(32)$.

A convenient way to solve this problem is to first express the
$D_{16}$ roots $\chi^{(D)}{}_I$ in terms of an orthonormal basis
$e^{(D)}{}_I$ in the standard way~\cite{Fuchs:1997jv},
\begin{equation}\label{eq:D16roots}
  \begin{split}
    \chi^{(D)}{}_1 &= e^{(D)}{}_1 + e^{(D)}{}_2,\\
    \chi^{(D)}{}_I &= e^{(D)}{}_I - e^{(D)}{}_{I-1},
    \quad\text{for}\quad I=2,\dots,16,
  \end{split}\
\end{equation}
where $(e^{(D)}_I,e^{(D)}_J) = -\d_{IJ}$, and then define
$\xi^{(D)}{}_a$ analogously to $\xi^{(A)}{}_a$,
\begin{equation}
  \xi^{(D)}_a = (\chi^{(D)\,\a},\chi^{(D)}{}_\a,e^{(D)}{}_I),
\end{equation}
so that
\begin{equation}
  \xi^{(D)}{}_a\cdot \xi^{(D)}{}_b 
  = \eta_{ab}.
\end{equation}
Here, $\eta_{ab}$ is as defined in Eq.~\eqref{eq:EtaDef}.  Then,
$[\xi^{(A)}{}_a] = V_a{}^b[\xi^{(D)}{}_b]$, where $V_a{}^{b}$
expressed in the new bases must be an $\SO(3,19)$ matrix preserving
$\eta_{ab}$,
\begin{equation}
  V\eta V^T = \eta.
\end{equation}
The task is to find a $V$ such that the $[\chi^{(D)}{}_a]$ are integer
classes in $H_2(\K3,\IZ)$.  

This problem is easily solved, with a little bit of inspiration from
the type IIA D6/O6 orientifold dual to M-theory on K3.  The solution
is
\begin{equation}\label{eq:Vofx}
  V_a{}^b = V_a{}^b(x) =
  \begin{pmatrix}
    1 & x^Tx & 2x^T\\
    0 & 1    & 0\\
    0 & x    & 1
  \end{pmatrix},
\end{equation}
where the $x^{I\a}$, for $I=1,\dots,16$ are given by
\begin{equation}\label{eq:xD16}
  \begin{split}
    \bx^1 &= (0,0,0),\\
    \bx^2 &= (0,0,0),\\
    \bx^3 &= (\half,0,0),\\
    \bx^4 &= (\half,0,0),\\
    \bx^5 &= (0,\half,0),\\
    \bx^6 &= (0,\half,0),\\
    \bx^7 &= (\half,\half,0),\\
    \bx^8 &= (\half,\half,0),
  \end{split},
  \quad
  \begin{split}
    \bx^{9\phantom{1}} &= (0,0,\half),\\
    \bx^{10} &= (0,0,\half),\\
    \bx^{11} &= (\half,0,\half),\\
    \bx^{12} &= (\half,0,\half),\\
    \bx^{13} &= (0,\half,\half),\\
    \bx^{14} &= (0,\half,\half),\\
    \bx^{15} &= (\half,\half,\half),\\
    \bx^{16} &= (\half,\half,\half),
  \end{split}
\end{equation}
in the notation $\bx^I=(x^{I1},x^{I2},x^{I3})$.  A total of 8
solutions are obtained by replacing these $\bx^I$ by $\bx^I-\bx_P$,
where $\bx_P$ is one of the $2^3=8$ $\IZ^2$ fixed point of $T^3$ with
all coordinates $0$ or $1/2$.  The proof that the resulting
$\chi^{(D)}{}_a$ are integral is given in App.~\ref{app:Splittings}.
There, it is also shown that in the $\bx_P=0$ case the integer
(co)homology lattice of K3 splits as $(-\Spin(32)/\IZ_2)\oplus
U_{3,3}$.  Here, $U_{3,3} = \langle D^\a_4,f_\a\rangle$ and the weight
lattice $(-\Spin(32)/\IZ_2) = \langle D^\a_4,f_\a\rangle^\perp$ is the
integer span of the $D_{16}$ roots $\chi^{(D)}_I$ together with the
chiral spinor weights differing from $\half\sum_{I=1}^{16}\xi^{(D)}_I$
by an even number of sign changes.\footnote{An equivalent statement
  holds for other choices of $\bx_P$, where shifting one of the three
  $x^\a_P$ from 0 to 1/2 replaces $D^\a_4$ by $D^\a_2$.  (Replacing
  $D^\a_4$ by $D^\a_3$ or $D^\a_2$ by $D^\a_1$ is less interesting and
  just reverses the sign of $e^{(A)}_I$ and $e^{(D)}_I$ for $I$ odd.
  It is a Weyl reflection of $(A_1)^{16}$ and $\Spin(32)/\IZ_2$.)}

In the type IIA dual, the $\bx^I$ become the locations of the 16
D6-branes on $T^3/\IZ_2$.  There are 8 equivalent $\Spin(32)/\IZ_2$
loci in moduli space in which all D6 branes coincide with an O6-plane
at one of the $2^3=8$ $\IZ_2$ fixed points $\bx_P$, and the dual K3
surface develops a $D_{16}$ singularity.  To move to the $(A_1)^{16}$
locus, one needs to move two D6 branes to each O6-plane, giving
``$D_2$'' $= (A_1)^2$ gauge symmetry at each.  The $\bx^I$ in
Eq.~\eqref{eq:xD16} are the D6-brane locations in this configuration.


\subsubsection{$(E_8)^2$ basis}
\label{sec:Ebasis}

To relate the $(A_1)^{16}$ (co)homology basis $[\chi^{(A)}{}_a]$ to
the $E_8\times E_8$ basis $[\chi^{(E)}{}_a]$ we proceed as in the
previous section.  We seek a linear transformation
\begin{equation}
    [\chi^{(A)}{}_a] = W_a{}^b[\chi^{(E)}{}_b]
\end{equation}
such that the $[\chi^{(E)}{}_a]$ are integer (co)homology classes with
inner product
\begin{equation}
  \chi^{(E)}{}_a\cdot \chi^{(E)}{}_b 
  = \eta^{(E)}{}_{ab} =
  \begin{pmatrix}
    0 & 2 & 0\\
    2 & 0 & 0\\
    0 & 0 & (-E_8)\oplus(-E_8)
  \end{pmatrix},
\end{equation}
where $\bigl(-E\bigr)_{IJ}$ is minus the Cartan matrix of $E_8$.

Again, a convenient way to solve this problem is to first express the
$(E_8)^2$ roots $\chi^{(D)}{}_I$ in terms of an orthonormal basis
$e^{(D)}{}_I$ in the standard way~\cite{Fuchs:1997jv},
\begin{subequations}\label{eq:E8xE8roots}
  \begin{align}
    \chi^{(E)}_1 &= e^{(E)}_1 + e^{(E)}_2,
    \quad
    \chi^{(E)}_I = e^{(E)}_I - e^{(E)}_{I-1}
    \ \ (2\le I\le7),\notag\\
    \chi^{(E)}_8 &= 
    \half\bigl(e^{(E)}_8+e^{(E)}_1-\sum_{I=2}^7e^{(E)}_I\bigr),
  \end{align}
and
  \begin{align}
    \chi^{(E)}_9 &= e^{(E)}_9 + e^{(E)}_{10},
    \quad
    \chi^{(E)}_I = e^{(E)}_I - e^{(E)}_{I-1}
    \ \ (10\le I\le15),\notag\\
    \chi^{(E)}_{16} &= 
    \half\bigl(e^{(E)}_{16}+e^{(E)}_9-\sum_{I=10}^{15}e^{(E)}_I\bigr),
  \end{align}
\end{subequations}
where $e^{(E)}_I\cdot e^{(E)}_J = -\d_{IJ}$.  Next, define
$\xi^{(E)}{}_a$ analogously to $\xi^{(A)}{}_a$,
\begin{equation}
  \xi^{(E)}_a = (\chi^{(E)\,\a},\chi^{(E)}{}_\a,e^{(E)}{}_I),
\end{equation}
so that
\begin{equation}
  \xi^{(E)}{}_a\cdot \xi^{(E)}{}_b 
  = \eta_{ab}.
\end{equation}
Here, $\eta_{ab}$ is again as defined in Eq.~\eqref{eq:EtaDef}.  Then,
$[\xi^{(A)}{}_a] = W_a{}^b[\xi^{(E)}{}_b]$, where $W_a{}^{b}$
expressed in the new bases must be an $\SO(3,19)$ matrix preserving
$\eta_{ab}$,
\begin{equation}
  W\eta W^T = \eta.
\end{equation}
The task is to find a $W$ such that the $[\chi^{(E)}{}_a]$ are integer
classes in $H_2(\K3,\IZ)$.  

Since the $E_8\times E_8$ point in the dual IIA D6/O6 orientifold
cannot be obtained solely by displacement of D6-branes, we do not
expect a transformation of the form~\eqref{eq:Vofx}.  The solution is
as follows.  For each $I$, let $\tx^I{}_\a$ denote coordinates on the
$T^3$ dual to that of $x^{I\a}$, and define an $\SO(3,3+16)$ matrix
$\tV(\tx)$ by
\begin{equation}\label{eq:Vtildeofxtilde}
  \tV_a{}^b(\tx) =
  \begin{pmatrix}
    1       & 0 & 0\\
    x^T\tx & 1 & 2\tx^T\\
    \tx     & 0 & 1
  \end{pmatrix}.
\end{equation}

As shown in App.~\ref{app:Splittings}, a (co)homology basis for the
$(D_8)^2$ locus in moduli space, at which the K3 surface develops two
$D_8$ singularities, can be obtained two equivalent ways.  Starting
from the $D_{16}$ locus, we can act on the $D_{16}$ basis
$\chi^{(D)}{}_b$ with $V_a{}^b(y)$, where
\begin{equation}
  y^{I3} = 
  \bigl(0^8;\half^8\bigr),
  \quad
  y^{I1}=y^{I2} = 0.
\end{equation}
This has the dual IIA D6/O6 interpretation of displacing 8 of the 16
D6-branes from the fixed point at $(0,0,0)$ to the fixed point at
$(0,0,\half)$.  

Alternatively, starting from the $(E_8)^2$ locus in moduli space, at
which the K3 surface develops two $E_8$ orbifold singularities, we can
act on the $(E_8)^2$ basis $\chi^{(E)}{}_b$ with $\tV_a{}^b(\ty)$,
where
\begin{equation}
  \ty^I{}_3 =
  \bigl(0^7,1;0^7,1),
  \quad
  \ty^I{}_1=\ty^I{}_2 = 0.
\end{equation}
Combining these results, the $(D_8)^2$ basis is given by
\begin{equation}\label{eq:D8basis}
  \xi_a = V_a{}^b(y)\xi^{(D)}{}_b = \tV_a{}^b(\ty)\xi^{(E)}{}_b.
\end{equation}
This agrees with the discussion of Wilson lines in Eqs.~(11.6.18) and
(11.6.19) of Ref.~\cite{Polchinski:1998rr}.\footnote{The context in
  Ref.~\cite{Polchinski:1998rr} is the T-duality between the
  $\Spin(32)/\IZ_2$ and $E_8\times E_8$ heterotic strings compactified
  on a circle.  The fact that the $\Spin(32)/\IZ_2$ and $E_8\times
  E_8$ Wilson lines are defined on T-dual circles is the same reason
  that transformations~\eqref{eq:Vofx} and \eqref{eq:Vtildeofxtilde}
  depend on dual moduli $y^{I3}$ and $\ty^I{}_3$.}  We deduce that
\begin{equation}
  \xi^{(E)}{}_b = W_a{}^b\xi^{(D)}{}_b,
  \quad\text{for}\quad
  W_a{}^b = \tV_a{}^c(-\ty)V_c{}^b(y),
\end{equation}
where we have used the fact that the inverse of $\tV(\ty)$ is $\tV(-\ty)$.

The $(E_8)^2$ basis has a simple interpretation along lines analogous
to those at the end of Sec.~\ref{sec:Dbasis}.  The integer cohomology
lattice of K3 splits as $(-E_8)\oplus(-E_8)\oplus U_{3,3}{}'$, where
\begin{equation}
  \begin{split}
    U_{3,3}' &= \langle D^1_4,D^2_4,D_{34},f_1,f_2,f^3\rangle,\\
    (-E_8)\oplus(-E_8)&= \langle D^1_4,D^2_4,D_{34},f_1,f_2,f^3\rangle^\perp,
  \end{split}
\end{equation}
i.e., by reversing the role of ``upper 3'' and ``lower 3'' in the
$(-\Spin(32)/\IZ_2)\oplus U_{3,3}$ splitting described at the end of
the previous section.  Other realizations of the
$(E_8)\oplus(-E_8)\oplus U_{3,3}'$ splitting are obtained by reversing
the role of upper/lower $\a=1$ or 2 instead of the 3, or by trading a
$D^\a{}_4$ or $D_{\a4}$ for a $D^\a{}_2$ or $D_{\a2}$.


\subsection{Moduli space of hyperk\"ahler structure}
\label{sec:ModHkahler}

The cohomology classes $[\xi^{(D)}_a] =
(\xi^{(D)\,\a},\xi^{(D)}{}_\a,e^{(D)}_I) \in \O_2(\K3,\half\IZ)$ form
a basis for $H^2(\K3,\IR)$.  From Eq.~\eqref{eq:EtaDef}, the pairing
\begin{equation}
  \a\cdot\b = \int_{\K3} a\w\b
\end{equation}
gives a signature $(3,19)$ inner product on $H^2(\K3,\IR)$.  A choice
of hyperk\"ahler structure on K3 is equivalent to a choice of positive
signature 3-plane in $H^2(\K3,\IR)$.  This is the choice of three
orthogonal k\"ahler classes $[J^\a]$, $\a=1,2,3$ of positive norm,
modulo rescalings and $\SO(3)$ rotations $[J^\a]\mapsto O^\a{}_\b [J^\b]$.
The choice can be parametrized as follows.

Let
\begin{equation}
  [\o_a] = V_a{}^b [\chi_b],
\end{equation}
where $V_a{}^b \in \SO(3,19)$ satisfies $V\eta V^T = \eta$.  Up to a
left $\SO(3)\times \SO(19)$ rotation, an arbitrary $\SO(3,19)$ matrix can
be written in the form
\begin{align}
  V(E,\b,x) &= V(E)V(\b)V(x)\notag\\
  &= 
  \begin{pmatrix}
    E & 0       & 0\\
    0 & E^{-1T} & 0\\
    0 & 0       & 1
  \end{pmatrix}\!\!
  \begin{pmatrix}
    1 & -\b & -0\\
    0 &  1 & 0\\
    0 &  0 & 1
  \end{pmatrix}\!\!
  \begin{pmatrix}
    1 & x^Tx & 2x^T\\
    0 & 1    & 0\\
    0 & x    & 1
  \end{pmatrix}\notag\\
  &=
  \begin{pmatrix}
    E & -E C     & 2Ex^T\\
    0 & E^{-1T} & 0\\
    0 & x       & 1
  \end{pmatrix}.\label{eq:SOvielbein}
\end{align}
Here, $E^\a{}_\b\in \GL(3,\IR)$, $\b^{\a\b}$ is antisymmetric
$3\times3$ matrix, $x^{I\a}$ is an arbitrary $16\times3$ matrix, and
$C^{\a\b} = \b^{\a\b} - \d_{IJ}x^{I\a}x^{J\b}$.  In components,
\begin{equation}\label{eq:OmegaVsXi}
  \begin{split}
    [\o^\a] &= [E^\a{}_\b]
    \bigl([\xi^{(D)\,\a}] - C^{\a\b}[\xi^{(D)}{}_\b] + 2x^{J\a}[\xi^{(D)}_J]\bigr),\\
    [\o_\a] &= (E^{-1T})_\a{}^\b[\xi^{(D)}{}_\b],\\
    [\o_I] &= [\xi_I] + x^{I\b}[\xi^{(D)}{}_\b].
  \end{split}
\end{equation}
A corresponding triple of K\"ahler classes is then
\begin{equation}
  [J^\a] = \sqrt{\frac{V_{\K3}}{2}} [\o_\a + \o^\a],\quad \a=1,2,3,
\end{equation}
where we have chosen the prefactor so that
\begin{equation}
  V_{\K3} = \frac12\int_{\K3}J^\a\w J^\a
  \quad\text{(no sum)}
\end{equation}
is the volume of $\K3$, for $\a=1,2,3$.  Note that the $\o_a$ have the
same intersections as the $\xi^{(D)}_a$,
\begin{equation}
  \int_{\K3} \o_a\w\o_b = \eta_{ab},
\end{equation}
with $\eta$ defined in Eq.~\eqref{eq:EtaDef}.  So far, we have defined
the classes $[\o_a]$, but not the differential forms themselves.  For
later convenience, we define the $\o_\a$ to be the harmonic
representatives of these (co)homology classes.  Whenever we refer to the
$\o_a$ below, we will have in mind this definition.

The identification of the triple $[J^\a]$ under $\SO(3)$ rotation is
equivalent to the identification of $E^\a{}_\b$ under $\SO(3)$ left
multiplication.\footnote{Note that $O^{-1T} = O$, for $O\in
  \SO(3,\IR)$, so that $\o^\a$ and $\o_\a$ transform identically under
  $\SO(3)$.}  Therefore, the moduli space of hyperk\"ahler structure
of $K3$ is the space of volumes $V_{K3}$ and of $\SO(3,3+16)$ matrices
$V(\a,\b,E)$ modulo $\SO(3)$ rotations of $E$.  Interpreting $E$ as a
vielbein for the metric $G_{\a\b} = (E^TE)_{\a\b}$, the hyperk\"ahler
moduli space is that of $V_{K3}$, $G_{\a\b}$, $\b^{\a\b}$, and
$x^{I\a}$.  We can view $V(E,\b,x)$ as a vielbein for an arbitrary
symmetric $\SO(3,3+16)$ matrix $M^{ab}$ satisfying
\begin{equation}
  M^T\eta M = \eta^{-1}.
\end{equation}
by writing
\begin{equation}
  M=V^TK^{-1}V,
\end{equation}
where $K_{ab}$ is defined below in Eq.~\eqref{eq:Kmatrix}.  Explicitly,
we have
\begin{equation}\label{eq:MabUpper}
  M =
    \begin{pmatrix}
      \half G     & -\half GC                      & Gx^T\\
      -\half C^TG & \half G^{-1}+\half C^TGC+x^Tx  & -C^TGx^T + x^T\\
      xG          & -xGC + x                       & 2xGx^T +1
    \end{pmatrix}.
\end{equation}

The moduli space of hyperk\"ahler structure is
\begin{equation}
  \CM_\text{HK} = \IR_{>0}\times
  \bigl(\SO(3)\times \SO(19)\bigr)\backslash \SO(3,19)/\G_{3,19},
\end{equation}
where the first factor is the overall volume $V_{\K3}$ and the second
factor can be interpreted as the moduli space of hypercomplex
structure.  Here, $\G_{3,19}$ is the group of lattice automorphisms of
the K3 integer (co)homology lattice.  The natural coset metric is
\begin{equation}\label{eq:CosetMetricI}
    ds^2 
    = \frac{1}{8}\Tr(M^{-1}dM M^{-1} dM),
\end{equation}
where we have chosen the normalization factor for later convenience.
In terms of $G$, $\b$, and $x$, this becomes
\begin{equation}\label{eq:CosetMetricII}
    ds^2 = \frac14 G_{\a\g}G_{\b\d}
    \bigl(d G^{\a\b}d G^{\g\d} + \tilde d\b^{\a\b}\tilde d\b^{\g\d}\bigr) 
    + 2\d_{IJ}G_{\a\b}d x^{I\a}d x^{J\b},
\end{equation}
where
\begin{equation}\label{eq:dtildeb}
  \tilde d\b^{\a\b} = d\b^{\a\b} - x^{I\a}d x^{I\b}+x^{I\b}d
  x^{I\a}.
\end{equation}
From lattice isomorphisms, the $x^{I\a}$ and $\b^{\a\b}$ can be shown
to be periodic with period $1$.  The space $\CM_\text{HK}$ can be
viewed as a ``fibration over a fibration.''  The moduli space of 3D
metrics $G$ is the base manifold $\SO(3)\backslash \GL(3)$.  The
$3\times 16 = 48$ periodic $x^{I\a}$ parametrize a $\U(1)^{48}$
fibration over this base.  And finally, the 3 periodic $\b^{\a\b}$
parametrize a $\U(1)^3$ fibration over the result.  The quantities
$dG^{\a\b}$, $dx^{I\a}$, and $\tilde d\b^{\a\b}$ are global 1-forms on
$\CM_\text{HK}$, but the $d\b^{\a\b}$ are not, since they shifts under
$x^{I\a}\to x^{I\a}+1$.  The $\tilde d\b^{\a\b}$ are the global 1-form
in the $\U(1)^3$ fiber directions, with connections
\begin{equation}\label{eq:betaConnection}
  A^{\a\b}{}_{I\g}dx^{I\g} = -x^{I\a}dx^{I\b}+x^{I\b}dx^{I\a}.
\end{equation}
These connections are dual to the abelian part the Chern-Simons term
of
\begin{equation}
  \tilde H_\text{Het} = dB 
  -\frac{(2\pi\ell)^2}{4} \Tr_f\Bigl(A\w dA - \frac23 A\w A\w A\Bigr),
\end{equation}
in the duality between M-theory on $K3$ and the heterotic string on
$T^3$.  Here $\ell=\sqrt{\a'}$.  In Sec.~\ref{sec:K3MetDefHarm}, we
will see that Eq.~\eqref{eq:CosetMetricII} agrees with the metric on
the metric moduli space of K3.


\subsection{Hodge duality and harmonic forms}
\label{sec:HodgeDuality}

Given a metric on $K3$, we can define the Hodge star operator, and a
positive definite inner product on $H^2(\K3,\IR)$,
\begin{equation}  
  (\l_1,\l_2) = \int_\K3 \l_1\w\star\l_2.
\end{equation}

If a 2-form is selfdual or anti-selfdual, $\star\l = \pm\l$, then
$\l\cdot\l = \pm (\l,\l)$ respectively.  Consequently, there is a
decomposition
\begin{equation}\label{eq:Hplusminus}
  H^2(\K3,\IR) = \CH^+\oplus\CH^-,
\end{equation}
where $\CH^+$ ($\CH^-$) denotes the (anti)-selfdual subspace of
$H^2(\K3,\IR)$.  By acting with $\pi_\pm=\half(1\pm\star)$, we can
project any 2-form to its selfdual or anti-selfdual component.
However, $\star$ need not map closed forms to closed forms, so that
the result of applying $\pi_\pm$ to an arbitrary representative of a
cohomology class is not necessarily closed.  Implicit in
Eq.~\eqref{eq:Hplusminus}, is that we must use harmonic
representatives.  The projector $\pi_\pm$ indeed maps harmonic forms
to harmonic forms, and with this restriction maps closed forms to
closed forms.

It is straightforward to show that the triple of K\"ahler forms
satisfy $\star J^\a = J^\a$.\footnote{We have $J = \half J_{ab} dx^a\w
  dx^b$, where $J_{ab} = J_a{}^cg_{cb}$.  In complex coordinates,
  $J_j{}^k = i\d_j{}^k$ and $J_\jbar{}^\kbar = -i\d_\jbar{}^\kbar$.
  For a Hermitian metric, $ds^2 = g_{i\jbar}dz^i dz^\jbar + g_{\ibar
    j}dz^\ibar dz^j$, with $g_{i\jbar} = (g_{\ibar j})^*$.  Thus, $J =
  ig_{i\jbar} dz^i\w dz^\jbar$. Applying the standard definition of
  Hodge duality, $\star J = J$ follows.}  Thus,
\begin{equation}
  \begin{split}
    \CH^+ &= \langle\half[\o_\a+\o^\a]\rangle,\\
    \CH^- &= \langle\half[\o_\a+\o^\a]\rangle^\perp
    = \langle\half[\o_\a-\o^\a],[\o_I]\rangle,
\end{split}
\end{equation}
where angle brackets denote ``span of'' and $\perp$ denotes the
orthogonal complement with respect to either of the two inner
products.  It is useful to introduce the notation
\begin{equation}
  \o^+_\a = \half(\o_\a + \o^a),\quad 
  \o^-_\a = \half(\o_\a - \o^a),\quad
  \star\o^\pm_\a = \pm\o^\pm_\a.
\end{equation}
Equivalently,
\begin{equation}
  \star[\o^\a] = [\o_\a],\quad
  \star[\o_\a] = [\o^\a],\quad
  \star[\o_I] = -[\o_I],
\end{equation}
from which
\begin{equation}\label{eq:Kmatrix}
  \int_\K3 [\o_a]\w\star[\o_b] = K_{ab},
  \quad\text{with}\quad
  K_{ab} = 
  \begin{pmatrix}
    2 & 0 & 0\\
    0 & 2 & 0\\
    0 & 0 & 1
  \end{pmatrix}.
\end{equation}
When expressed in terms of the cohomology classes $[\xi^{(D)}{}_a]$,
the last equation becomes
\begin{equation}
 \int_{\K3} [\xi^{(D)}{}_a]\w\star[\xi^{(D)}{}_b] = M_{ab},
\end{equation}
where $M_{ab}$ is the inverse of the moduli matrix $M^{ab}$ of
Eq.~\eqref{eq:MabUpper}.


\subsection{K3 metric moduli space}
\label{sec:K3MetricMod}

In this final section of our review of the geometry of K3, we describe
the metric on the moduli space of K3 metrics.  We require that this
metric be invariant under diffeomorphisms $\K3\to\K3$.  To achieve the
diffeomorphism invariance, compensating vector fields are introduced
in Sec.~\ref{sec:DiffInvMet} to project generic metric deformations to
transverse traceless gauge.  Sec.~\ref{sec:K3MetDefHarm} relates the
transverse traceless deformations to harmonic forms, using the
hypercomplex structure to convert differential forms to symmetric
tensors.

Background material accompanying this chapter can be found in the
Appendices.  App.~\ref{app:Lichnerowicz} reviews the Lichnerowicz
operator, which acts on a metric deformation to give the deformation
of the Ricci tensor.  As explained in App.~\ref{app:Lichnerowicz}, for
a K\"ahler manifold, the complex structure relates the action of the
Laplace-de Rham operator on $(1,1)$-forms to the action of the
Lichnerowicz operator on symmetric tensors.  For a Calabi-Yau
$n$-fold, the holomorphic $(n,0)$-form achieves the same map for
$(1,2)$-forms.  App.~\ref{app:MetDefHarm} applies this observation to
the correspondence between harmonic forms and transverse traceless
metric deformations preserving the Ricci-flatness of K3.


\subsubsection{Diffeomorphism invariant metric on moduli space}
\label{sec:DiffInvMet}

Given a $d$-dimensional manifold $X$ and a family of metrics,
\begin{equation}
  ds^2 = g_{mn}(x;\m) dx^m dx^n,
\end{equation}
parametrized by a moduli space $\CM$ with coordinates $\m^i$, a
natural guess for the metric on moduli space is
\begin{align}
  &ds^2_\CM \stackrel{?}{=} \frac14\int_Xd^dx\sqrt{g}g^{mp}g^{nq}
  \d g_{mn}\d g_{pq}
  = K_{ij} \d\m^i\d\m^j,\notag\\
  &\text{where}\quad
  K_{ij} = \frac14\int_X d^dx\sqrt{g}g^{mp}g^{nq}\pd_ig_{mn}\pd_jg_{pq},
  \label{eq:NaiveModMetric}
\end{align}
with $\pd_ig_{mn} = \pd g_{mn}/\pd\m^i$.  However, this guess is not
correct, since we would like the moduli space metric to be purely
horizontal in the space of gauge orbits, i.e., diffeomorphism
invariant.  The expression~\eqref{eq:NaiveModMetric} is not
diffeomorphism invariant.

For simplicity, let us restrict to metrics on $X$ of fixed overall
volume $V_X$.  Then solution is as follows.  One finds that the
transverse traceless gauge of metric deformations is the distinguished
purely horizontal
gauge~\cite{Douglas:2008jx,Singer:1981xw,Manton:1981mp}.  For a
coordinate chart on $X$ such that the $\pd_ig_{mn}$ defined above are
transverse and traceless,
\begin{equation}
  \nabla^m \pd_ig_{mn} = 0,\quad g^{mn}\pd_ig_{mn} = 0,
\end{equation}
Eq.~\eqref{eq:NaiveModMetric} indeed gives the correct metric on
moduli space.  More generally, a metric deformation $\d g_{mn} =
\pd_ig_{mn}(x;\m)\d\m^i$ must be combined with a moduli-dependent
diffeomorphism
\begin{equation}\label{eq:NDiff}
  y^m = x^m - N_i^m(x;\m)\d\m^i
\end{equation}
to ensure that this is the case.  The quantity $N_i^m\d\mu^i$ is known
as the \emph{compensating vector field} on~$X$ (or simply, the
\emph{compensator}) corresponding to a change $\d\m^i$ in moduli.  If
we define the $\pd_ig_{mn}$ in the privileged coordinates $y^m$, and
then pullback to the original coordinates $x^m$ via
\begin{equation}
  g'_{mn}(x,\m) = g_{pq}(y(x),\m)
  \frac{\pd y^p}{\pd x^m}\frac{\pd y^q}{\pd x^n},
\end{equation}
we find that the corresponding transverse traceless metric deformation
$\d_i^\perp g_{mn} = \pd_ig'_{mn}(x;\m)$ is
\begin{equation}
  \begin{split}
    \d^\perp_ig_{mn}
    &= \pd_i g_{mn} -\CL_{N_i\d\m^i}g_{mn}\\
    &= \pd_i g_{mn} - \d\m^i\bigl(\nabla_mN_{in} + \nabla_nN_{im}\bigr).
  \end{split}
\end{equation}
Here, in the first equality, $\CL_{N_i\d\m^i}$ denotes the Lie
derivative with respect to the compensating vector field.  In the
second equality, $N_{im} = g_{mn}N_i^n$.  The diffeomorphism invariant
metric on moduli space is
\begin{align}
  &ds^2_\CM = \frac14\int_X d^dx\sqrt{g}g^{mp}g^{nq}
  \d^\perp g_{mn}\d^\perp g_{pq}
  = K^\perp_{ij} \d\m^i \d\m^j,\notag\\
  &\text{where}\quad
  K^\perp_{ij} = \frac14\int_X d^dx\sqrt{g}g^{mp}g^{nq}\d^\perp_ig_{mn}\d^\perp_jg_{pq}.
  \label{eq:ModMetric}
\end{align}
Compensators are a necessary feature in all but the simplest string
theory compactifications.  They modify the naive product metric ansatz
in such a way that the moduli kinetic terms come with the correct
horizontal moduli space
metric~\eqref{eq:ModMetric}~\cite{Douglas:2008jx,Singer:1981xw,Manton:1981mp}.


\subsubsection{K3 metric deformations and harmonic forms}
\label{sec:K3MetDefHarm}

The relation between harmonic forms and metric deformations for a
Calabi-Yau $n$-fold is reviewed in App.~\ref{app:MetDefHarm}.  The K3
case $n=2$ is special in that $(1,1)$-forms generate both K\"ahler and
complex structure deformations.  As we have already noted,
$H^2(\K3,\IR)$ has signature $(3,19)$ with respect to the inner
product $(\a,\b) = \int \a\w *\b$\@.  The hyperk\"ahler 2-forms
$J^\a$,
\begin{equation}
  J^1 = -\re\O,\quad J^2 = \im\O,
  \quad\text{and}\quad
  J^3 = J
\end{equation}
span the selfdual subspace $\CH^+$ of signature $(3,0)$.  Their
orthogonal complement in $H^2(\K3,\IR)$, of signature $(0,19)$ is the
anti-selfdual subspace $\CH^-$ of primitive
$(1,1)$-forms.\footnote{Recall that a cohomology class $[\o]$ is said
  to be \emph{primitive} when it is not of the form $[J]\w[\o']$ for
  some $\o'$.  The class $[J]$ itself is nonprimitive, since it is of
  the form $[J]\w[1]$.)}


\paragraph{Metric deformations.} We can think of the 58 metric
deformations of a K3 surface in at least two ways:

\begin{enumerate}


\item[i.] \emph{K\"ahler and complex deformations.}  There are 20
  (real) K\"ahler deformations and 19 (complex) complex structure
  deformations generated by $J$ together with the 19 primitive
  $(1,1)$-forms.  Since the complex structure deformation generated by
  $J$ leads to vanishing metric deformation, we have a total of $20 +
  2\times 19 = 58$ real metric deformations from
  \begin{equation}\label{eq:K3kahlerdef}
    h_{mn}(\CJ,\o) = -\frac12\left(\CJ_m{}^p\o_{pn} + \CJ_n{}^p\o_{pm}\right)
    \quad\text{(K\"ahler defs)}
  \end{equation}
  with $\o$ a $(1,1)$-form and 
  \begin{equation}\label{eq:K3cpxdef}
    h_{mn}(\O,\o) = -\frac12\left(\O_m{}^p\o_{pn}+\O_n{}^p\o_{pm}\right)
    \quad\text{(cpx.~str. defs)}
  \end{equation}
  with $\omega$ a primitive $(1,1)$-form.
  
  Note that from $h_{i\jbar} = -i\o_{i\jbar}$ in the K\"ahler case, we
  obtain a metric deformation $\d g_{i\jbar} \propto g_{i\jbar}$ for $\o
  = J$.  Thus the K\"ahler deformation associated to $J$ simply scales
  the overall volume of the K3 surface.  As a consequence of the
  discussion in App.~\ref{app:MetDefHarm}, all of the deformations
  \eqref{eq:K3kahlerdef} and \eqref{eq:K3cpxdef} are transverse, and all
  but the overall volume deformation are traceless.


\item[ii.] \emph{Hypercomplex and volume deformations.} There are
  $3\times 19$ (real) hypercomplex metric deformations generated by
  the 19 primitive $(1,1)$-forms $\o$,
  \begin{equation}\label{eq:hypercomplexdef}
    h_{mn}(\CJ^\a,\o) = -\frac12\left((\CJ^\a)_m{}^p\o_{pn} + (\CJ^\a)_n{}^p\o_{pm}\right),
  \end{equation}
  plus 1 overall volume deformation, for a total of $1 + 3\times19 = 58$
  metric deformations.

\end{enumerate}


\paragraph{Moduli space metric from harmonic forms.}

Let $\o_A$, for $A=1,\dots,19$, denote a basis for the space $\CH^-$ of
anti-selfdual harmonic forms on K3.  Then, a general volume-preserving
metric deformation in transverse traceless gauge can be written
\begin{equation}
  \d(ds^2) = \d\m_\a^A h(\CJ^\a,\o_A),
  \quad\text{where}\quad h = h_{mn}dx^m dx^n,
\end{equation}
in terms of $3\times19$ deformation parameters $\d\m^\a_A$.  It is
possible to show that
\begin{equation}\label{eq:ModMetricFromHarmonicI}
  \int_{\K3}d^4x \sqrt{g} h_{mn}(\CJ^\a,\o_A) h^{mn}(\CJ^\b,\o_B)
  = \d^{\a\b}\int_{\K3}\o_A\w\star\o_B.
\end{equation}
Therefore, in this parametrization, the moduli space metric is
\begin{equation}\label{eq:ModMetricFromHarmonicII}
  ds^2_\CM = \d^{\a\b}K_{AB}\d\m_\a^A \d\m_\b^B,
  \quad\text{for}\quad
  K_{AB} = \frac14\int_{\K3}\o_A\w\star\o_B.
\end{equation}

In the notation of Secs.~\ref{sec:ModHkahler} and
\ref{sec:HodgeDuality}, the space of anti-selfdual harmonic forms
$\CH^-$ is spanned by $\o^-_\a$ and $\o_I$.  Therefore, a basis of
transverse traceless metric deformations is
\begin{subequations}\label{eq:hbasis}
  \begin{align}
    (h^G)_{\a\b} &= \frac12 (E^{-1T})_\a{}^{\a'} (E^{-1T})_\b{}^{\b'} 
    h(\CJ^{\phantom{-}}_{(\a'|},\o^-_{|\b')}),\\
    (h^\b)_{\a\b} &= \frac12 (E^{-1T})_\a{}^{\a'} (E^{-1T})_\b{}^{\b'} 
    h(\CJ^{\phantom{-}}_{[\a'|},\o^-_{|\b']}),\\
    (h^x)_{I\a} &= (E^{-1T})_\a{}^{\a'}h(\CJ_{\a'},\o_I),
\end{align}
\end{subequations}
where $J_\a = \d_{\a\b}J^\b = J^\a$, and where for convenience below,
the factors of $E^{-1T}$ have been included to convert 3D frame
indices to coordinate indices.  Then, an arbitrary transverse
traceless metric deformation of K3 can be parametrized as
\begin{equation}
  \d(ds^2) = \d G^{\a\b}(h^G)_{a\b} + \tilde\d\b^{\a\b}(h^\b)_{a\b} 
  + \d x^{I\a} (h^x)_{I\a},
\end{equation}
where we define $\tilde\d\b^{\a\b}$ as in Eq.~\eqref{eq:dtildeb}.  In
this parametrization, results~\eqref{eq:ModMetricFromHarmonicI} and
\eqref{eq:ModMetricFromHarmonicII} give moduli space metric
\begin{equation}\label{eq:K3ExactModMetric}
  ds^2_\CM 
  = \frac14 G_{\a\a'}G_{\b\b'}\bigl(\d G^{\a\b}\d G^{\a'\b'} 
  + \tilde\d\b^{\a\b}\tilde\d\b^{\a'\b'}\bigr)
  + 2\d_{IJ}G_{\a\b}\d x^{I\a}\d x^{J\b},
\end{equation}
which is the $\SO(3,19)$ coset metric~\eqref{eq:CosetMetricII}.


\paragraph{Metric deformations generated by nonharmonic forms.}

As a final generalization, let $\o$ denote an anti-selfdual harmonic
2-form on $\K3$ and consider another representative $\o' = \o + d\l$
of the same cohomology class in $H^2(\K3,\IR)$.  Then, it is possible
to show that
\begin{equation}\label{eq:hshift}
  h_{mn}(\CJ^\a,\o') = h_{mn}(\CJ^\a,\o) - \nabla_mN_n - \nabla_nN_m,
\end{equation}
where $N_m = - (\CJ^\a)_m{}^n\l_n$, so that the exact part of $\o'$
generates a diffeomorphism.  Thus, general metric deformations,
\begin{equation}
  \d(ds^2) = \d G^{\a\b}(h^G)_{a\b} + \tilde\d\b^{\a\b}(h^\b)_{a\b} 
  + \d x^{I\a} (h^x)_{I\a}\\
  + \text{diffeomorphisms,}
\end{equation}
not necessarily transverse, are generated by cohomology
representatives that are not necessarily harmonic.


\section{K3 metric in the Gibbons-Hawking approximation}
\label{sec:K3MetricGH}

Having reviewed the geometry of K3, we now turn the explicit, but
approximate description of this geometry in terms of the $\IZ_2$
quotient of a metric of Gibbons-Hawking form.  The latter describes a
$U(1)$ principal bundle over $T^3$.  It is hyperk\"ahler and
Calabi-Yau, and positive definite away from neighborhoods of the $2^3$
$\IZ_2$ fixed points on $T^3$.  These neighborhoods becomes
arbitrarily small in the large hypercomplex structure limit of small
$U(1)$ fiber and large base.  The exact K3 metric differs from the
approximate one by replacing these neighborhoods with Atiyah-Hitchin
spaces, which smoothly excise the regions in which the metric becomes
negative.  The results of this section are as follows.

Let us write the metric on K3 as
\begin{equation}\label{eq:K3fromUnitMetric} 
  ds^2_{\K3} = (2V_{\K3})^{1/2} \bar G_{mn}dx^m dx^n,
\end{equation}
where $\bar G_{mn}$ is the metric for a ``unit'' K3 of volume $\half$,
obtained from the resolution of $T^4/\IZ_2$ for $T^4$ of volume~1.

From the duality between M-theory on K3 and type IIA on the
$T^3/\IZ_2$ orientifold, truncated to the tree-level type IIA
supergravity description, we obtain a first-order K3
metric~\eqref{eq:K3fromUnitMetric} of Gibbons-Hawking form
\begin{equation}\label{eq:K3UnitMetric}
\bar G_{mn} dx^m dx^n = 
\D^{-1} ZG_{\a\b} dx^\a dx^\b + \D Z^{-1}(dx^4+A)^2.
\end{equation}
This metric is derived in Sec.~\ref{sec:LiftToK3}, after a review of
Gibbons-Hawking multicenter metrics in Sec.~\ref{sec:GHmulticenter}
and M-theory/type IIA duality in Sec.~\ref{sec:MtypeIIA}.  Here,
coordinates $x^m$, for $m=1,2,3,4$, have periodicity~1, and are
subject to an additional $\IZ_2$ identification under the involution
\begin{equation}
  \CI_4\colon\ x^m\mapsto -x^m.
\end{equation}
This metric is that of a circle bundle over $T^3$, quotiented by
$\IZ_2$.

The quantity $G_{\a\b}$, for $\a,\b=1,2,3$, is a constant metric on
$T^3$ of volume $\D = \det^{1/2}G$.  The function $Z$ on $T^3$
satisfies the Poisson equation
\begin{equation}\label{eq:PoissonZ}
 -\nabla^2 Z = \sum_\text{sources $s$} Q_s \delta^3(\bx-\bx^s),
\end{equation}
where the index $s$ runs over (i) 16 points $\bx^I$ on $T^3$ with
$Q_I=1$, (ii) 16 image points $\bx^{I'} = -\bx^I$ with $Q_{I'} = 1$,
and (iii) the $2^3=8$ fixed points $\bx^{O_i}$ of $\CI_3\colon\
x^\a\to\-x^\a$ on $T^3$ with $Q_{O_i} = -4$.  The additive constant in
$Z$ is chosen so that when two $\bx^I$ (and their two images
$\bx^{I'}$) coincide with each $\bx^{O_i}$, we have $Z=1$, and the
metric~\eqref{eq:K3UnitMetric} becomes the orbifold metric on
$T^4/\IZ_2$.

The connection $A$ is defined by
\begin{equation}
  dA = \star_G dZ,
\end{equation}
where we choose a gauge condition $G^{\a\b}\pd_\a A_\b=0$.  This
determines $A$ only up to a closed 1-form.  We fix this ambiguity by
setting $A$ equal to an arbitrary constant 1-form $\b_\a dx^\a$ at the
$T^4/\IZ_2$ orbifold locus $\bx^I=0$, and
\begin{equation*}
  \d A = \d x^{I\b}\Bigl((F_I-F_I')_{\a\b}+\e_{\a\b\g}x^{I\g}\Bigr) dx^\a
  +\d\b_\a dx^\a
\end{equation*}
away from this locus.  Here, $F_I = dA_I$ and $F_{I'}=dA_{I'}$ are
defined in Eqs.~\eqref{eq:dAIdZI} and \eqref{eq:FIequalsdAI} of
Sec.~\ref{sec:LiftToK3}.

As $\bx$ approaches $x^I$, the $S^1$ fiber shrinks and the metric is
locally that of a smooth Taub-NUT space.  On the other hand, near the
fixed points $\bx^{O_i}$, the metric~\eqref{eq:K3UnitMetric} is
locally that of Taub-NUT space of \emph{negative} mass parameter.
This space is not itself well behaved, and is the large distance
approximation to a smooth Atiyah-Hitchin
space~\cite{Sen:1997kz,Atiyah:1985dv,Atiyah:1985fd,Atiyah:1988jp}.
The latter is obtained locally in the full nonperturbative lift of
type IIA string theory to \mbox{M-theory} \cite{Sen:1997kz}.  For
$\D\ll1$, the metric~\eqref{eq:K3UnitMetric} closely approximates the
exact K3 metric everywhere except in a small neighborhood of each
$\bx^{O_i}$, as explained in more detail in Secs.~\ref{sec:D6O6lift}
and \ref{sec:T3modZ2lift} below.

In Sec.~\ref{sec:HarmApprox}, we write down the hyperk\"ahler forms
and a basis of harmonic forms in this metric, showing that the basis
approximates that of Sec.~\ref{sec:ModHkahler} and has the same inner
product.  Therefore, the moduli space of hyperk\"ahler structure is
identical to that of Sec.~\ref{sec:ModHkahler}.

Finally we turn to the metric moduli space in
Sec.~\ref{sec:MetricModGH}.  To describe this space, it is useful to
write $\b^{\a\b} = \e^{\a\b\g}\b_\g$.  Then, the approximate metric
depends on parameters $G_{\a\b}$, $\b^{\a\b}$, $\bx^I$, and the
overall volume modulus $V_{\K3}$.  It is clear that these parameters
should \emph{determine} the moduli as defined in
Secs.~\ref{sec:ModHkahler} and \ref{sec:K3MetDefHarm},\footnote{The
  choice of metric defines Hodge duality and harmonicity of
  differential forms.} and we have suggestively given them the same
names.  In Sec.~\ref{sec:Method1}, we show that the metric
deformations from small changes in the quantities $G_{\a\b}$,
$\b^{\a\b}$, $\bx^I$ parametrizing the approximate metric precisely
agree with the metric deformations generated by the harmonic forms of
Sec.~\ref{sec:HarmApprox}, in the manner described in
Sec.~\ref{sec:K3MetDefHarm}.  It follows from this and the results of
the previous paragraph that the moduli space metric of the approximate
K3 metric is the same $\IR_{>0}\times \bigl(\SO(3)\times
\SO(19)\bigr)\backslash \SO(3,19)$ coset metric of the exact
discussion in Sec.~\ref{sec:ReviewK3}.

As discussed in Sec.~\ref{sec:DiffInvMet}, we require that metric on
metric moduli space be invariant under diffeomorphisms $\K3\to\K3$.
To achieve the diffeomorphism invariance, compensating vector fields
were introduced in Sec.~\ref{sec:DiffInvMet} to project generic metric
deformations to transverse traceless gauge.  In
Sec.~\ref{sec:Method2}, we consider the moduli space metric from naive
dimensional reduction of the $D$-dimensional Einstein-Hilbert action
on a $d$-dimensional manifold $X$, and ask when this gives the correct
result, without compensators.  This metric differs from the previous
one in two ways: there is no projection of metric deformations to
their transverse traceless part, and there are additional terms in the
metric obtained by integrating $(\Gbar^{mn}\d G_{mn})^2$.  We find
that Gibbons-Hawking metrics are special in that their naive moduli
space metrics precisely agree with the diffeomorphism invariant moduli
space metrics of Sec.~\ref{sec:DiffInvMet}.  Therefore compensators
are not necessary, and one can equivalently use the naive metrics from
dimensional reduction.  We show explicitly that the naive moduli space
metric for K3 in the Gibbons-Hawking approximation exactly reproduces
the $\IR_{>0}\times \bigl(\SO(3)\times \SO(19)\bigr)\backslash
\SO(3,19)$ coset metric of Sec.~\ref{sec:ReviewK3}.


\subsection{Gibbons-Hawking multicenter metrics}
\label{sec:GHmulticenter}

Gibbons-Hawking multicenter
metrics~\cite{Gibbons:1979zt,Eguchi:1978gw} are gravitational
multi-instanton solutions to general relativity discovered in the late
1970s.  Here, the word \emph{instanton} indicates that they are
solutions to the 4D Euclidean (rather than Lorentzian) vacuum Einstein
equations $R_{mn}=0$.  They have selfdual curvature 2-forms, and in
this sense are analogs of Yang-Mills instantons.  As discussed in
Sec.~\ref{sec:HkahlerMan}, they are noncompact hyperk\"ahler
4-manifolds, and they are distinguished by a $\U(1)$ isometry.  The
Gibbons-Hawking metric take the form
\begin{subequations}
  \begin{equation}
    ds^2 = Z^{-1}(d\psi + \bomega\cdot d\bx)^2 + Z d\bx\cdot d\bx,
  \end{equation}
  where $\phi\cong\phi+8\pi M$, and  
  \begin{equation}
    \grad Z = \grad\times\bomega
    \quad\text{and}\quad
    Z = \e + 2M\sum_{I=1}^N \frac1{\bigl|\bx-\bx^I\bigr|}.
  \end{equation}
\end{subequations}
Here, $\e$ takes the values 0 or 1, and we will refer to the quantity
$M>0$ as the mass parameter.

For $\e=0$, the metric describes flat $\IR^4/\IZ_N$ at large $x$ and
the space is asymptotically locally Euclidean (ALE)\@.  As such, it is
a model for the resolution/deformation of the singular space
$\IR^4/\IZ_N\cong \IH^1/\IZ_N$.  Generalizations exist for the other
ADE discrete subgroups $\G\subset\Sp(1)$.  The case $N=2$ gives the
Eguchi-Hanson space~\cite{Eguchi:1978xp}.  Since $\IZ_1$ is trivial,
the case $N=1$ should be asymptotic to $\IR^4$, and is in fact
globally $\IR^4$.

For $\e=1$, we have the multi-Taub-NUT metric~\cite{Hawking:1976jb}.
The metric describes flat $\IR^3\times S^1$ at large $x$ and the space
is asymptotically locally flat (ALF)\@.  This is the case relevant to
the applications below.

It is useful to give two other presentations of the Gibbons-Hawking
metric.  Let $R_4 = 8\pi M$ denote the length of the $\psi$ circle,
and define $x^4 = \psi/R^4$, so that $x\cong x+1$ has unit period.
Then, the connection 1-form becomes $A = \bomega\cdot d\bx/R_4$, and
we have
\begin{subequations}
  \begin{equation}
    ds^2 = R_4{}^2 Z^{-1}(dx^4+A)^2 + Z g_{\a\b}dx^\a dx^\b,
  \end{equation}
  where
  \begin{equation}
    dA = R_4{}^{-1}\star_{\displaystyle g}\,dZ, \quad
    Z = \e + \frac{R_4}{4\pi}\sum_{I=1}^N \frac1{\bigl|\bx-\bx^I\bigr|}.
  \end{equation}
\end{subequations}
Here $\star_{\displaystyle g}$ denotes the Hodge star operator in the
metric $g_{\a\b}$.  Although we have derived this expression for the
special case $g_{\a\b}=\d_{\a\b}$, it remains valid for arbitrary
constant $\IR^3$ metric, provided we understand $\bigl|\bx-\bx^I\bigr|$ to be
the length in this metric.

Finally, define a 4D volume modulus $V = \sqrt{g}R_4$ and a ``unit''
4D metric $G_{\a\b} = (R_4/\sqrt{g})g_{\a\b}$.  In terms of these
variables and $\D = \sqrt{G}$, we have
\begin{subequations}
  \begin{equation}\label{eq:GHmetric3rdform}
    ds^2 = \sqrt{V}\bigl(\D Z^{-1}(dx^4+A)^2 + \D^{-1}Z G_{\a\b} dx^\a dx^\b\bigr),
  \end{equation}
  where
  \begin{equation}\label{eq:ZA3rdform}
    dA = \star_G\,dZ,
    \quad
    Z = \e + \frac{\D}{4\pi}\sum_{I=1}^N \frac1{\bigl|\bx-\bx^I\bigr|},
  \end{equation}
\end{subequations}
and $\bigl|\bx-\bx^I\bigr|$ is now the length in the metric $G_{\a\b}$.


\subsection{M-theory/type IIA duality}
\label{sec:MtypeIIA}

The duality between M-theory and type IIA string theory is an exact
equivalence, a truncation of which relates the classical 11D and 10D
type IIA supergravity actions.  In our conventions, the bosonic terms
in these supergravity actions are given by
\begin{equation}
  \frac{(2\pi\ell)^9}{2\pi}S_{11} = 
  \int dV_{11}
  \biggl(R^{(11)}-\frac12\frac1{4!} F_{(4)}^2\biggr)
  -\frac16\int A_{(3)}\w F_{(4)}\w F_{(4)},
\end{equation}
and
\begin{multline}
  \frac{(2\pi\ell)^8}{2\pi}S_\text{10} = 
  \int dV_{10}\,
  e^{-2\Phi}\biggl(R^{(10)} + 4 (\pd\Phi)^2
  -\frac12\frac1{3!} H^2\biggr)\\
  -\int dV_{10}\biggl(
  \frac12\frac1{2!}F_{(2)}^2 +\frac12\frac1{4!}\tilde F_{(4)}^2 
  \biggr)
  -\frac12\int B\w F_{(4)}\w F_{(4)},
\end{multline}
where $dV_n = d^n x(-G_{(n)})^{1/2}$.  In these conventions,
$\ell$ is the 11D Planck length \emph{and} the 10D string length
$\sqrt{\a'}$, and the square of a differential form is defined by
contracting all indices with the metric.  Here, field strengths are
related to $p$-form potentials via
\begin{subequations}
  \begin{align}
    &F_{(4)} = dA_{(3)},\\
    &H = dB,\ F_{(2)} = dC_{(1)},\ F_{(4)} = dC_{(3)} - C_{(1)}\w H,
  \end{align}
\end{subequations}
in 11D and 10D, respectively.  If we assume a $\U(1)$ isometry in 11D,
the 11D supergravity action reduces to the 10D type IIA supergravity
action.  The identification between 11D and 10D fields is as follows:
\begin{equation}\label{eq:LiftIdentification}
  \begin{gathered}
    ds_\text{11}^2 = e^{-2\Phi/3}ds^2_\text{IIA} + (2\pi\ell)^2e^{4\Phi/3}(dy+A)^2,\\
    A = \frac1{2\pi\ell}C_{(1)},\quad
    A_{(3)} = C_{(3)}+ (2\pi\ell) dy\w B,
  \end{gathered}
\end{equation}
with coordinate periodicity $y\cong y+1$.

In the presence of a D$p$-brane, the 10D type IIA supergravity action
has an additional source term,
\begin{equation}
  S_\text{source} = \frac{2\pi}{(2\pi\ell)^{p+1}}\int C_{(p+1)}.
\end{equation}


\subsection{The lift of the type IIA $T^3/\IZ_2$ orientifold to
  M-theory on K3}
\label{sec:LiftToK3}

In this section, we derive an approximate K3
metric~\eqref{eq:K3UnitMetric} of Gibbons-Hawking form from the
M-theory lift of the tree-level supergravity description of the type
IIA $T^3/\IZ_2$ orientifold.  We do this in three steps, considering
the M-theory lift of a collection of $N$ D6-branes transverse to
$\IR^3$ in Sec.~\ref{sec:D6lift}, then adding an O6-plane in
Sec.~\ref{sec:D6O6lift}, and then finally compactifying the transverse
$\IR^3$ to $T^3$ and lifting the 16 D6-branes and 8 O6-planes of
$T^3/\IZ_2$ in Sec.~\ref{sec:T3modZ2lift}.  The discussion closely
follows that in Ref.~\cite{Schulz:2004tt} by one of the authors, which
in turn relies heavily on Ref.~\cite{Sen:1997kz}.


\subsubsection{M-theory lift of a collection of $N$ D6-branes}
\label{sec:D6lift}

The type IIA supergravity solution corresponding to $N$ D6 branes
located at transverse locations $\bx^I$ on a space of topology
$\IR^{6,1}\times \IR^3$ with arbitrary constant product metric 
\begin{equation}\label{eq:WithoutD6s}
  ds^2 = 
  g_{(7)\m\n} dy^\m dy^\n + g_{(3)\a\b} dx^\a dx^\b
\end{equation}
is
\begin{subequations}\label{eq:WithD6s}
  \begin{gather}
    ds^2_{10} = 
    Z^{-1/2}g_{(7)\m\n} dy^\m dy^\n + Z^{1/2} g_{(3)\a\b} dx^\a dx^\b,
    \label{eq:WithD6sMetric}\\
    e^\Phi= e^{4\phi}
    \left(\frac{\sqrt{\vphantom{1}g_{(3)}}}{(2\pi\ell)^3}\right)^{3/2}Z^{-3/4},
    \quad
    F_{(2)} = \star_3 dZ,
    \label{eq:WithD6sDilF}
  \end{gather}
\end{subequations}
where $\m,\n=0,\dots,6$ and $\a,\b=1,2,3$.  Here, $\phi$ is an
integration constant\footnote{In the compact setting of
  Sec.~\ref{sec:T3modZ2lift}, $\phi$ can be identified with the 7D
  dilaton.  The effective field theory is discussed in
  Ref.~\cite{SchulzTammaroA}.} and $Z(\bx)$ satisfies the 3D Poisson
equation
\begin{equation}
  -\nabla_{(3)}{}^2 Z =
  2\pi\ell\sum_{I=1}^N\frac{\d^3(\bx-\bx^I)}{\sqrt{\vphantom{1}g_{(3)}}}.
\end{equation}

Carrying out the lift to an 11D solution via the
identification~\eqref{eq:LiftIdentification}, and defining
\begin{equation}
  G_{\a\b} = e^\Phi
  \frac{(2\pi\ell)Z^{1/2}g_{(3)\a\b}}{\det^{1/2}(Z^{1/2}g_{(3)})}
  \quad\text{and}\quad
  \D = {\det}^{1/2}G,
\end{equation}
we obtain a product metric
\begin{equation}\label{eq:ND6lift}
  ds^2_{11} = ds^2_7 + V^{1/2}d{\bar s}^2_4,\quad
  V=(2\pi\ell)^4e^{-4\phi/3},
\end{equation}
where $ds^2_7$ is a constant metric on $\IR^{6,1}$ and $d{\bar s}_4^2$
is a multicenter metric
\begin{equation}
    d{\bar s}^2_4 = \bar G_{mn}dx^m dx^n
    = \D^{-1}ZG_{\a\b}dx^a dx^b + \D Z^{-1}(dx^4 + A)^2
\end{equation}
of Gibbons-Hawking form.  Note that $\phi$ gives the overall 4D volume
modulus and $\det{\bar G} = Z$.  In the metric $G_{\a\b}$, we have
\begin{equation}
  dA = \star_G dZ
  \quad\text{and}\quad
  -\nabla^2 Z = \sum_{I=1}^N\d^3(\bx-\bx^I),
\end{equation}
with solution
\begin{equation}
  Z = 1 + \sum_{I=1}^N Z_I,
  \quad\text{where}\quad
  Z_I = \frac{\D}{4\pi\bigl|\bx-\bx^I\bigr|}.
\end{equation}
Here, $\bigl|\bx-\bx^I\bigr|$ is the distance computed in the metric
$G_{\a\b}$.  Here, we have chosen an integration constant of 1 in the
definition of $Z$, for agreement between the metrics
\eqref{eq:WithoutD6s} and \eqref{eq:WithD6s} at large $x$, far from
the D6-branes.  Therefore, the metric is a multi-Taub-NUT
metric~\cite{Hawking:1976jb}.

The $L^2$ harmonic forms in this metric are
known~\cite{Ruback:1986ag}.  They are
\begin{align}
  \o_I &=  \Bigl(\frac{Z_I}{Z}\Bigr)_{,\,\a} 
  \Bigl(dx^\a\w(dx^4+A)-
  \frac{Z}2 G^{\a\a'}\e_{\a'\b\g}dx^\b\w dx^\g\Bigr)\notag\\
  &= -d\Bigl(A_I - \frac{Z_I}{Z}(dx^4+A)\Bigr),
\end{align}
satisfying
\begin{equation}
  \int \o_I\w \o_J = -\d_{IJ},
\end{equation}
as reviewed in App.~\ref{app:Integrals}.  Here, $A_I$ is the magnetic
dual of $Z_I$ in the metric $G_{\a\b}$,
\begin{equation}\label{eq:dAIdZI}
  dA_I = \star dZ_I.
\end{equation}

The compact homology of this space is generated by 2-spheres $S_{IJ}$,
where $S_{IJ}$ is swept out by the expanding and shrinking
$x^4$-circle fibration over the line segment from $x^I$ to $x^J$.
Since $Z^{-1}\to0$ at the source locations $x^I$, the sphere smoothly
caps off at the endpoints.  In the lift from type IIA to M-theory, a
fundamental string stretched between the $I$th and $J$th D6-brane
becomes an M-theory membrane wrapped on this sphere.  The intersection
pairing on $H_2$ is minus the Cartan matrix of $A_{N-1}$.  To identify
the compact homology lattice with the root lattice of $A_{N-1}$ in
terms of its standard orthonormal basis, we can associate the $I$th
D6-brane, or $I$th center in M-theory, with a unit vector $e_I$, and
the string stretched from $I$th to $J$th D6-brane, or its M-theory
lift $S_{IJ}$, with the root $e_J-e_I$~\cite{Sen:1997kz}.


\subsubsection{M-theory lift of an O6-plane and a collection of $N$
  D6-branes}
\label{sec:D6O6lift}

If we instead begin with $N$ D6-branes and an O6-plane in type IIA,
the story is very similar.  A new feature is the $\IZ_2$ orientifold
involution $\O(-1)^{F_L}\CI_3$, where $\O$ is worldsheet orientation
reversal, $(-1)^{F_L}$ is left-moving fermion parity,\footnote{The
  factor $(-1)^{F_L}$ is needed to ensure a supersymmetric spectrum.
  For the supergravity fields, it acts as $-1$ on left-moving Ramond
  sector states and $+1$ on left-moving Neveu-Schwarz sector states.}
and $\CI_3\colon\ x^\a\mapsto -x^\a$, for $\a=1,2,3$, is inversion of
$\IR^3$.  This truncates the type IIA supergravity action to the
fields
\begin{equation}
  G_{\m\n},\ G_{\a\b},\ B_{\a\m},\ \Phi,\ C_{(1)\a},\ C_{(3)\a\b\m}.
\end{equation}

In the lift to 11D supergravity, the orientifold $\IZ_2$ becomes an
orbifold $\IZ_2$ acting as $\CI_4\colon\ x^m\mapsto -x^m$ for
$m=1,2,3,4$.  Correspondingly, we define the modulus $V$ of
Eq.~\eqref{eq:ND6lift} with a factor of 2,
\begin{equation}\label{eq:ND6O6lift}
  ds^2_{11} = ds^2_7 + (2V)^{1/2}d{\bar s}^2_4,\quad
  2V=(2\pi\ell)^4e^{-4\phi/3},
\end{equation}
Aside from the $\IZ_2$ identification of coordinates, the only aspect
of the previous discussion that is modified is the definition of $Z$.
The Poisson equation for $Z$ becomes
\begin{equation}
  -\nabla^2 Z = \sum_\text{sources $s$}Q_s\,\d^3(\bx-\bx^s),
\end{equation}
where $s$ runs over $I=1,\dots,N$, $I'=1',\dots,N$, and $O$, with (i)
$Q_I = 1$ from a D6-brane source at $\bx^I$ in type IIA; (ii) $Q_{I'}
= 1$ from an image D6-brane source at $\bx^{I'} = -\bx^I$ in type IIA;
(iii) $Q_O = -4$ from an O6-plane at the $\CI_3$ fixed point at the
origin of $\IR^3$ in type IIA.\footnote{An O6-plane exactly cancels
  the RR charge of 2 D6-branes and 2 image D6-branes.}  The solution
is
\begin{equation}
  Z = 1 +Z_O + \sum_{I=1}^N(Z_I+Z_{I'}),
\end{equation}
where
\begin{equation*}
  Z_I = \frac{\D}{4\pi\bigl|\bx-\bx^I\bigr|},\quad
  Z_{I'} = \frac{\D}{4\pi\bigl|\bx+\bx^I\bigr|},\quad
  Z_O  = -4 \frac{\D}{4\pi|\bx|}. 
\end{equation*}

The $L^2$ harmonic forms are 
\begin{align}
  \o_I &=  \Bigl(\frac{Z_I-Z_{I'}}{Z}\Bigr)_{,\,\a} 
  \Bigl(dx^\a\w(dx^4+A)
  -\frac{Z}2 G^{\a\a'}\e_{\a'\b\g}dx^\b\w dx^\g\Bigr)\notag\\
  &= -d\Bigl((A_I-A_{I'}) - \frac{(Z_I-Z_{I'})}{Z}(dx^4+A)\Bigr),\label{eq:L2forms}
\end{align}
satisfying
\begin{equation}
  \int \o_I\w\o_{J} = -\half(\d_{IJ}+\d_{I'J'}) = -\d_{IJ},
\end{equation}
where the $\half$ is due to the $\IZ_2$ quotient.  It is convenient to
define
\begin{equation}\label{eq:FIequalsdAI}
  F_I = dA_{I}
  \quad\text{and}\quad
  F_{I'} = dA_{I'}.
\end{equation}
Then, in terms of cohomology classes with compact support, $[\o_I] =
-[F_I-F_{I'}]$.

Whereas the lift of a collection of D6 branes gives a smooth positive
definite multi-Taub-NUT metric, the lift of the tree-level type IIA
supergravity description of an O6-plane gives a singular metric that
changes sign near $x=0$.  The interpretation is as follows.  The field
identification of Eq.~\eqref{eq:LiftIdentification} assume a $\U(1)$
isometry of the M-theory circle.  That is because higher Fourier modes
around the M-theory circle correspond to states with D0-charge in type
IIA, and these are not included in the classical supergravity
action~\cite{Schulz:2004tt}.

Recall that the mass of a D0-brane is proportional to $e^{-\Phi}$, and
for the tree-level D6/O6 solution we have just described,
$e^{\Phi}\propto Z^{-3/4}$.  Since a D6 brane has positive Taub-NUT
mass parameter, we see that $Z\to\infty$ and D0-branes become
infinitely heavy near a D6 brane.  Thus, they do not affect the local
description of a D6 brane, and the lift of a D6-brane to M-theory
truly has an isometry around the M-theory circle.

On the other hand, the tree-level supergravity description of an
O6-plane gives $-4$ times the Taub-NUT mass parameter of a D6-brane in
the lift to M-theory.  Thus, as we approach an O6-plane, $Z$ decreases
to 0 at finite distance, and then becomes negative in a region around
the O6 plane.  Correspondingly, D0-branes and their bound states
become light as they approach this region, and need to be included in
the low energy type IIA supergravity theory~\cite{Schulz:2004tt}.  The
exact M-theory lift does \emph{not} have an isometry around the
M-theory circle, a fact suggested already by the presence of the
$\IZ_2$ identification, which breaks this isometry at fixed points.
The negative mass Taub-NUT space, together with the $\IZ_2$
identification $(\bx,x^4)\cong -(\bx,x^4)$, defines the large distance
approximation to an Atiyah-Hitchin
space~\cite{Sen:1997kz,Atiyah:1985dv,Atiyah:1985fd,Atiyah:1988jp}. It
is singular at small $\bx$, however this just reflects the fact that
we have discarded all higher Fourier modes around the M-theory circle.
The complete Atiyah-Hitchin geometry is smooth, and excises the $Z<0$
regions.

The compact homology of the space obtained from the exact M-theory
lift is similar to that of the previous section.  We again have
2-spheres $S_{IJ}$ arising from the M-theory lift of a string
stretched from the $I$th to $J$th D6-brane.  Equivalently, these are
the $\IZ_2$-invariant combinations $S_{IJ}$ - $S_{I'J'}$ on the
$\IZ_2$ covering space.  However, we also obtain new cycles from
$S_{II'} + S_{JJ'}$ on the covering space.  The latter arise from
pairs of strings connected between D6-branes and their $\IZ_2$ images
on the covering space, or equivalently, between pairs of D6-branes and
an O6-plane on the quotient space.  In the compact homology lattice,
the new cycles add roots $e_I+e_J$ to the previous lattice spanned by
$e_J-e_I$, enhancing the $A_{N-1}$ lattice to $D_N$.  The intersection
pairing on $H_2$ is minus the Cartan matrix of
$D_N$~\cite{Sen:1997kz}.


\subsubsection{M-theory lift of $T^3/\IZ_2$}
\label{sec:T3modZ2lift}

The $T^3/\IZ_2$ orientifold is the analog of the previous section with
$y^\a$ valued on $T^3$ instead of $\IR^3$.  It is the background
T-dual to type I on $T^3$, via dualization of all three $T^3$
directions. This gives $N=16$ pairs of D6-branes and image D6-branes
and $2^3=8$ O6-planes located at the fixed points of $\CI_3$ on $T^3$.
Choosing coordinate periodicity $x^\a\cong x^\a+1$, the fixed points
are $\bx^{O_i}$, for $i=1,\dots,8$, are the $2^3$ points on $T^3$
where $x^\a = 0$ or $\half$ for each $\a=1,2,3$.  This gives the
metric described at the beginning of Sec.~\ref{sec:K3MetricGH}:
\begin{equation*}
  ds^2_{\K3} = (2V_{\K3})^{1/2} \bar G_{mn}dx^m dx^n,
\end{equation*}
where $\bar G_{mn}$ is an approximate metric of Gibbons-Hawking form
for a ``unit'' K3 of volume~$\half$, obtained from the resolution of
$T^4/\IZ_2$ for $T^4$ of volume~1,
\begin{equation*}
\bar G_{mn} dx^m dx^n = 
\D^{-1} ZG_{\a\b} dx^\a dx^\b + \D Z^{-1}(dx^4+A)^2.
\end{equation*}

The solution to the Poisson equation~\eqref{eq:PoissonZ} for $Z$ can
be expressed formally as
\begin{equation}\label{eq:ZforK3}
  Z = 1 + \sum_{i=1}^{8} Z_{O_i} + \sum_{I=1}^{16}(Z_I+Z_{I'})
\end{equation}
where $Z_s = Q_s K(\bx,\bx^s)$.  Here, the Green's function
$K(\bx,\bx')$ satisfies
\begin{subequations}
  \begin{equation}
    -\nabla^2 K(\bx,\bx') = \d^3(\bx-\bx') - 1,
  \end{equation}
  where the delta function is a periodic delta function on $T^3$, and
  where the constant of integration is fixed by requiring
  \begin{equation}
    \int_{T^3}d^3x K(\bx,\bx')=0.
  \end{equation}
\end{subequations}
The leading constant of unity in Eq.~\eqref{eq:ZforK3} ensures that
$Z$ drops out of the metric in the limit that two $\bx^I,\bx^{I'}$
pairs coincide with each $\bx^{O_i}$ and the sources locally cancel.
This is the $T^4/\IZ_2$ orbifold limit of K3.  From $\int_{T^3} d^3x Z
= 1$, the metric $\bar G_{mn}$ indeed gives a ``unit'' K3 metric of
volume $\int_{\K3}d^4x Z = \half\int_{T^4}d^4x Z =\half$, whose double
cover is a $T^4$ of volume 1.

In this compact setting, it is meaningful to define a lower
dimensional dilaton after reducing to 7D.  The parameter $\phi$ of
Eq.~\eqref{eq:WithD6sDilF} can be identified with the 7D dilaton from
the IIA point of view~\cite{SchulzTammaroA}.  From
Eq.~\eqref{eq:ND6O6lift}, this is the same as the overall K3 volume in
11D,
\begin{equation}
  e^{-4\phi/3} = \frac{2V_{\K3}}{(2\pi\ell)^4}.
\end{equation}

This approximate K3 metric suffers from the same pathology described
at the end of the previous section.  Near $\bx=\bx^I$, the local
geometry is that of a smooth Taub-NUT space.  However, in a region
around the $\IZ_2$ fixed points $\bx^{O_i}$, the metric gives a
negative mass Taub-NUT space which is the large distance approximation
to the Atiyah-Hitchin
space~\cite{Sen:1997kz,Atiyah:1985dv,Atiyah:1985fd,Atiyah:1988jp},
obtained in the exact M-theory lift.  In the limit of small
$x^4$-circle, $\D=\sqrt{G}\to0$, the pathological regions become
smaller and smaller, and the metric~\eqref{eq:K3UnitMetric} well
approximates the exact K3 metric except in an arbitrarily small
neighborhood of each fixed point $\bx^{O_i}$.

The homology lattice is spanned by the cycles inherited from
$T^4/\IZ_2$, i.e., the 2-tori $f_\a,f^\a$ and 2-spheres $D_{\a
  s},D^\a_s$ of Sec.~\ref{sec:Homology}, together with the $D_{16}$
lattice described at the end of the previous section, from the lift of
the D6-branes and any one of the $O_i$.  Focusing on the $O_i$ at the
origin, the arguments of Sec.~\ref{sec:Dbasis} and
App.~\ref{app:Splittings} show that the $D_{16}$ lattice of the
previous section is enlarged to the weight lattice of
$\Spin(32)/\IZ_2$ and that the homology lattice splits as the sum of
this weight lattice and a signature $(3,3)$ lattice $\langle
D^\a_4,f_\a\rangle$.


\subsection{Harmonic forms in the approximate K3 metric}
\label{sec:HarmApprox}

In this section, we describe the coframe, hyperk\"ahler forms, and a
basis of harmonic forms of K3 in the approximate
metric~\eqref{eq:K3fromUnitMetric}.  We show that this basis of
harmonic forms approximates the basis of Sec.~\ref{sec:ModHkahler} in
the sense that the cohomology classes agree, even though the harmonic
representatives of these classes obviously differ in the approximate
and exact metrics.  As already noted, in the large hypercomplex
structure limit $\D\ll1$, the metric~\eqref{eq:K3UnitMetric} closely
approximates the exact K3 metric everywhere except in a small
neighborhood of each $\bx^{O_i}$.  In the same limit, we expect the
harmonic forms of this section to closely approximate the exact
harmonic forms.


\paragraph{Frame.}

Let $E^\a{}_\b$ be a vielbein for the 3D metric $G_{\a\b}$ appearing
in the approximate K3 metric~\eqref{eq:K3UnitMetric}.  Then,
\begin{equation*}
  d{\bar s}^2 = \d_{mn}\bar\th^m\bar\th^n,
  \quad\text{with volume form}\quad
  \bar\th^1\w\bar\th^2\w\bar\th^3\w\bar\th^4.
\end{equation*}
where $\bar\th^m$ is the coframe
\begin{equation}
  \begin{split}
    \bar\th^\a &= \D^{-1/2}Z^{1/2}(E^{-1})^\a{}_\b dx^\b,\\
    \bar\th^4 &= \D^{1/2}Z^{-1/2}(dy^4+A).
  \end{split}
\end{equation}


\paragraph{Hyperk\"ahler forms.}

In terms of this coframe, the metric~\eqref{eq:K3fromUnitMetric}
admits a triple of selfdual harmonic forms
\begin{equation}
  J^\a = \sqrt{\frac{V_{\K3}}2}\Bigl(
  2\bar\th^\a\w\bar\th^4 + \d^{\a\a'}\e_{\a'\b\g}\bar\th^\b\w\bar\th^\g\Bigr),
\end{equation}
where $\e_{\a\b\g}$ is the antisymmetric tensor with $\e_{123}=1$.
The selfduality is manifest from the form of $J^\a$, and the closure
$dJ^\a=0$ follows from $dA = *dZ$.  In the coordinate basis, this
becomes
\begin{equation}
  J^\a = \sqrt{\frac{V_{\K3}}2}\Bigl(2dx^\a\w(dx^4+A) 
  + Z G^{\a\a'}\e_{\a'\b\g}dx^\b\w dx^\g\Bigr).
\end{equation}
These are the hyperk\"ahler forms of K3 in the approximate
metric~\eqref{eq:K3fromUnitMetric}.


\paragraph{Basis of harmonic forms.}

Next, observe that $J^\a$ can be written as the sum of two closed 2-forms
\begin{equation}
  J^\a = \sqrt{\frac{V_{K3}}2}(\o^\a + \o_\a),
\end{equation}
where
\begin{equation}
  \begin{split}
  \o^\a &= 2\bar\th^\a\w\bar\th^4 +
  \Bigl(\frac{Z-1}{Z}\Bigr)\e_{\a\b\g}\bar\th^\b\w\bar\th^\g
  + d\bar\l_\a,\\
  \o_\a &= \frac1Z \e_{\a\b\g}\bar\th^\b\w\bar\th^\g-d\bar\l_\a,
\end{split}
\end{equation}
or equivalently, in the coordinate basis,
\begin{equation}
  \o^\a = E^\a{}_\b\z^\b,\quad \o_\a = (E^{-1T})_\a{}^\b\z_\b,
\end{equation}
where
\begin{equation}\label{eq:ZetaLower}
  \begin{split}
    \z^\a &= 2dx^\a\w(dx^4+A) +
    G^{\a\a'}\e_{\a'\b\g}(Z-1) dx^\b\w dx^\g + G^{\a\a'}d\l_{\a'},\\
    \z_\a &= \e_{\a\b\g}dx^\b\w dx^\g - d\l_\a.
  \end{split}
\end{equation}
Here, we have included an exact term $d\bar\l_\a=(E^{-1T})_\a{}^\b
d\l_\b$ in the definition of $\o^\a$ and $\o_\a$.  We define
$\bar\l_\a$ so that $\o^\a$ and $\o_\a$ are harmonic.  Then
$\bar\l_\a$ satisfies
\begin{equation}
  \sqrt{\frac{V_{\K3}}{2}}\bigl(d\bar\l_\a + \star d\bar\l_\a\bigr) 
  = \frac{1-Z}{Z}J^\a.
\end{equation}

Finally, as in Sec.\ref{sec:D6O6lift}, we obtain anti-selfdual
harmonic forms
\begin{align}
  \o_I &=  \Bigl(\frac{Z_I-Z_{I'}}{Z}\Bigr)_{,\,\a} 
  \Bigl(dx^\a\w(dx^4+A)
  -\frac{Z}2 G^{\a\a'}\e_{\a'\b\g}dx^\b\w dx^\g\Bigr)\notag\\
  &= -d\Bigl((A_I-A_{I'}) - \frac{(Z_I-Z_{I'})}{Z}(dx^4+A)\Bigr),
\end{align}
for $I=1,\dots,16$.

In summary, a basis of harmonic 2-forms is
\begin{equation}\label{eq:ApproxHarmBasis}
  \o_a = (\o^\a,\o_\a,\o_I),
\end{equation}
and it is straightforward to show that
\begin{equation}
  \int \o_\a\w\o_b = \eta_{ab},
\end{equation}
with $\eta_{ab}$ defined in Eq.~\eqref{eq:EtaDef}.


\paragraph{Identification with the basis of
  Sec.~\ref{sec:ModHkahler}.}

Working in the approximate metric, it is natural to seek to identify
the forms~\eqref{eq:ApproxHarmBasis} with those of
Sec.~\ref{sec:ModHkahler}.  At the $T^4/\IZ_2$ orbifold locus with two
$\bx^I$ at each fixed point (i.e., $\bx^I$ given by
Eq.~\eqref{eq:xD16}), we have $Z=1$ and $A$ equal to a constant 1-form
$\b_\a dx^\a$, and we find $\l^\a = 0$.  In this case, comparing to
Sec.~\ref{sec:Abasis}, we see that the
2-forms~\eqref{eq:ApproxHarmBasis} precisely coincide with those of
Sec.~\ref{sec:ModHkahler} for all $E^\a{}_\b$ and $\b^{\a\b}$.
Working in the approximate metric, we would like to identify the forms
for all $x^{I\a}$ as well.  To do so, it suffices to show that the
expression for $\d[\o_a]$ under a small change $\d x^{I\a}$ takes the
same form here as in Eq.~\eqref{eq:OmegaVsXi}.

From Eq.~\eqref{eq:OmegaVsXi}, we should have
\begin{equation}
  \begin{split}
    \d[\z^\a] &= -\tilde\d\b^{\a\b}[\z_\b]
    +2\d x^{I\a}[\o_I],\\
    \d[\z_\a] &= 0,\\
    \d[\o_I] &= \d x^{I\b}[\z_\b],
    \end{split}
\end{equation}
where
\begin{equation}
  \tilde\d\b^{\a\b} = \d\b^{\a\b} - x^{I\a}\d x^{I\b}+x^{I\b}\d
  x^{I\a}.
\end{equation}
Since $Z_I$ and $Z_{I'}$ depend on $x^{I\b}$ only through the
combinations $x^a \pm x^{I\a}$, under a small change $\d x^{I\b}$, we
have
\begin{equation}
  \d Z = \d x^{I\b}\frac{\pd}{\pd x^{I\b}}(Z_I + Z_{I'})
  = - \d x^{I\b}\pd_\b(Z_I-Z_{I'}).
\end{equation}
Moreover, since $A$ satisfies $d A = \star_G dZ$ in the 3D metric
$G_{\a\b}$, we have
\begin{subequations}
  \begin{equation}
    \d A = \d x^{I\a}\d(A_{I\a}-A_{I'\a}),
  \end{equation}
  where
  \begin{equation}
    d(\d A_I) = \star_G d(\d Z_I),
    \quad
    d(\d A_{I'}) = \star_G d(\d Z_{I'}).
  \end{equation}
\end{subequations}
Given the form of $\d Z$ above, an obvious solution is $\d A_\a = -\d
x^{I\b}\pd_\b (A_I-A_{I'})_\a$.  However, a gauge equivalent and more
convenient choice is
\begin{equation}
  \d A_\a = \d x^{I\b}(F_I-F_{I})_{\a\b} + c_\a,
\end{equation}
where
\begin{equation}
  F_I = dA_I
  \quad\text{and}\quad
    F_{I} = dA_{I'}.
\end{equation}
Here, $c$ is a moduli-dependent constant 1-form on $T^3$.  We choose
$c$ so that
\begin{equation}\label{eq:dAGaugeChoice}
  \d A = \Bigl(\d x^{I\b}(F_I-F_I')_{\a\b}
  +\half\e_{\a\b\g}\tilde\d\b^{\b\g}\Bigr) dx^\a.
\end{equation}
Then, Eq.~\eqref{eq:ZetaLower} gives, for example
\begin{equation}
    \d\z^3 = \Bigl(2\d x^{I\b}(F_I-F_{I'})_{\d\b} 
    + \e_{\d\b\g}\tilde\d\b^{\b\g}\Bigr)dx^3\w dx^\d
    -\d x^{I\b}G^{3\a'}\e_{\a'\g\d}\pd_\b(Z_I-Z_{I'})dx^\g\w dx^\d.
\end{equation}
In cohomology, it is possible to show that the last equation
simplifies as follows:
\begin{align}
    [\d\z^3] &= \Bigl[\Bigl(2\d x^{I\b}(F_I-F_{I'})_{\d\b} 
    +\e_{\d\b\g}\tilde\d\b^{\b\g}\Bigr)dx^3\w dx^\d\notag\\
    &\qquad
    -\d x^{I\b}G^{3\a}\pd_\a(Z_I-Z_{I'})\e_{\b\g\d}dx^\g\w dx^\d\Bigr]\notag\\
    &= \Bigl[\Bigl(2\d x^{I\b}(F_I-F_{I'})_{\d\b} 
    +\e_{\d\b\g}\tilde\d\b^{\b\g}\Bigr)dx^3\w dx^\d\notag\\
    &\qquad
    -\d x^{I\b}(F_I-F_{I'})_{12}\e_{\b\g\d}dx^\g\w dx^\d\Bigr]\notag\\
    &= \Bigl[\Bigl(-2\d x^{I3}(F_I-F_{I'}) 
    +\e_{\d\b\g}\tilde\d\b^{\b\g}\Bigr)dx^3\w dx^\d\Bigr]\notag\\
    &= -2\d x^{I3}[F_I-F_{I'}] 
    -\tilde\d\b^{3\b}[\o_\b].
\end{align}
Noting that $[\o_I] = -[F_I-F_{I'}]$ and generalizing from $[\d\z^3]$
to $[\d\z^\a]$, we obtain
\begin{equation*}
  [d\z^\a] = 2\d x^{I\a}[\o_I] - \tilde\d\b^{\a\b}[\o_\b],
\end{equation*}
as desired.


\subsection{Metric moduli space in the Gibbons-Hawking
  approximation}
\label{sec:MetricModGH}

In this section we describe the metric moduli space of the approximate
K3 metric from two points of view.  First, in Sec.~\ref{sec:Method1},
we determine the diffeomorphism invariant moduli space metric of
Sec.~\ref{sec:DiffInvMet} applied to the approximate metric.  We find
agreement with the exact coset moduli space metric of
Sec.~\ref{sec:ReviewK3}.  Then, in Sec.~\ref{sec:Method2}, we consider
the naive moduli space metric from dimensional reduction, and show
that it agrees with the diffeomorphism invariant metric.  This special
property is due to the Gibbons-Hawking form of the approximate metric.

The approximate metric depends on parameters $G_{\a\b}$, $\b^{\a\b}$,
$\bx^I$, and the overall volume modulus $V_{\K3}$.  We have
suggestively given these parameters the same names as the moduli as
defined in Secs.~\ref{sec:ModHkahler} and \ref{sec:K3MetDefHarm}.  To
identify the two, we show in Sec.~\ref{sec:Method1}, that the metric
deformations due to small changes in the quantities $\bx^I$
parametrizing the approximate metric precisely agree with the metric
deformations generated by the harmonic forms of
Sec.~\ref{sec:HarmApprox}, in the manner described in
Sec.~\ref{sec:K3MetDefHarm}.  We expect the identification to hold for
$G_{\a\b}$ and $\b^{\a\b}$ as well.  Since the inner products of the
harmonic forms~\eqref{eq:ApproxHarmBasis} in the approximate metric
are the same as those~\eqref{eq:OmegaVsXi} in the exact metric, it
follows that the moduli space metric of the approximate K3 metric is
the same $\IR_{>0}\times \bigl(\SO(3)\times \SO(19)\bigr)\backslash
\SO(3,19)$ coset metric of the exact discussion in
Sec.~\ref{sec:ReviewK3}.

To achieve the diffeomorphism invariance of the metric on metric
moduli space, compensating vector fields were introduced in
Sec.~\ref{sec:DiffInvMet} to project generic metric deformations to
transverse traceless gauge.  In Sec.~\ref{sec:Method2}, we consider
the moduli space metric from naive dimensional reduction of the
$D$-dimensional Einstein-Hilbert action on a $d$-dimensional manifold
$X$.  This metric differs from the previous one in two ways: there is
no projection of metric deformations to their transverse traceless
part, and there are additional terms in the metric obtained by
integrating $(\Gbar^{mn}\d G_{mn})^2$.  For the approximate K3
metric~\eqref{eq:K3fromUnitMetric}, we find that the naive moduli
space metrics precisely agrees with the diffeomorphism invariant
moduli space metrics of Sec.~\ref{sec:DiffInvMet}.  Thus, the subtlety
of compensators and projection to transverse traceless gauge is not
necessary for the approximate K3 metric, and one can equivalently use
the naive moduli space metric from dimensional reduction.


\subsubsection{Method 1:  moduli space metric using compensators}
\label{sec:Method1}

The unit K3 metric in the Gibbon-Hawking
approximation~\eqref{eq:K3UnitMetric} 
\begin{equation*}
  \Gbar_{mn}(x;G,\b,x)dx^m dx^n
\end{equation*}
depends on moduli $G_{\a\b}$, $\b^{\a\b}$, and $x^{I\a}$, which we
would like to identify with the like-named hyperk\"ahler structure
moduli of Sec.~\ref{sec:Hyperkahler}.  To make this identification, it
is necessary to show that metric deformations
\begin{equation}
  \begin{split}
    \d(ds^2) 
    & = \d G^{\a\b}\frac{\pd}{\pd G^{\a\b}}\Gbar_{mn}dx^m dx^n\\
    &\quad + \tilde\d\b^{\a\b}\frac{\pd}{\pd\b^{\a\b}}\Gbar_{mn}dx^m dx^n\\
    &\quad + \d x^{I\a}\frac{\pd}{\pd x^{I\a}}\Gbar_{mn}dx^m dx^n
  \end{split}
\end{equation}
agree with those of Eq.~\eqref{eq:hbasis} up to compensating
diffeomorphism~\eqref{eq:NDiff}.  That is,
\begin{subequations}\label{eq:WithWithoutCompensators}
  \begin{align}
    \frac{\pd}{\pd G^{\a\b}}\Gbar_{mn} &= \bigl((h^G)_{\a\b})_{mn} 
    + 2\nabla_{(m|}\bigl((N^G)_{\a\b}\bigr)_{|n)},\\
    \frac{\pd}{\pd\b^{\a\b}}\Gbar_{mn} &= \bigl((h^\b)_{\a\b})_{mn} 
    + 2\nabla_{(m|}\bigl((N^\b)_{\a\b}\bigr)_{|n)},\\
    {\frac{\tilde\pd}{\pd x^{I\a}}}\Gbar_{mn} &= \bigl((h^x)_{I\a})_{mn} 
    + 2\nabla_{(m|}\bigl((N^x)_{I\a}\bigr)_{|n)},
    \label{eq:xequality}
  \end{align}
\end{subequations}
for the appropriate choice of compensating vector fields
\begin{subequations}
  \begin{align}
    (\d N^G)^n &= \d G^{\a\b}\bigl((N^G)_{\a\b}\bigr)^n,\\
    (\d N^\b)^n &= \tilde\d\b^{\a\b}\bigl((N^\b)_{\a\b}\bigr)^n,\\
    (\d N^x)^n &= \d x^{I\a}\bigl((N^x)_{I\a}\bigr)^n.
  \end{align}
\end{subequations}
Here,
\begin{equation}
  {\frac{\tilde\pd}{\pd x^{I\a}}} 
  = {\frac{\pd}{\pd x^{I\a}}}
  +\bigl(x^{I\b}\d^\g_\a -x^{I\g}\d^\b_\a\bigr)\frac{\pd}{\pd\b^{\b\g}},
\end{equation}
where the quantity appearing in parentheses is $-A^{\b\g}{}_{I\a}$,
with $A$ the connection defined in Eq.~\eqref{eq:betaConnection}.

For the deformations $\d x^{I\a}$, Eq.~\eqref{eq:xequality} can be
shown to hold with the choice
\begin{equation}
  (\d N^x)^\a = -\frac{Z_I}{Z} \d x^{I\a},\quad (\d N^x)^4 = 0. 
\end{equation}
This is proven in App.~\ref{app:GHdefs} in the simpler context of the
Gibbons-Hawking multicenter space from the lift of $N$ D6 branes.  The
computation is analogous for K3 in the Gibbons-Hawking approximation.

Equivalently, the metric deformations agree, provided that the
harmonic forms $\o_I$ generating the deformations are shifted by an
exact 2-form.  Since
\begin{equation}
  (\d N^x)_\b dx^\b = \D^{-1}ZG_{\a\b} (\d N^\a) dx^\b,
\end{equation}
with
\begin{equation}
  \D^{-1}Z dx^\b = (E^{-1T})^\b{}_{\b'}\CJ^{\b'}(dx^4+A),
\end{equation}
we see from Eq.~\eqref{eq:hshift} and the definition~\eqref{eq:hbasis}
of $(h^x)_{I\a}$ that the appropriate shift is
\begin{equation}
  \o_I\mapsto \o_I - d\Bigl(\frac{Z_I}{Z}(dx^4+A)\Bigr) = -(F_I-F_{I'}),
\end{equation}
where in the last equality, we have used Eqs.~\eqref{eq:FIequalsdAI}.
We expect that appropriate diffeomorphisms can also be found for the
deformations $\d G^{mn}$ and $\tilde\d\b^{mn}$, but we leave this as
an exercise for the future.

As a consequence of Eq.~\eqref{eq:WithWithoutCompensators}, the moduli
space metric~\eqref{eq:ModMetric} from transverse traceless
deformations of the K3 metric in the Gibbons-Hawking approximation
agrees with the exact moduli space metric~\eqref{eq:K3ExactModMetric}
discussed in Secs.~\ref{sec:ModHkahler} and~\ref{sec:K3MetricMod}.


\subsubsection{Method 2: moduli space metric without compensators}
\label{sec:Method2}


\paragraph{Generalities.}

We now ask the question: under what conditions are the compensators
unnecessary?  In the case of $D$-dimensional pure gravity, if we
ignore the subtlety of compensators and dimensionally reduce on a
$d$-dimensional Ricci flat manifold $X$ using the metric ansatz
\begin{equation}\label{eq:Dtod}
  ds^2 = (V_X)^{-\frac2{D-d-2}}g_{E\,\m\n}(y)dy^\m dy^\n
  + (2\pi\ell)^2(nV_X)^{\frac2d}\bar G_{mn}(\m(y),x) dx^m dx^n,
\end{equation}
we find that that $D$ dimensional Einstein-Hilbert action
\begin{equation}
  S^{(D)}_\text{EH} = \frac{2\pi}{(2\pi\ell)^{D-2}}\int d^Dx\sqrt{-G^{(D)}}R^{(D)},
\end{equation}
reduces to a $(D-d)$-dimensional action
\begin{equation}
  S^{(D-d)} = S^{(D-d)}_\text{EH} + S^{(D-d)}_{\bar G} + S^{(D-d)}_V,
\end{equation}
where, writing 
\begin{equation}
  S = \frac{2\pi}{(2\pi\ell)^{D-d-2}}\int d^{D-d}y\sqrt{-g_E}\,\CL,
\end{equation}
we have
\begin{subequations}
  \begin{align}
    \CL^{(D-d)}_\text{EH} &= R_E,\\
    \CL^{(D-d)}_{\bar G} &= n\int_X d^dx\sqrt{\Gbar}\Bigl(-\tfrac14
    \Gbar^{mp}\Gbar^{nq}\pd_\m\Gbar_{mn}\pd_E^\m\Gbar_{pq}
    +\tfrac14(\pd_E\log\Gbar)^2\Bigr),\label{eq:NaiveMetricGbar}\\
    \CL^{(D-d)}_V &= -\tfrac{D-2}{(D-d-2)d}\pd_\m(\log V)\pd^\m_E(\log V).
    \label{eq:NaiveMetricV}
  \end{align}
\end{subequations}
Here, $\Gbar_{mn}$ denotes a ``unit'' metric, satisfying
\begin{equation}
  n\int_X d^dx \sqrt{\Gbar} = 1,
\end{equation}
for some normalization constant $n$, and $(2\pi\ell)V_X$ is the volume
of $X$.  We have introduced the parameter $n$ so that for $X$ an
orbifold $Y/\G$, we can choose the convention $\int_Y
d^dx\,\sqrt{\Gbar} = 1$ on the covering space $Y$ by setting $n$ equal
to the dimension of the orbifold group $\G$.  For example, for
$\K3\cong T^4/\IZ_2$, we would set $n=2$.

From the kinetic terms~\eqref{eq:NaiveMetricGbar}
and~\eqref{eq:NaiveMetricV}, we can read off the naive metric on
moduli space, ignoring the subtlety of compensators.  Writing
$\d\log\Gbar = \Gbar^{mn}\d\Gbar_{mn}$ in
Eq.~\eqref{eq:NaiveMetricGbar}, the naive moduli space metric for
constant volume deformations is
\begin{equation}\label{eq:NaiveModMet}
    ds^2_\text{$\CM$, naive} 
    = n \int_X d^dx \sqrt{\Gbar}\Bigl(
      \tfrac14\Gbar^{mp}\Gbar^{nq}\d\Gbar_{mn}\d\Gbar_{pq}
      - \tfrac14 (\Gbar^{mn}\d\Gbar_{mn})^2\Bigr),
\end{equation}
and that for the overall volume modulus is
\begin{equation}\label{eq:Vmetric}
  ds^2_V = \frac{D-2}{(D-d-2)d}\Bigl(\frac{\d V_X}{V_X}\Bigr)^2.
\end{equation}
The former can be compared to the actual moduli space
metric~\eqref{eq:ModMetric} for constant volume deformations from
Sec.~\ref{sec:DiffInvMet}, which in the present notation becomes
\begin{equation}\label{eq:ActualModMetric}
    ds^2_\CM = n\int_X d^dx\sqrt{\Gbar}\,\tfrac14
      \Gbar^{mp}\Gbar^{nq}\d^\perp\Gbar_{mn}\d^\perp\Gbar_{pq}.
\end{equation}

If the moduli space metrics~\eqref{eq:NaiveModMet} and
\eqref{eq:ActualModMetric} agree, then compensators are unnecessary.
For example, whenever the volume form
\begin{equation}
  \Vol_X = nV_X\sqrt{\Gbar}dx^1\w\dots\w dx^d,
\end{equation}
is constant, the two moduli space metrics agree.  This is the case for
constant metrics on tori, whose volume preserving deformations are
automatically transverse and traceless.


\paragraph{Application to a metric of  Gibbons-Hawking form.}

A 4D metric of Gibbons-Hawking form,
\begin{equation}\label{eq:GHform}
  \bar G_{mn} dx^m dx^n
  = \D^{-1}ZG_{\a\b} dx^\a dx^\b
  + \D Z^{-1} (dx^4+A)^2,
\end{equation}
with
\begin{equation}
  dA = \star_G dZ,
  \quad \D=\sqrt{\Gbar},
\end{equation}
is another special case in which the two moduli space metrics agree
and compensators are unnecessary.  For a metric of the
form~\eqref{eq:GHform}, we have
\begin{equation}
  \sqrt{\Gbar} = Z,
  \quad
  -\tfrac14(\Gbar^{mn}\d\Gbar_{mn}) = -(\d\log Z)^2,
\end{equation}
and the net effect of the projection from $\d\Gbar_{mn}$ to
$\d^\perp\Gbar_{mn}$ after integration in
Eq.~\eqref{eq:ActualModMetric} is to ``ignore'' variations in $Z$ when
moduli are varied.  Correspondingly, in the naive moduli space
metric~\eqref{eq:NaiveModMet}, all $Z$ derivatives cancel between the
first and second terms, leaving the same result (up to multiplicative
factors of $Z$) as if the metric were
\begin{equation*}
  \bar G_{mn} dx^m dx^n = 
  \D^{-1} G_{\a\b} dx^\a dx^\b + \D (dx^4+A)^2,
\end{equation*}
A further convenient simplification is that the resulting
$\Gbar^{mp}\Gbar^{nq}\d\Gbar_{mn}\d\Gbar_{pq}$ terms in the naive
moduli space metric are just right to cancel all explicit $\D$
derivatives, leaving the same result (again, up to multiplicative
factors of $Z$) as if the metric were
\begin{equation*}
  \Gbar_{mn} dx^m dx^n = 
  G_{\a\b} dx^\a dx^\b + (dx^4+A)^2.
\end{equation*}
What remains is
\begin{equation}\label{eq:GHmodmetric}
  ds^2_\text{$\CM$, naive} = n\int_X d^4x\Bigl(
  \tfrac14 Z\,G^{\a\g}G^{\b\d}\d G_{\a\b}\d G_{\g\d}\\ 
  +Z^{-1}G^{\a\b}\d A_{\a}\d A_{\b}\Bigr).
\end{equation}


\paragraph{Application to K3 in the Gibbons-Hawking approximation.}

We are now in a position to evaluate the naive moduli space metric of
K3 in the Gibbons-Hawking approximation~\eqref{eq:K3UnitMetric}.
Specializing to $X=\K3\cong T^4/\IZ_2$, we set $n=2$ and write
$2\int_{T^4/\IZ_2} = \int_{T^4}$, so that
\begin{equation}
  \int_{T^4}d^4x Z = 1,
\end{equation}
and the naive moduli space metric becomes
\begin{equation}\label{eq:K3NaiveModMetric}
  ds^2_\text{$\CM$, naive} = \int_{T^4} d^4x\Bigl(
  \tfrac14 Z\,G^{\a\g}G^{\b\d}\d G_{\a\b}\d G_{\g\d}
  +Z^{-1}G^{\a\b}\d A_{\a}\d A_{\b}\Bigr)
\end{equation}
for volume-preserving deformations, and
\begin{equation}
  ds^2_V = \frac9{20} \Bigl(\frac{\d V_{\K3}}{V_{\K3}}\Bigr)^2
\end{equation}
for the overall K3 volume modulus (from $D=11$ and $d=4$ in
Eq.~\eqref{eq:Vmetric}).

In evaluating the first integral in Eq.~\eqref{eq:K3NaiveModMetric},
the quantity $\int_{T^4}d^4xZ = 1$ trivially factors out, leaving
\begin{equation*}
  \int_{T^4} d^4x \tfrac14 Z\,G^{\a\g}G^{\b\d}\d G_{\a\b}\d G_{\g\d}
  = \tfrac14 G^{\a\g}G^{\b\d}\d G_{\a\b}\d G_{\g\d}.
\end{equation*}
To evaluate the second integral, we recall from
Eq.~\eqref{eq:dAGaugeChoice} that $\d A$ has two terms, one
proportional to $\d x^{I\a}$ and the other proportional to
$\tilde\d\b^{\a\b} = \d\b^{\a\b}-x^{I\a}d x^{I\b} + x^{I\b}dx^{I\a}$:
\begin{equation*}
  \d A = \Bigl(\d x^{I\a}(F_I-F_I')_{\g\a}
  +\half\e_{\a\b\g}\tilde\d\b^{\a\b}\Bigr) dx^\g.
\end{equation*}
The two terms contribute orthogonally to the naive moduli space
metric~\eqref{eq:K3NaiveModMetric}.  The contribution from $\d
x^{I\a}$ term can be evaluated using
\begin{equation}
  \int_{T^4}d^4x Z^{-1}(F_I-F_I')_{\g\a}(F_J-F_J')^\g{}_{\b}
  = (\d_{IJ}+\d_{I'J'})G_{\a\b}
  = 2\d_{IJ}G_{\a\b},
\end{equation}
which follows from Eq.~\eqref{eq:Fintegral} of
App.~\ref{app:Integrals}.  The contribution from the
$\tilde\d\b^{\a\b}$ term gives an integral in which $\int_{T^4}d^4xZ =
1$ again trivially factors out.

Collecting all terms, we find
\begin{equation}
  ds^2_\text{$\CM$, naive}
  = \tfrac14 G_{\a\g}G_{\b\d}
  \Bigl(\d G^{\a\b}\d G^{\g\d} + \tilde\d\b^{\a\b}\tilde\d\b^{\g\d}\Bigr)\\
  +2 G_{\a\b}\d x^{I\a}\d x^{I\b}.
\end{equation}
This naive moduli space metric precisely matches the diffeomorphism
invariant moduli space metric $ds_\CM^2$ obtained in the previous
section.  So, indeed, compensators are an unnecessary machinery for
computing the K3 moduli space metric in the Gibbons-Hawking
approximation, and one can equivalently use the naive moduli space
metric~\eqref{eq:NaiveModMet} from dimensional reduction.


\section{Conclusions}
\label{sec:Conclusions}

We have studied the geometry of K3 surfaces through the lens of a
soluble model for the K3 metric, of Gibbons-Hawking form.  The metric
well approximates the K3 metric almost everywhere in the large
hypercomplex structure limit ($\D\ll 1$ in the language of
Sec.~\ref{sec:LiftToK3}).

In Sec.~\ref{sec:ReviewK3}, we reviewed the holonomy, hyperk\"ahler
structure, and homology of K3 surfaces.  In our discussion of the K3
homology, we provided an explicit description of the
$(-E_8)\oplus(-E_8)\oplus(U_{1,1})^3$ and
$\Spin(32)/\IZ_2\oplus(U_{1,1})^3$ splittings of the K3 homology
lattice in terms of the natural homology basis from the Kummer
resolution of $T^4/\IZ_2$.  The latter consists of the six 2-tori
inherited from $T^4$ and 16 exceptional divisors of the Kummer
resolution of $T^4/\IZ_2$.  This account fills a void in the
literature and we hope that it will serve as a useful reference to
others.  (Our discussion explicitly relates the $\bigl(A_1\bigr)^{16}$
Kummer basis to the natural homology bases at $D_{16}$ and $(E_8)^2$
orbifold points and subsequently to the desired splittings).  The
remainder of Sec.~\ref{sec:ReviewK3} focused on harmonic forms and the
moduli space of hyperk\"ahler structure, as well as the relation
between metric deformations and harmonic forms.  We derived the coset
form of the diffeomorphism invariant metric on the moduli space of K3
metrics.  In our discussion of the metric on moduli space, we
described the need for ``compensators.'' The compensators project to
transverse traceless gauge to ensure diffeomorphism invariance of the
metric on moduli space.

Sec.~\ref{sec:K3MetricGH} focused on the Gibbons-Hawking model for the
K3 metric~\eqref{eq:K3UnitMetric}, which we obtained from the
\mbox{M-theory} lift of the tree-level type IIA supergravity
description of the $T^3/\IZ_2$ orientifold.  In the large hypercomplex
structure limit, the metric closely approximates the exact K3 metric
everywhere except in a small neighborhood of 8 points corresponding to
the lifts of the orientifold planes.  The full nonperturbative lift
replaces these regions with smooth Atiyah-Hitchin spaces and gives the
exact K3 metric.  We described the coframe, hyperk\"ahler forms, and
basis of harmonic forms in the approximate metric, identifying the
cohomology classes of this basis with known classes in the exact
description.  The metric components are simple functions of the
parameters $G^{\a\b}$, $\b^{\a\b}$, and $x^{I\a}$ defining the
$\bigl(\SO(3)\times\SO(19)\bigr)\backslash SO(3,19)$ transformation
from an integer cohomology basis to the harmonic basis, i.e.,
parametrizing the hyperk\"ahler moduli space.  Finally, we studied the
diffeomorphism invariant metric on metric moduli space, as well as the
naive moduli space metric from dimensional reduction.  We found that
the Gibbons-Hawking form of the approximate metric leads to the novel
property that these moduli space metrics agree.  Moreover, both
coincide with the exact $\IR_{>0}\times \bigl(\SO(3)\times
\SO(19)\bigr)\backslash \SO(3,19)$ coset metric of K3.

As described in the Introduction, this project forms a part of a
larger investigation with the goal of providing a duality derivation
of the procedure for warped Kaluza-Klein
reduction~\cite{SchulzTammaroA,SchulzTammaroB} through duality to
conventional compactifications, and a secondary goal of shedding light
on compactification of type II string theory on 6D manifolds of
$SU(2)$ structure~\cite{Schulz:2012uj}, of which the abelian surface
fibered Calabi-Yau 3-folds of Refs.~\cite{Schulz:2004tt,Donagi:2008ht}
are examples.  While not themselves realistic, the warped backgrounds
of interest here are simplified analogs of the type IIA intersecting
\mbox{D6-brane} models and type IIB flux compactifications that
feature prominently in model building.  The $T^3/\IZ_2$ orientifold,
upon further compactification on $T^3$, the can be thought of as a
baby version of a type IIA intersecting D6-brane model: it based on a
$T^6$ rather than Calabi-Yau 3-fold; the orientifold involution
preserves a $T^3$ of $T^6$ instead of special Lagrangian 3-cycles of
the Calabi-Yau; and all D6-branes and O6-planes are parallel instead
of intersecting.  Similarly, the type IIB $T^6/\IZ_2$ orientifold with
$\CN=2$ flux shares much in common with $\CN=1$ or $0$ flux
compactifications based on F-theory compactifications or Calabi-Yau
orientifolds; the main difference here is that there are solely
D3-branes and O3-planes, rather than 7-branes wrapping holomorphic
cycles.  Studying the warped Kalaza-Klein reduction of these simple
models~\cite{SchulzTammaroA,SchulzTammaroB} should offer lessons for
the more realistic compactifications of phenomenological interest, and
provide useful examples for probing the formalism developed in
Refs.~\cite{Shiu:2008ry,Douglas:2008jx,Frey:2008xw,Frey:2009qb,
  Underwood:2010pm}.


\bigskip\centerline{\bf Acknowledgments}\nobreak\medskip\nobreak

M.S. is grateful to the Center for Theoretical Physics at MIT for its
hospitality during a Junior Faculty Research Leave, and Bryn Mawr
College for its support in making this leave possible.  In addition,
he is indebted to the KITP Scholars program and the University of
Pennsylvania for their continued hospitality.  E.T. thanks the ICTP
for a stimulating 2012 Trieste Spring School.  This material is based
upon work supported by the National Science Foundation under Grant
No.~PHY09-12219.  This research was supported in part by the National
Science Foundation under Grant No.~PHY11-25915.


\appendix


\section{Hyperk\"ahler structure on $T^4$}
\label{app:HkahlerT}

A choice of hyperk\"ahler structure on $T^4$ is analogous to a choice
of complex structure on $T^2$.  Let us first review the latter in a
way that makes the generalization natural, and then go on to discuss
hyperk\"ahler structure on $T^4$.  This review first appeared as
App.~A of Ref.~\cite{Cvetic:2007ju}, but (as stated there) was written
with the application to the duality between \mbox{M-theory} on K3 and
IIA on $T^3/\IZ_2$ in mind.  Minor changes have been made to conform
to the notation of the present paper.

On $T^2$, we can express the metric as
\begin{equation}\label{eq:Ttwoframemetric}
  ds^2_{T^2} = \th^1\otimes \th^1 + \th^2\otimes \th^2,
\end{equation}
where $\th^m = \th^m{}_n dx^n$ in terms of a vielbein $\th^m{}_n$.
The complex structure is defined by a tensor $\CJ_m{}^n$ which we view
as a map\footnote{The usual convention in the math literature is the
  transpose of this: $\CJ$ has index structure $\CJ^m{}_n$, so that
  $\CJ$ acts from the left on the tangent space, and $\CJ^T$ acts from
  the left on the cotangent space.  However, if we require that (i)
  $\CJ$ with holomorphic (antiholomorphic) indices be $+i$ ($-i$), as
  is customary in both the math and physics conventions, and (ii) the
  K{\"a}hler form $J$ be obtained by lowering one index of the tensor
  $\CJ$, with no sign change, then we are uniquely led to the
  conventions used in this paper.}  $\CJ\colon\ T^*\to T^*$, such that
\begin{equation}\label{eq:Jmap}
  \CJ\colon\quad \th^2\to \th^1,\quad \th^1\to -\th^2.
\end{equation}
Lowering the upper index of $\CJ_m{}^n$ gives the K\"ahler form $J$ on
$T^2$.  By $SL(2,\IZ)$ change of lattice basis for the lattice
$\Lambda$ of $T^2 = \IR^2/\Lambda$, we can always write
\begin{equation}\label{eq:Ttwovielbein}
  \begin{split}
    \th^1 &= R^1(dx^1+a^1{}_2 dx^2),\\
    \th^2 &= R^2 dx^2,
    \quad\text{where}\quad 
    x^m\cong x^m+1.
  \end{split}
\end{equation}
The holomorphic 1-form is
\begin{equation}\label{eq:HoloOneform}
  \th^z = \th^1+i\th^2 = R^1(dx^1+\t_1 dx^2),
\end{equation}
where $\t_1 = a^1{}_2 + i\frac{R^2}{R^1}$ is the complex structure
modulus.

Likewise, we can express the metric on $T^4$ as
\begin{equation}\label{eq:Tfourframemetric}
  ds^2_{T^4} = 
  \th^1\otimes \th^1 + \th^2\otimes \th^2 + \th^3\otimes \th^3 + \th^4\otimes \th^4,
\end{equation}
where, again, $\th^m = \th^m{}_n dx^m$ in terms of a vielbein
$\th^m{}_n$ The hyperk\"ahler structure is defined by a triple of
tensors $(\CJ^\a)_m{}^n$, $\a=1,2,3$, which we view as maps
$J^\a\colon\ T^*\to T^*$, such that
\begin{equation}\label{eq:JHKmap}
  \begin{split}
    \CJ^1\colon\quad &
    \th^4\to \th^1,\quad \th^3\to \th^2,\quad \th^1\to -\th^4,\quad \th^2\to -\th^3,\\
    \CJ^2\colon\quad &
    \th^4\to \th^2,\quad \th^1\to \th^3,\quad \th^2\to -\th^4,\quad \th^3\to -\th^1,\\
    \CJ^3\colon\quad &
    \th^4\to \th^3,\quad \th^2\to \th^1,\quad \th^3\to -\th^4,\quad \th^1\to -\th^2.
  \end{split}
\end{equation}
The $(\CJ^\a)_m{}^n$ satisfy 
\begin{equation*}
\CJ^1 \CJ^2 = - \CJ^2 \CJ^1 = -\CJ^3,\qquad (\CJ^1)^2 = -1,
\end{equation*}
plus cyclic permutations.  Lowering the upper index on
$(\CJ^\a)_m{}^n$ gives a triple of K\"ahler forms $J^\a_{mn}$.  The
quaternionic 1-form is
\begin{equation}\label{eq:QuatOneform}
  \th^q = \th^4 - {\bf i} \th^1 - {\bf j} \th^2-{\bf k} \th^3,
\end{equation}
where the quaternions ${\bf i},{\bf j},{\bf k}$ satisfy the same
algebra as $-\CJ^\a$:\footnote{Here, $-\CJ^\a$ rather than $\CJ^\a$
  satisfies the quaternion algebra for the reason discussed in the
  previous footnote.  Note that the tangent space map $(\CJ^\a)^T$
  satisfies the quaternion algebra with no minus sign.}
\begin{equation}\label{eq:QuatAlg}
  {\bf i}{\bf j} = {\bf k},
  \quad {\bf j}{\bf k} = {\bf i},
  \quad {\bf k}{\bf i} = {\bf j},
  \quad {\bf i}^2 = {\bf j}^2 = {\bf k}^2 = -1.
\end{equation}

A choice of complex structure on $T^4$ is then a choice of $i$ on the
${\bf i},{\bf j},{\bf k}$ unit sphere.  By a $SL(4,\IZ)$ change of
lattice basis for the $T^4$, we can write, in addition to
Eq.~\eqref{eq:Ttwovielbein},
\begin{equation}\label{eq:Tfourvielbein}
  \begin{split}
    \th^3 &= R^3(dx^3 + a^3{}_1 dx^1 + a^3{}_2 dx^2),\\
    \th^4 &= R^4(dx^4 + a^4{}_1 dx^1 + a^4{}_2 dx^2 + a^4{}_3 dx^3),
  \end{split}
\end{equation}
where $x^m\cong x^m+1$.  So, for example, if we choose complex
structure $i={\bf k}$, then the complex pairing that follows from $\CJ
= \CJ^3$ is
\begin{equation}\label{eq:Tfourcpx}
  \begin{split}
    \th^{z^1} &= R_1 \th^1 + i R_2 \th^2 = R^1 (dx^1 + \t_1 dx^2),\\
    \th^{z^2} &= R_4 \th^4 - i R_3 \th^3 = R^4 (dx^4 +\t_2{}^{-1} dx^3 + \ldots),
  \end{split}
\end{equation}
where $\t_2{}^{-1} = a^4{}_3 - i\frac{R_3}{R_4}$ and the ``\dots'' is a
1-form on $T^2(x^1,x^2)$, which can be interpreted as the connection
for a trivial fibration of $T^2(x^3,x^4)$ over $T^2(x^1,x^2)$.  The
holomorphic $(2,0)$ form in this case is
\begin{equation}\label{eq:twozeroform}
  \O_{(2,0)} = \th^{z^1}\w \th^{z^2} = -J^1 + iJ^2.
\end{equation}

If we write the metric on $T^4$ as
\begin{equation}\label{eq:Tfourmetric}
  ds^2_{T^4} = R_4{}^2 (dx^4 + a^4{}_\a dx^\a)^2 + g_{\a\b} dx^a\ dx^\b
\end{equation}
$(\a=1,2,3)$, then the choice of hyperk\"ahler structure is the choice
of $T^4$ volume $V_{T^4} = \sqrt{g}R_4$ together with the choice of
hypercomplex structure.  The latter is the choice of $\b_\a =
a^4{}_\a$ together with the dimensionless metric $G_{\a\b} =
(R_4/\sqrt{\mathstrut g})g_{\a\b}$.  In terms of $V_{T^4}$,
$G_{\a\b}$, and $\b_\a$, we can write the $T^4$ metric as
\begin{equation}
  ds^2 = \sqrt{V_{T^4}}
  \Bigl(\D(dx^4 + \b_\a dx^\a)^2 + \D^{-1}G_{\a\b}dx^\a dx^\b\Bigr).
\end{equation}
Let us define $\b^{\a\b} = \e^{\a\b\g}\b_\g$, where $\e^{123} = 1$,
and let $E^\a{}_\b$ be a vielbein for the metric $G_{\a\b}$.  Then,
the choice of $E^\a{}_\b$ and $\b^{\a\b}$ parametrizes the
$\bigl(\SO(3)\times \SO(3)\bigr)\backslash \SO(3,3)/\G_{3,3}$
truncation of the coset \eqref{eq:SOvielbein}, with vielbein
\begin{equation}\label{eq:SOvbTrunc}
  V = 
  \begin{pmatrix}
    E & -E\b\\
    0 & E^{-1T}
  \end{pmatrix}.
\end{equation}
This coset can be interpreted as the choice of positive signature
3-plane spanned by $J^1,J^2,J^3$ in $H^2(T^4,\IR) = \IR^{3,3}$, modulo
lattice isomorphisms of $H^2(T^4,\IZ)$.


\section{The homology lattice of $\K3$}
\label{app:HomologyLattice}

For completeness, we review the integer homology lattice of $\K3$ via
its interpretation as the resolution of the orbifold $T^4/\IZ_2$.
App.~\ref{app:Resolution} is based primarily on
Refs.~\cite{Denef:2005mm} and \cite{Wendland:2000ry}, and has been
reproduced from App.~B of Ref.~\cite{Cvetic:2007ju}, with minor
notational changes.  App.~\ref{app:Splittings} derives the
$(-E_8)\oplus(-E_8)\oplus U_{1,1}{}^{\oplus3}$ and
$(-\Spin(32)/\IZ_2)\oplus U_{1,1}{}^{\oplus3}$ splittings of this
lattice.


\subsection{Resolution of $T^4/\IZ_2$}
\label{app:Resolution}

\begin{figure}[ht]
  \begin{center}
    \includegraphics[width=3.0truein]{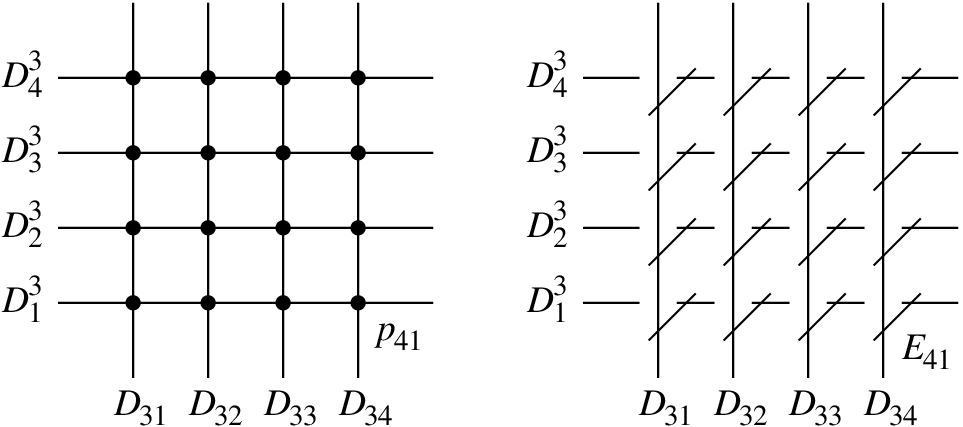}
  \end{center}
  \caption{\small (a) In the singular $T^4/\IZ_2$ (left), each of the
    sixteen $A_1$ singularities is the ``half point'' of intersection,
    $p_{st}$, of two fixed $\IP^1$s, $D_{3s}$ and $D^3_t$.  (b) In the
    resolved $K3$ (right), each $p_{st}$ is blown up to an exceptional
    divisor $E_{st}$.  After resolution, $D_{3s}$ and $D^3_t$ no
    longer intersect, but each intersects $E_{st}$ in a point.  In the
    figures above, only $p_{41}$ and its blow up $E_{41}$ are labeled
    explicitly.}
  \label{fig:K3Fig1}
\end{figure}

Let us view $T^4$ as $T^2_{(1)}(x^1,x^2) \times T^2_{(2)}(x^3,x^4)$
with complex pairing $dz_1 = dx^1+\t_1 dx^2$ and $dz_2 = dx^3+\t_2
dx^4$.  Now consider $T^4/\IZ_2$.  There are $2^4=16$ points of local
geometry $\IC^2/\IZ_2$ (16 $A_1$ singularities), located at the fixed
points where each of the four coordinates is equal to $0$ or $1/2$.
There are also $4+4=8$ fixed lines $\IP^1$ with a simple description
in this complex structure: let $D_{3s}$, $s=1,2,3,4$ label the
divisors $\IP^1 = T^2_{(2)}/\IZ_2$ located at each of the four fixed
points in $(x^1,x^2)$ and $D^3_t$, $t=1,2,3,4$ denote the divisors
$\IP^1 = T^2_{(1)}/\IZ_2$ located at the four fixed point in
$(x^3,x^4)$.  The intersections of these $\IP^1$s in the singular
geometry is illustrated schematically in
Fig.~\ref{fig:K3Fig1}~(a).

The homology classes of the $D_{3s}$ and $D^3_t$ in the singular
geometry are
\begin{equation}\label{eq:TfourSingRels}
  D_{3s} = \half f_3,\quad
  D^3_t = \half f^3,\quad\text{independent of $s,t$,}
\end{equation}
where $f_3$ is the class of $T^2_{(2)}$ and $f^3$ is the class of
$T^2_{(1)}$.  Let us focus on the singularity at the ``half
point''\footnote{This ``half point'' is the interpretation of
  $\int_{K3}(dx^1\w dx^2)\w (dx^3\w dx^4) = \half\int_{T^4} dx^1\w
  dx^2\w dx^3\w dx^4 = 1/2$.} $p_{st} = D_{3s}\cap D^3_t$, and
consider the local model $\IC^2/\IZ_2$ at this point.

\begin{figure}[ht]
  \begin{center}
    \includegraphics[width=3.0truein]{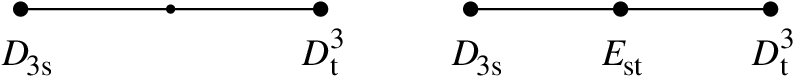}
  \end{center}
  \caption{\small (a) The fan for the local model $\IC^2/\IZ_2$ at the
    singular point $p_{st}$ in $T^4/\IZ_2$ (left), and (b)~the fan for
    the resolution (right), with the point $p_{st}$ blown up to the
    exceptional divisor $E_{st}$.}
  \label{fig:K3Fig2}
\end{figure}
\smallskip

Fig.~\ref{fig:K3Fig2}~(a) gives the fan for the toric description
of $\IC^2/\IZ_2$.  There is a single two dimensional fan of volume 2
generated by the lattice vectors $D_{3s}=(0,1)$ and $D^2_j=(2,1)$,
each of which represents a divisor of $T^4/\IZ_2$.  If we take
$p_{st}$ to be the origin of $\IC^2/\IZ_2$, then these divisors are
$D_{3s} = \{z_1=0\}$ and $D^3_t = \{z_2=0\}$.  In the toric
description, to resolve the singularity, we subdivide the original
singular cone into two cones of volume 1 by introducing a new divisor
$E_{st}$.  $E_{st}$ is the exceptional divisor obtained by blowing up
the origin of $\IC^2/\IZ_2$.  This notation differs slightly from the
notation $E_I$ given in the body of the paper, however, in either
notation a unique $(s,t)$ or $I$ characterizes each of the 16 singular
points of $T^4/\IZ_2$.

Let us make this more explicit.  To each of the lattice components
$r$, we associate a monomial $U_r=\prod_{i=1}^2 z_i{}^{(V_i)_r}$,
where $(V_i)_r$ is the $r$th component of the lattice vector $V_i$ in
the fan.  The toric variety is then given by the set of all
$(z_1,z_2)$ not in the excluded set $F$, modulo rescalings that leave
the $U_r$ invariant.  The excluded set $F$ consists of all points that
have simultaneous zeros of coordinates whose corresponding $V_i$ do
not lie in the same cone.  For the unresolved fan of Fig.~3~(a), there
is just a single two dimensional cone, so $F = \emptyset$.  The only
rescaling that leaves $U_1,U_2$ invariant is $\IZ_2\colon\
(z_1,z_2)\to(-z_1,-z_2)$.  So, the toric variety is indeed
$\{(z_1,z_2)\}/\IZ_2 = \IC^2/\IZ_2$.

For the resolved fan, we include the lattice vector $E_{st}=(1,1)$ as
well, as shown in Fig.~\ref{fig:K3Fig2}~(b).  In this case, $U_1 =
z_2{}^2 w$ and $U_2 = z_1z_2w$, where $w$ is the new coordinate
associated to $E_{st}$.  The excluded set is $F=\{z_1=z_2=0\}$.  The
rescaling symmetry of $U_1,U_2$ is $\IC^*\colon\ (z_1,z_2,w)\to(\l
z_1,\l z_2,\l^{-2}w)$.  Away from $w=0$, this gives
$(z_1,z_2,1)/\IZ_2=\IC^2/\IZ_2$ with the $z$-origin deleted.  At
$w=0$, we obtain the exceptional $\IP^1$, $E_{st} =
\{(z_1,z_2,0)\,\backslash\, (0,0,0)\}/\IC^*$.

Divisors can always be represented in patches as the vanishing loci of
local meromorphic functions.  However, divisors that globally have
such a representation are homologically trivial and have trivial
intersection with other divisors.  (See, for example,
Ref.~\cite{Griffiths:1978}.)

In our toric model for the resolution of $\IC^2/\IZ_2$, a basis of
such global meromorphic functions is $U_1,U_2$.  The corresponding
homologically trivial divisors are $2D_{3s}+E_{st}$ (from $U_2{}^2/U_1
= 0$) and $2D^3_t+E_{st}$ (from $U_1= 0$).  In the compact $K3$ (as
explained for $T^6/(\IZ_2\times\IZ_2)$ in Ref.~\cite{Denef:2005mm}),
these relations become
\begin{align}\label{eq:fKthree}
  f_3 &= 2D_{3s}+\sum_{t=1}^4 E_{st}
  \quad\text{independent of $s$,}\cr
  f^3 &= 2D^3_t+\sum_{s=1}^4 E_{st}
  \quad\text{independent of $t$,}
\end{align}
where the divisors $f_3$ and $f^3$ are not homologically trivial, but
instead correspond to ``sliding divisors'' that can be moved away from
the (resolved) singularities.  They have trivial intersection with the
exceptional divisors $E_{st}$ and represent the tori
$f_3=\{z_1=c_1\}\cup\{z_1=-c_1\}$ and
$f^3=\{z_2=c_2\}\cup\{z_2=-c_2\}$ on the $T^4$ covering space, where
$c_1,c_2$ are non fixed points.  The corresponding Poincar\'e dual
cohomology classes are $f_3=2dx^1\w dx^2$ and $f^3=2dx^3\w dx^4$,
respectively.

The cycles in $K3$ described so far are those that are particularly
simple in the complex structure $\CJ^3$.  In the same way, in the
complex structure $\CJ^1$ we obtain homology classes $f_1$ and $f^1$
from elliptic curves $T^2_{(4)}$ and $T^2_{(3)}$ located at non fixed
points in $(x^2,x^3)$ and $(x^1,x^4)$, respectively.  In the complex
structure $\CJ^2$ we obtain homology classes $f_2$ and $f^2$ from
elliptic curves $T^2_{(6)}$ and $T^2_{(5)}$ located at non fixed
points in $(x^3,x^1)$ and $(x^2,x^4)$.  Likewise, we obtain divisors
$D_{1s},D^1_t$ and $D_{2s},D^2_t$ by setting the corresponding pairs
of coordinates equal to their $\IZ_2$ fixed values before the
resolution.  The homology lattice of $K3$ is the integer span of the
overcomplete basis given by the $6 f$, $24$ $D$ and 16 $E$ divisors.


\subsection{Splittings of the homology lattice}
\label{app:Splittings}

Returning to the notation of the body of the paper, let us label the
16 exceptional divisors $E_I$ and their corresponding fixed points
$(x^1,x^2,x^3,x^4)\in\half\IF_2^{4}\subset T^4/\IZ_2$ as follows:
{\small
\begin{equation}
  \begin{split}
    E_1 &\qquad (0,0,0,0),\\
    E_2 &\qquad (\half,0,0,0),\\
    E_3 &\qquad (0,\half,0,0),\\
    E_4 &\qquad (\half,\half,0,0),\\
    E_5 &\qquad (0,0,\half,0),\\
    E_6 &\qquad (\half,0,\half,0),\\
    E_7 &\qquad (0,\half,\half,0),\\
    E_8 &\qquad (\half,\half,\half,0),
  \end{split}
  \qquad\qquad
  \begin{split}
    E_{9\phantom1} &\qquad (0,0,0,\half),\\
    E_{10} &\qquad (\half,0,0,\half),\\
    E_{11} &\qquad (0,\half,0,\half),\\
    E_{12} &\qquad (\half,\half,0,\half),\\
    E_{13} &\qquad (0,0,\half,\half),\\
    E_{14} &\qquad (\half,0,\half,\half),\\
    E_{15} &\qquad (0,\half,\half,\half),\\
    E_{16} &\qquad (\half,\half,\half,\half).
  \end{split}
\end{equation}
} 
Again, recall that we label the divisors $D$ so that $D^1_t$, for
$t=1,2,3,4$, denote the four $D$ at fixed $(x^1,x^4)$, and $D_{1s}$,
for $s=1,2,3,4$, denote the four $D$ at fixed $(x^2,x^3)$, with
analogous definitions obtained by cyclic permutation of $1,2,3$.
Then, from Eq.~\eqref{eq:fKthree}, the 24 divisors $D$ and their
locations in $T^4/\IZ_2$ are
{\small
\begin{align*}
    (x^1,x^4) &= (0,0)         & D^1_1 &= \half f^1 -\half(E_1 + E_5 + E_9 + E_{13}),\\
    (x^1,x^4) &= (0,\half)     & D^1_2 &= \half f^1 -\half(E_2 + E_6 + E_{10} + E_{14}),\\
    (x^1,x^4) &= (\half,0)     & D^1_3 &= \half f^1 -\half(E_3 + E_7 + E_{11} + E_{15}),\\
    (x^1,x^4) &= (\half,\half) & D^1_4 &= \half f^1 -\half(E_4 + E_8 + E_{12} + E_{16}),\\
    & & & \\
    (x^2,x^4) &= (0,0)         & D^2_1 &= \half f^2 -\half(E_1 + E_3 + E_9 + E_{11}),\\
    (x^2,x^4) &= (0,\half)     & D^2_2 &= \half f^2 -\half(E_2 + E_4 + E_{10} + E_{12}),\\
    (x^2,x^4) &= (\half,0)     & D^2_3 &= \half f^2 -\half(E_5 + E_7 + E_{13} + E_{15}),\\
    (x^2,x^4) &= (\half,\half) & D^2_4 &= \half f^2 -\half(E_6 + E_8 + E_{14} + E_{16}),\\
    & & & \\
    (x^3,x^4) &= (0,0)         & D^3_1 &= \half f^3 -\half(E_1 + E_3 + E_5 + E_7),\\
    (x^3,x^4) &= (0,\half)     & D^3_2 &= \half f^3 -\half(E_2 + E_4 + E_6 + E_8),\\
    (x^3,x^4) &= (\half,0)     & D^3_3 &= \half f^3 -\half(E_9 + E_{11} + E_{13} + E_{15}),\\
    (x^3,x^4) &= (\half,\half) & D^3_4 &= \half f^3 -\half(E_{10} + E_{12} + E_{14} + E_{16}),
\end{align*}
\begin{align*}
    (x^2,x^3) &= (0,0)         & D_{11} &= \half f_1 -\half(E_1 + E_2 + E_3 + E_4),\\
    (x^2,x^3) &= (0,\half)     & D_{12} &= \half f_1 -\half(E_9 + E_{10} + E_{11} + E_{12}),\\
    (x^2,x^3) &= (\half,0)     & D_{13} &= \half f_1 -\half(E_5 + E_6 + E_7 + E_8),\\
    (x^2,x^3) &= (\half,\half) & D_{14} &= \half f_1 -\half(E_{13} + E_{14} + E_{15} + E_{16}),\\
    & & & \\
    (x^3,x^1) &= (0,0)         & D_{21} &= \half f_2 -\half(E_1 + E_2 + E_5 + E_6),\\
    (x^3,x^1) &= (0,\half)     & D_{22} &= \half f_2 -\half(E_3 + E_4 + E_7 + E_8),\\
    (x^3,x^1) &= (\half,0)     & D_{23} &= \half f_2 -\half(E_9 + E_{10} + E_{13} + E_{14}),\\
    (x^3,x^1) &= (\half,\half) & D_{24} &= \half f_2 -\half(E_{11} + E_{12} + E_{15} + E_{16}),\\
    & & & \\
    (x^1,x^2) &= (0,0)         & D_{31} &= \half f_3 -\half(E_1 + E_2 + E_9 + E_{10}),\\
    (x^1,x^2) &= (0,\half)     & D_{32} &= \half f_3 -\half(E_5 + E_6 + E_{13} + E_{14}),\\
    (x^1,x^2) &= (\half,0)     & D_{33} &= \half f_3 -\half(E_3 + E_4 + E_{11} + E_{12}),\\
    (x^1,x^2) &= (\half,\half) & D_{34} &= \half f_3 -\half(E_7 + E_8 + E_{15} + E_{16}).
\end{align*}
} 
In terms of the orthonormal basis $e_I$, defined by
\begin{align}
  & e_1 = \half(E_2-E_1),\quad e_2 = \half(E_2+E_1),
  \quad\dots,\quad\notag\\
  & e_{15} = \half(E_{16}-E_{15}),\quad e_{16} = \half(E_{16}+E_{15}),
\end{align}
this becomes
{\small
\begin{subequations}\label{eq:Dfe}
  \begin{equation}
    \begin{split}
      D^1_1 &= \half f^1 -\half(- e_1 + e_2 - e_5 + e_6 - e_9 + e_{10} - e_{13} + e_{14}),\\
      D^1_2 &= \half f^1 -\half(  e_1 + e_2 + e_5 + e_6 + e_9 + e_{10} + e_{13} + e_{14}),\\
      D^1_3 &= \half f^1 -\half(- e_3 + e_4 - e_7 + e_8 - e_{11} + e_{12} - e_{15} + e_{16}),\\ 
      D^1_4 &= \half f^1 -\half(  e_3 + e_4 + e_7 + e_8 + e_{11} + e_{12} + e_{15} + e_{16}),\\ 
      & \\
      D^2_1 &= \half f^2 -\half(- e_1 + e_2 - e_3 + e_4 - e_9 + e_{10} - e_{11} - e_{12}),\\
      D^2_2 &= \half f^2 -\half(  e_1 + e_2 + e_3 + e_4 + e_9 + e_{10} + e_{11} + e_{12}),\\ 
      D^2_3 &= \half f^2 -\half(- e_5 + e_6 - e_7 + e_8 - e_{13} + e_{14} - e_{15} + e_{16}),\\
      D^2_4 &= \half f^2 -\half(  e_5 + e_6 + e_7 + e_8 + e_{13} + e_{14} + e_{15} + e_{16}),\\
      & \\
      D^3_1 &= \half f^3 -\half(- e_1 + e_2 - e_3 + e_4 - e_5 + e_6 - e_7 + e_8),\\
      D^3_2 &= \half f^3 -\half(  e_1 + e_2 + e_3 + e_4 + e_5 + e_6 + e_7 + e_8),\\
      D^3_3 &= \half f^3 -\half(- e_9 + e_{10} - e_{11} + e_{12} - e_{13} + e_{14} - e_{15} + e_{16}),\\  
      D^3_4 &= \half f^3 -\half(  e_9 + e_{10} + e_{11} + e_{12} +
      e_{13} + e_{14} + e_{15} + e_{16}),
    \end{split}
  \end{equation}
  \begin{equation}
    \begin{split}
      & \\
      D_{11} &= \half f_1 - e_2 - e_4,\\
      D_{12} &= \half f_1 - e_{10} - e_{12},\\
      D_{13} &= \half f_1 - e_6 - e_8,\\
      D_{14} &= \half f_1 - e_{14} -e_{16},\\
      & \\
      D_{21} &= \half f_2 - e_2 - e_6,\\
      D_{22} &= \half f_2 - e_4 - e_8,\\
      D_{23} &= \half f_2 - e_{10} - e_{14},\\
      D_{24} &= \half f_2 - e_{12} - e_{16},\\
      & \\
      D_{31} &= \half f_3 - e_2 - e_{10},\\
      D_{32} &= \half f_3 - e_6 - e_{14},\\
      D_{33} &= \half f_3 - e_4 - e_{12},\\
      D_{34} &= \half f_3 - e_8 - e_{16}.
    \end{split}
  \end{equation}
\end{subequations}
} 
Here, the $E_I$ and $e_I$ coincide with $\chi^{(A)}{}_I$ and
$e^{(A)}_I$ of Sec.~\ref{sec:Dbasis}, and we have dropped the
superscripts in this Appendix for notational simplicity.

The K3 (co)homology lattice is the integer span of
$f^\a,f_\a,D^\a_s,D_{\a s},E_I$.  We now obtain lattice bases
realizing the splittings
\begin{align}
  H_2(\K3,\IZ)
  &\cong (-\Spin(32)/\IZ_2)\oplus(U_{1,1})^{\oplus3}\notag\\
  &\cong (-E_8)\oplus(-E_8)\oplus(U_{1,1})^{\oplus3},
\end{align}
where $(-\Spin(32)/\IZ_2)$ denotes the weight lattice of
$\Spin(32)/\IZ_2$ with opposite sign inner product, $(-E_8)$ denotes
the $E_8$ root lattice with opposite sign inner product, and $U_{1,1}$
denotes the even selfdual lattice of signature $(1,1)$ with inner
product $\bigl(\begin{smallmatrix} 0 & 1\\ 1 &
  0\end{smallmatrix}\bigr)$.\footnote{For $E_8$, the weight lattice is
  the same as the root lattice.  For $\Spin(32)$, the ratio of the two
  is $\IZ_2\times\IZ_2$.  See Footnote~\ref{footnote:RootsWeights} for
  the relation between the weights of $\Spin(32)$, $\Spin(32)/\IZ_2$
  and $SO(32)$.}


\subsubsection{$(-\Spin(32)/\IZ_2)\oplus(U_{1,1})^{\oplus3}$ splitting}
\label{app:Spin32splitting}

To realize a $D_{16}\oplus U_{3,3}$ splitting of $H_2(\K3,\IZ)$, we
define
\begin{equation}
  U_{3,3} = \langle f_\a,D^\a_4\rangle,
\end{equation}
and ask what the orthogonal complement $U_{3,3}^\perp$ is.  The map
from the (co)homology lattice to $U_{3,3}^\perp$ takes $v$ to $v^\perp
= v - a^\a f_\a - b_\a D^\a_4$ with $a^\a$ and $\b_\a$ chosen so that
$v^\perp\cdot f_\a = v^\perp\cdot D^4_\a = 0$.  With this definition,
we find
{\small
\begin{equation}\label{eq:eperpD}
  \begin{split}
    e_1^\perp &= e_1,\\
    e_2^\perp &= e_2,\\
    e_3^\perp &= e_3 -\half f_1,\\
    e_4^\perp &= e_4 -\half f_1,\\
    e_5^\perp &= e_5 -\half f_2,\\
    e_6^\perp &= e_6 -\half f_2,\\
    e_7^\perp &= e_7 -\half f_1 -\half f_2,\\
    e_8^\perp &= e_8 -\half f_1 -\half f_2,
   \end{split}
  \qquad\qquad
  \begin{split}
    e_9^\perp &= e_9 -\half f_3,\\
    e_{10}^\perp &= e_{10} -\half f_3,\\
    e_{11}^\perp &= e_{11} -\half f_1 -\half f_3,\\
    e_{12}^\perp &= e_{12} -\half f_1 -\half f_3,\\
    e_{13}^\perp &= e_{13} -\half f_2 -\half f_3,\\
    e_{14}^\perp &= e_{14} -\half f_2 -\half f_3,\\
    e_{15}^\perp &= e_{15} -\half f_1 -\half f_2 -\half f_3,\\
    e_{16}^\perp &= e_{16} -\half f_1 -\half f_2 -\half f_3,
   \end{split}
\end{equation}
} 
which can be summarized as
\begin{equation}\label{eq:eperpexf}
  e^\perp_I = e_I - x^{I\a}f_\a,
\end{equation}
with $x^{I\a}$ given by Eq.~\eqref{eq:xD16}.  Identifying the
$e^\perp_I$ of this section with $e^{(D)}{}_I$ of
Sec.~\ref{sec:Dbasis}, this extends to the
transformation~\eqref{eq:Vofx} of the full basis from $\xi^{(A)}{}_\a$
to $\xi^{(D)}{}_\a$.  It is straightforward to check that
$e^\perp_I\pm e^\perp_J$ is an integer linear combination of
$f^\a,f_\a,D^\a_s,D_{\a s}$ for all $e^\perp_J$ and $e^\perp_J$, and
that the span of the $e^\perp_I\pm e^\perp_J$ is contained in
$U_{3,3}{}^\perp$.  Comparing to Eq.~\eqref{eq:D16roots}, we see that
$U_{3,3}{}^\perp$ contains the lattice $(-D_{16})$, where $(-D_{16})$
denotes the $D_{16}$ root lattice with the sign of the inner product
reversed.  In fact, $U_{3,3}^\perp$ is even larger than this.  Summing
the $e^\perp_I$, we find
\begin{align}
  \half\sum_{I=1}^{16} e^\perp_I 
  &= \half\bigl(\sum_{I=1}^{16}e_I\bigr) - 2(f_1+f_2+f_3)\notag\\
  &= f^3 - D^3_2 + D^3_4 - 2(f_1 + f_2 + f_3),
\end{align}
so that $\half\sum_{I=1}^{16} e^\perp_I$ is also an integer lattice
vector in $U_{3,3}{}^\perp\subset H_2(\K3,\IZ)$.  By subtracting
$D_{16}$ roots of the form $e^\perp_I+e^\perp_J$ from this lattice
vector, we see that $U_{3,3}$ contains not only the $D_{16}$ roots,
but also the lattice vectors differing from $\half\sum_{I=1}^{16}
e^\perp_I$ by an even number of sign flips.  The latter are the
weights of the chiral spinor representation of $\Spin(32)$.  Together,
the roots and chiral spinor weights span $U_{3,3}{}^\perp$ and form
the weight lattice of $\Spin(32)/\IZ_2$.\footnote{See
  Footnote~\ref{footnote:RootsWeights}.}  Therefore,
\begin{equation}
  U_{3,3}{}^\perp\cong(-\Spin(32)/\IZ_2),
\end{equation}
where $(-\Spin(32)/\IZ_2)$ denotes the weight lattice of
$\Spin(32)/\IZ_2$ with opposite sign inner product.  One can check
that $U_{3,3}\oplus U_{3,3}{}^\perp$ is indeed the whole homology
lattice.\footnote{Over the reals, this is guaranteed.  Over the
  integers, this simply requires checking that that no integer
  (co)homology class decomposes into half integer classes in $U_{3,3}$
  and $U_{3,3}^\perp$.  In contrast, for the $(A_1)^{16}$ basis, note
  that the sum of \unexpanded{$\langle f_\a,f^\a\rangle$} and
  \unexpanded{$\langle f_\a,f^\a\rangle^\perp = \langle E_I\rangle$}
  is a sublattice of the (co)homology lattice of order 2 that misses
  the $D^\a_s$ and $D_{\a,i}$, which have half integer coefficients.}
Therefore,
\begin{equation}
  H_2(K3,\IZ)\cong(-\Spin(32)/\IZ_2)\oplus U_{3,3}.
\end{equation}

If we exchange one or more of the $D^\a_4$ for $D^\a_2$ in the
definition of $U_{3,3}$, then Eq.~\eqref{eq:eperpexf} becomes
\begin{equation}\label{eq:eperpDgeneral}
  e^\perp_I = e_I - (x^{I\a}-x_P)f_\a,
\end{equation}
where $x_P$ is the $\IZ_2$ fixed point with $x^\a$ equal to $0$ for
each $D^\a_4$ and $\half$ for each $D^\a_2$.  Exchanging a $D^\a_4$
for a $D^\a_3$ or a $D^\a_2$ for a $D^\a_1$ is less interesting.  It
reverses the sign of $f_\a$ in Eq.~\eqref{eq:eperpDgeneral} for odd
$I$, which can be undone by the automorphism (Weyl reflection)
$e_I,e^\perp_I\to -e_I,-e^\perp_I$ for odd $I$.

Finally, we note that the lattice $U_{3,3}$ splits as
\begin{equation}
  U_{3,3}=U_{1,1}\oplus U_{1,1}\oplus U_{1,1}.
\end{equation}
For definiteness, consider choice $U_{3,3} = \langle
f_\a,D^\a_4\rangle$.  The basis $D^\a_4,f_\a$ does not realize this
splitting, since $D^\a_4\cdot D^\b_4 = -1$ for $\a\ne\b$.  However,
the basis
\begin{equation}
  D^1_4 + f_1 + f_2 + f_3,\ f_1;
  \ D^2_4 + f_2 + f_3,\ f_2;
  \ D^3_4,\ f_3,
\end{equation}
indeed has intersection form $\bigl(\begin{smallmatrix} 0 & 1\\1 &
  0\end{smallmatrix}\bigl)^{\oplus\,3}$.


\subsubsection{$(-E_8)\oplus(-E_8)\oplus(U_{1,1})^{\oplus3}$
  splitting}
\label{app:E8E8splitting}

To realize a $-(E_8)\oplus(-E_8)\oplus U_{3,3}$ splitting of
$H_2(\K3,\IZ)$, we observe that the expressions for $D^1_s$, $D^2_s$,
and $D_{3s}$ in Eq.~\eqref{eq:Dfe} do not mix $e_1,\dots,e_8$ with
$e_9,\dots,e_{16}$.  Therefore, let us exchange the roles of
$D^3_s,f_3$ and $D_{3s},f^3$ in the $\Spin(32)/\IZ_2$ discussion, and
define
\begin{equation}
  U_{3,3}{}' = \langle f_1,f_2,f^3,D^1_4,D^2_4,D_{34}\rangle.
\end{equation}
What is the orthogonal complement $U_{3,3}{}'^\perp$?  We find
{\small
\begin{equation}\label{eq:eperpE}
  \begin{split}
    e_1^\perp &= e_1,\\
    e_2^\perp &= e_2,\\
    e_3^\perp &= e_3 -\half f_1,\\
    e_4^\perp &= e_4 -\half f_1,\\
    e_5^\perp &= e_5 -\half f_2,\\
    e_6^\perp &= e_6 -\half f_2,\\
    e_7^\perp &= e_7 -\half f_1 -\half f_2,\\
    e_8^\perp &= e_8 -\half f_1 -\half f_2 - f^3,
   \end{split}
  \qquad\qquad
  \begin{split}
    e_9^\perp &= e_9,\\
    e_{10}^\perp &= e_{10},\\
    e_{11}^\perp &= e_{11} -\half f_1,\\
    e_{12}^\perp &= e_{12} -\half f_1,\\
    e_{13}^\perp &= e_{13} -\half f_2,\\
    e_{14}^\perp &= e_{14} -\half f_2,\\
    e_{15}^\perp &= e_{15} -\half f_1,\\
    e_{16}^\perp &= e_{16} -\half f_1 - f^3.
   \end{split}
\end{equation}
}
Using $(D^1_4)^\perp = (D^2_4)^\perp = (D_{34})^\perp = 0$ to
determine $(f^1)^\perp,(f^2)^\perp,(f_3)_\perp$, we find from
Eq.~\eqref{eq:Dfe} that in the $e^\perp_I$ basis,
{\small
\begin{align*}
    D^{1\perp}_1 &= (\half,-\half,\half,\half,\half,-\half,\half,\half;\half,-\half,\half,\half,\half,-\half,\half,\half)\\
    D^{1\perp}_2 &= (-\half,-\half,\half,\half,-\half,-\half,\half,\half;-\half,-\half,\half,\half,-\half,-\half,\half,\half)\\
    D^{1\perp}_3 &= (0,0,1,0,0,0,1,0;0,0,1,0,0,0,1,0)\\ 
    D^{1\perp}_4 &= 0\\
    & \\
    D^{2\perp}_1 &= (\half,-\half,\half,-\half,\half,\half,\half,\half;\half,-\half,\half,-\half,\half,\half,\half,\half)\\
    D^{2\perp}_2 &= (-\half,-\half,-\half,-\half,\half,\half,\half,\half;-\half,-\half,-\half,-\half,\half,\half,\half,\half)\\
    D^{2\perp}_3 &= (0,0,0,0,1,0,1,0;0,0,0,0,1,0,1,0;)\\
    D^{2\perp}_4 &= 0\\
    & \\
    D^{3\perp}_1 &= (\half,-\half,\half,-\half,\half,-\half,\half,-\half;0^8)\\
    D^{3\perp}_2 &= (-\half,-\half,-\half,-\half,-\half,-\half,-\half,-\half;0^8)\\
    D^{3\perp}_3 &= (0^8;\half,-\half,\half,-\half,\half,-\half,\half,-\half)\\
    D^{3\perp}_4 &= (0^8;-\half,-\half,-\half,-\half,-\half,-\half,-\half,-\half)\\
\end{align*}

\begin{align*}
    D^\perp_{11} &= (0,-1,0,-1,0,0,0,0;0^8)\\
    D^\perp_{12} &= (0^8;0,-1,0,-1,0,0,0,0)\\
    D^\perp_{13} &= (0,0,0,0,0,-1,0,-1;0^8)\\
    D^\perp_{14} &= (0^8;0,0,0,0,0,-1,0,-1)\\
    & \\
    D^\perp_{21} &= (0,-1,0,0,0,-1,0,0;0^8)\\
    D^\perp_{22} &= (0,0,0,-1,0,0,0,-1;0^8)\\
    D^\perp_{23} &= (0^8;0,-1,0,0,0,-1,0,0)\\
    D^\perp_{24} &= (0^8;0,0,0,-1,0,0,0,-1)\\
    & \\
    D^\perp_{31} &= (0,-1,0,0,0,0,0,1;0,-1,0,0,0,0,0,1)\\
    D^\perp_{32} &= (0,0,0,0,0,-1,0,1;0,0,0,0,0,-1,0,1)\\
    D^\perp_{33} &= (0,0,0,-1,0,0,0,1;0,0,0,-1,0,0,0,1)\\
    D^\perp_{34} &= 0.
\end{align*}
} 
These $D^\perp$, together with the $\chi_{2i-1}^\perp = e_{2i}^\perp -
e_{2i-1}^\perp$ and $\chi_{2i}^\perp = e_{2i}^\perp + e_{2i-1}^\perp$
for $i=1,\dots,8$, span the lattice $(-E_8)\oplus(-E_8)$.  Therefore,
we identify the $e^\perp_I$ here with $e^{(E)}_I$ of
Sec.~\ref{sec:Dbasis}.  Recall that the roots of the lattice $(-E_8)$
are given by all permutations of
\begin{equation}
  (1,1,0^6),\quad (1,-1,0^6),\quad (-1,-1,0^6),
\end{equation}
together with all roots obtained from
\begin{equation}
  (\half,\half,\half,\half,\half,\half,\half,\half)
\end{equation}
by an even number of sign flips.\footnote{The former gives the root
  lattice of $\SO(16)$ and the latter the weights of the spinor of
  $\SO(16)$, which is indeed how the root lattice of $E_8$ decomposes.}

Finally, on general grounds, the even selfdual lattice $U_{3,3}{}'$
should split as $U_{3,3}{}' = U_{1,1}{}^{\oplus3}$.  This was
demonstrated in the last section for the $(-\Spin(32)/\IZ_2)\oplus
U_{3,3}$ splitting of $H_2(\K3,\IZ)$.  For the
$(-E_8)\oplus(-E_8)\oplus U_{3,3}{}'$ splitting now under discussion,
the lattice $U_{3,3}{}'$ decomposes in the same way, with $D^3_4,f_3$
replaced by $D_{34},f^3$.


\subsubsection{Relation to $(-D_{16})^{\oplus2}$}

Let $e^{(D)}_I$ denote the $e^\perp_I$ of Eq.~\ref{eq:eperpD} and
$e^{(E)}_I$ denote the $e^\perp_I$ of Eq.~\ref{eq:eperpE}.  Define
\begin{equation}
  U_{3,3}{}'' = \langle f_1,f_2f_3,D^1_4,D^2_4,f^3\rangle.
\end{equation}
What is the orthogonal complement of $\displaystyle U_{3,3}{}''$?  We
find
{\small
\begin{equation}
  \begin{split}
    e_1^\perp &= e_1,\\
    e_2^\perp &= e_2,\\
    e_3^\perp &= e_3 -\half f_1,\\
    e_4^\perp &= e_4 -\half f_1,\\
    e_5^\perp &= e_5 -\half f_2,\\
    e_6^\perp &= e_6 -\half f_2,\\
    e_7^\perp &= e_7 -\half f_1 -\half f_2,\\
    e_8^\perp &= e_8 -\half f_1 -\half f_2,
   \end{split}
  \qquad\qquad
  \begin{split}
    e_9^\perp &= e_9,\\
    e_{10}^\perp &= e_{10},\\
    e_{11}^\perp &= e_{11} -\half f_1,\\
    e_{12}^\perp &= e_{12} -\half f_1,\\
    e_{13}^\perp &= e_{13} -\half f_2,\\
    e_{14}^\perp &= e_{14} -\half f_2,\\
    e_{15}^\perp &= e_{15} -\half f_1,\\
    e_{16}^\perp &= e_{16} -\half f_1.
   \end{split}
\end{equation}
} 
The $e^\perp_I\pm e^\perp_J$, for $I,J=1,\dots8$, and for
$I,J=9,\dots16$ span two distinct $D_8$ root lattices.  We have
\begin{equation}
  U_{3,3}{}''^\perp = -(D_8)\oplus(-D_8),
\end{equation}
where $(-D_8)$ denotes the $D_8$ root lattice of with opposite sign
inner product.  Compared to the $e^{(D)}_I$ basis~\eqref{eq:eperpD},
the $e^\perp_I$ here differ by a shift of $e^{(D)}_9$ though
$e^{(D)}_{13}$ by $\half f_3$.  Compared to the $e^{(E)}_I$
basis~\eqref{eq:eperpE}, they differ by a shift of $e^{(E)}_8$ and
$e^{(E)}_{16}$ by $f^3$.  Identifying the $e^\perp_I$ of this section
with the $\xi_I$ of Sec.~\ref{sec:Ebasis}, these relations extend to
the transformations~\eqref{eq:D8basis} of the full basis from
$\xi^{(D)}{}_a$ to $\xi_a$ and from $\xi^{(E)}{}_a$ to $\xi_a$.


\section{The Lichnerowicz operator}
\label{app:Lichnerowicz}

Given a metric $g_{mn}$ on a manifold $X$, the change in the Ricci
tensor $R_{mn}$ under an infinitesimal metric deformation $\d g_{mn}$ is
\begin{equation}\label{eq:deltaR}
  \d R_{mn} = \ha (\D_L \d g)_{mn},
\end{equation}
where 
$\D_L$ is the \emph{Lichnerowicz operator},
\begin{equation}\label{eq:CAichnerowicz}
  -(\D_L \d g)_{mn} = \nabla^p\nabla_p\d g_{mn}
  - 2R^p{}_{mn}{}^q\d g_{pq}
  - R_m{}^p\d g_{pn} - R_n{}^p\d g_{mp}.
\end{equation}
Here $\nabla_p$ is the metric connection.

The Lichnerowicz operator can be thought of as a symmetric tensor
version of the Laplace-de~Rham operator, $\Delta = dd^\dagger +
d^\dagger d$.  The latter acts on differential forms (i.e.,
antisymmetric tensors).  Here, $d^\dagger$ is the codifferential,
defined by $^\star d^\star$ (with $\star$ the Hodge star) up to a
convention-dependent sign.  Harmonic forms are the zero eigenfunctions
of the Laplace-de Rham operator.  The harmonic forms on $X$ are in
one-to-one correspondence with the cohomology of $X$, since there is a
unique harmonic representative of each cohomology class.

Explicitly, the Laplace-de~Rham operator acts on 2-forms and 3-forms
as
\begin{equation}
  -(\D\o)_{mn} = \nabla^p\nabla_p\omega_{mn}
  -2R^{pq}{}_{mn}\o_{pq}
  -R_m{}^p\o_{pn} -R_n{}^p\o_{mp},
\end{equation}
and
\begin{multline}
  -(\D\o)_{mnp} = \nabla^p\nabla_p\omega_{mnp}
  -2R^{qr}{}_{mn}\o_{qrp}
  -2R^{qr}{}_{mp}\o_{qnr}-2R^{qr}{}_{np}\o_{mqr}\\
  -R_m{}^q\o_{qnp} -R_n{}^q\o_{mqp}-R_p{}^q\o_{mnq}.
\end{multline}

On a K\"ahler manifold, the complex structure $\CJ_m{}^n$ is
covariantly constant $\nabla_p \CJ_m{}^n = 0$ (as is any other version
of the same tensor with raised or lowered indices, such as the
K\"ahler form $J_{mn} = \CJ_m{}^pg_{pn}$, since $\nabla$ is metric
compatible).  This leads to the result that
\begin{equation}
  (\D_L h)_{mn} = \half\bigl(\CJ_m{}^p(\D\o)_{pn}+\CJ_n{}^p(\D\o)_{pm}\bigr)
\end{equation}
where $h_{mn} = \half(\CJ_m{}^p\o_{pn} + \CJ_n{}^p\o_{pm})$, and
$\o_{mn}$ is any $(1,1)$-form.  The proof uses the symmetry relations
\begin{equation}
  \begin{split}
    R_{mnpq} &= -R_{nmpq} = -R_{mnqp},\\
    R_{mnpq} &= R_{pqmn},\\
    R_{[mnp]q} &=0,
  \end{split}
\end{equation}
and is most easily performed in complex coordinates.  Recall that the
only nonvanishing components of the Riemann tensor on a K\"ahler
manifold are $R_{i\bar\jmath k\bar\ell}$ or those related to these by
symmetries.  Likewise, the Ricci tensor has components
$R_{i\bar\jmath} = R_{\bar\jmath i}$.

If in addition, the manifold is Calabi-Yau, then $R_{i\bar\jmath}=0$
and there exists a covariantly constant $(3,0)$-form $\O_{ijk}$.  In
this case, we similarly have
\begin{equation}
  (\D_L h)_{mn} = \half\bigl(\O_m{}^{pq}(\D\chi)_{pqn}+
  \O_n{}^{pq}(\D\chi)_{pqm}\bigr),
\end{equation}
where $h_{mn} = \half\bigl(\O_{m}{}^{pq}\chi^{\vphantom{pq}}_{pqn}+
\O_{n}{}^{pq}\chi^{\vphantom{pq}}_{pqm}\bigr)$ and $\chi_{mnp}$ is any
$(1,2)$-form.  However, note that \emph{primitive} $(1,2)$-forms $J\w
v$ lead to vanishing $h_{mn}$.


\section{Metric deformations and harmonic forms}
\label{app:MetDefHarm}

Given a Ricci flat metric $g_{mn}$ on $X$, the metric deformations
that preserve Ricci flatness are
\begin{equation}
  \d g_{mn} = \d\m^i h_{(i)mn},
\end{equation}		
where the $\d\m^i$ are small parameters and $h_{(i)mn}$ are a complete
set of symmetric tensor fields annihilated by the Lichnerowicz
operator\footnote{See App.~\ref{app:Lichnerowicz} for a review of the
  Lichnerowicz operator.  Compared to App.~\ref{app:Lichnerowicz},
  this section focuses only on the zero modes of the Lichnerowicz and
  Laplace-de Rham operators.} on $X$.  On manifolds of special
holonomy, these are closely related to harmonic forms.


\subsection{K\"ahler manifolds}
\label{app:KahlerDef}

From the results of App.~\ref{app:Lichnerowicz}, on a complex K\"ahler
manifold ($dJ=0$), the covariant constancy of $\CJ_m{}^n$ implies that
every harmonic $(1,1)$-form $\o$ leads to a metric deformation
\begin{equation}\label{eq:kahlerdef}
  h_{mn} = -\frac12\left(\CJ_m{}^p\o_{pn} + \CJ_n{}^p\o_{pm}\right),
\end{equation}
annihilated by the Lichnerowicz operator.  This metric deformation is
transverse $\nabla^m h_{mn}=0$ as a consequence of the harmonicity of
$\o$ and the covariant constancy of $\CJ$.  The deformation is
traceless $g^{mn}h_{mn}=0$ when $\o$ is primitive.  Recall that a
harmonic form is said to be primitive when it cannot be written as $\o
= J\w\o'$ for some $\o'$.  This is equivalent to the condition that
$J^{mn}\o_{mn\dots p} =0$.


\subsection{Calabi-Yau $n$-folds}
\label{app:CYdef}

Similarly, on a Calabi-Yau 3-fold, the existence of a holomorphic
$(3,0)$ form $\O$ allows us to associate a metric deformation
\begin{equation}\label{eq:cpxdef}
  h_{mn} = -\frac12\frac1{2!}\left(\O_m{}^{pq}\chi_{pqn}+\O_n{}^{pq}\chi_{pqm}\right),
\end{equation}
annihilated by the Lichnerowicz operator, to each harmonic
$(1,2)$-form $\chi$.  In complex coordinates,
Eqs.~\eqref{eq:kahlerdef} and \eqref{eq:cpxdef} become, respectively,
\begin{equation}
  \begin{split}
    h_{i\jbar} &= -i\o_{i\jbar}\\
    h_{ij} &= -\frac1{2!}\O_i{}^{\bar k\bar l}\chi_{\bar k\bar l j}.
  \end{split}
\end{equation} 		

The story is very similar for a Calabi-Yau $n$-fold, $n\ge3$ with the
holomorphic $(n,0)$ form replacing the $(3,0)$-form, and
$(1,n-1)$-forms replacing $(1,2)$-forms:
\begin{equation}
  \begin{split}
    h_{i\jbar} &= -i\o_{i\jbar}\\
    h_{ij} &= -\frac1{(n-1)!}\O_i{}^{\bar k_1\cdots\bar k_{n-1}}
    \chi_{\bar k_1\cdots\bar k_{n-1} j}.
  \end{split}
\end{equation} 		
The first class of deformations are the K\"ahler deformations $\d
g_{i\bar\jmath}$ with indices of mixed type.  The second class of
deformations are the complex structure deformations $\d g_{ij}$, with
indices of the same type.  In the latter case, the metric deformation
must be combined with a change in the definition of the complex
coordinates in order to preserve the hermiticity of the metric.  For
now, we leave this coordinate redefinition implicit.  This metric
deformation is transverse as a consequence of the harmonicity of $\o$
and the covariant constancy of $\O$.  It is traceless as a consequence
of the difference in Hodge type between $\chi$ and $\O$.

For $n=2,1$, we have a torus or K3 surface.  We treat the torus case
below.  The K3 case is treated in Sec.~\ref{sec:K3MetDefHarm}.


\subsection{Tori}
\label{app:Tori}

For $T^{2n}$, there are $\frac12 (2n)(2n+1) = n(2n+1)$ real metric
moduli.  On the other hand, naively, there are $n^2$ real K\"ahler
deformations (the number of real degrees of freedom in~$g_{i\jbar}$),
and $n^2$ complex complex-structure deformations (the number of
complex degrees of freedom in $\tau^i{}_j$ of $dz^i = dx^i + \t^i{}_j
dy^j$), for a total of $3n^2$ real moduli.  The apparent conflict is
resolved by observing that some of the complex structure deformations
lead to vanishing metric deformation.  In particular, $\d h_{mn} = 0$
for \emph{nonprimitive} $(1,n-1)$-forms $\chi=J\w \o$, where $\o$ is a
$(0,n-2)$-form.

The reader is invited to check this explicitly for $T^6$, with 21 real
metric moduli.  In this case, there are 9 K\"ahler deformations and 9
complex structure deformations.  The 3 complex structure deformations
generated by $(1,2)$-forms $J\w d\bar z^\ibar$, for $i=1,\dots 3$
vanish, leaving 6 nontrivial metric deformations from complex
structure for a total of $9\text{ (K\"ahler)} + 2\times 6\text{
  (complex)} = 21$ real metric moduli, as desired.


\section{Gibbons-Hawking multicenter metric deformations and harmonic
  forms}
\label{app:GHdefs}

The metric deformations and harmonic forms of the approximate K3
metric~\eqref{eq:K3UnitMetric} are closely related to those of the
Gibbons-Hawking multicenter metric, so it is helpful to review the
latter.  Let us write the multicenter
metric~\eqref{eq:GHmetric3rdform} as
\begin{equation}
  \begin{split}
    ds^2 &= \bar G_{mn}dx^m dx^n\\
    &= \D^{-1}ZG_{\a\b}dx^\a dx^\b + \D Z^{-1} (dx^4+A)^2\\ 
    &= \d_{\mhat\nhat}\th^\mhat\th^\nhat,
  \end{split}
\end{equation}
where the $\th^\mhat$ form a coframe
\begin{equation}\label{eq:coframehat}
  \begin{split}
    \th^\ahat &= \D^{-1/2}Z^{1/2}E^\ahat{}_\b dx^\b,\\
    \th^{\4hat} &= \D^{1/2}Z^{-1/2}(dx^4+A),
  \end{split}
\end{equation}
and $E^\ahat{}_\b$ is a vielbein for $G_{\a\b}$.  In this Appendix, hats
denote frame indices.  In the body of the paper, this is clear from
context, so we suppress the hats to simplify notation.  The quantities
$Z$ and $A$ are defined in Eq.~\eqref{eq:ZA3rdform}.


\subsection{Metric deformations from explicit moduli dependence}
\label{app:DefExplicit}

The metric depends on $G_{\a\b}$, the source locations $\bx^I$, and
shifts of $A$ by a constant 1-form $\b_\a dx^\a$.  As in
Sec.~\ref{sec:K3MetricGH}, we define a covariantized deformation
\begin{equation*}
  \tilde\d\b^{\a\b} = \e^{\a\b\g}\b_\g - x^{I\a}\d x^{I\b} + x^{I\b}\d x^{I\a}. 
\end{equation*}

Let focus on the moduli $\bx^I$, and consider a small change $\d\bx^I$
at $\d G^{\a\b} = \tilde\d\b^{\a\b} = 0$.  Since $Z$ depends on
$\bx^I$ only through $Z_I$ and only through the combination
$\bx-\bx^I$, we have
\begin{equation}
  \d Z = \d\bx^I\cdot \frac{\pd}{\pd\bx^I} Z
    = -\d\bx^I\cdot\boldsymbol{\nabla}Z_I,
\end{equation}
where $\boldsymbol{\nabla}$ is the 3D gradient operator.  The
corresponding metric deformation is
\begin{equation}
  \begin{split}\label{eq:ExplicitMetricDef}
    \d(ds^2) &= \frac{\d Z}{Z}\left[\D^{-1}Z g_{\a\b} dx^\a dx^\b -
      \D Z^{-1}R^2(dx^4+A)^2 \right]
    + 2\D Z^{-1}\d A_\a dx^\a\\
    &= \frac{\d Z}{Z}\left[ (\th^{\1hat})^2 + (\th^{\2hat})^2 
      + (\th^{\3hat})^2 - (\th^{\4hat})^2 \right]
    +2 \th^{\4hat} \d A^{\4hat}{}_\ahat \th^\ahat,
  \end{split}
\end{equation}
where $\d A^{\4hat} = \D^{1/2}Z^{-1/2}\d A$.  Equivalently,
\begin{equation}
  \begin{split}
    \d \th^\ahat &= \frac12\frac{\d Z}{Z} \th^\ahat,\\
    \d \th^{\4hat} &= -\frac12\frac{\d Z}{Z} \th^{\4hat} 
    + \d A^{\4hat}{}_\ahat \th^\ahat.
  \end{split}
\end{equation}

Since $A$ satisfies $d A = *_3 dZ$, we have
\begin{equation}
  \d A = \d x^{I\a}\d A_{I\a},
  \quad\hbox{where}\quad
  d(\d A_I) = *_3 d(\d Z_I).
\end{equation}
Given the form of $\d Z$ above, an obvious solution is $\d A_\a = -\d
x^{I\b}\pd_\b A_{I\a}$.  However, a gauge equivalent and more
convenient choice is
\begin{equation}
  \d A_\a = \d x^{I\b}F_{I\a\b},
  \quad
  F_{I\a\b} = \pd_\a A_{I\b}-\pd_\b A_{I\a}.
\end{equation}
With this choice, the metric deformation becomes

\begin{align}
  \d(ds^2) &= -\d x^{I\a}\frac{\pd_\b Z_I}{Z}\left[\D^{-1} Z G_{\a\b} dx^\a dx^\b -
    \D Z^{-1}(dx^4+A)^2 \right]
  + 2\d x^{I\b}\D Z^{-1}\d F_{I\,\a\b} dx^\a\notag\\
  &= -\d x^{I\ahat}\frac{\pd_\ahat Z_I}{Z}\left[ 
    (\th^{\1hat})^2+(\th^{\2hat})^2+(\th^{\3hat})^2-(\th^{\4hat})^2\right]
  +2 \d x^{I\bhat} \th^{\4hat}F^{\4hat}_{I\,\ahat\bhat} \th^\ahat,
\end{align}
where $F^4_I = \D^{1/2}Z^{-1/2}F_I$ and $F_I = dA_I$.  Equivalently,
\begin{equation}
  \begin{split}
    \d \th^\ahat &= -\frac12 \d x^{I\bhat}\frac{\pd_\bhat Z_I}{Z} \th^\ahat,\\
    \d \th^{\4hat} &= \frac12 \d x^{I\bhat}\frac{\pd_\bhat Z_I}{Z} \th^{\4hat} 
    + \d x^{I\bhat}F^{\4hat}_{I\,\ahat\bhat} \th^\ahat.
  \end{split}
\end{equation}


\subsection{Metric deformations generated by harmonic forms}
\label{app:DefHarmonic}

For each $I$, we have the anti-selfdual harmonic 2-form
\begin{equation}
  \begin{split}
    \o_I &= \Bigl(\frac{Z_I}{Z}\Bigr)_{,\ahat}
    \left(\th^\ahat\w \th^{\4hat} 
      - \d^{\ahat\ahat'}\e_{\ahat\bhat\ghat}\th^\bhat\w \th^\ghat \right)\\
    &= -d\Bigl(A_I - \frac{Z_I}{Z}(dx^4 + A)\Bigr).
\end{split}
\end{equation}
The remaining six harmonic 2-forms analogous to $\o_\a$ and $\o^\a$ of
Sec.~\ref{sec:HarmApprox} are not square integrable on the multicenter
space.  The triple of hyperk\"ahler 2-forms is
\begin{equation}
  J^\ahat  = \th^\ahat\w \th^4 + \d^{\ahat\ahat'}\e_{\ahat\bhat\ghat} 
  \th^\bhat\w \th^\ghat, 
  \quad\text{for}\quad
  \ahat=\1hat,\2hat,\3hat.
\end{equation}
By raising the second 2-form index of $J^\ahat$, we obtain a triple of
complex structures $(\CJ^\ahat)_m{}^n$ satisfying $\CJ^\ahat\CJ^\bhat
= -\d^{\ahat\bhat} - \d^{\ahat\ahat'}\e_{\ahat\bhat\ghat}\CJ^\ghat$.

The hyperk\"ahler metric deformations generated by the $\o_I$ and
deformations parameters $\d x^{I\ahat}$ are
\begin{equation}\label{eq:HarmonicMetricDef}
  \d(ds^2) = \d x^{I\ahat} (h_{I\ahat})_{mn} dx^m dx^n, 
\end{equation}
where
\begin{equation}
  (h_{I\ahat})_{mn} = -\frac12\Bigl((\CJ^\ahat)_m{}^p \o_{I\,pn}
  +(\CJ^\ahat)_n{}^p \o_{I\,pm}\Bigr).
\end{equation}
After some algebra, we find
\begin{equation}
  (h_{I\1hat})_{\mhat\nhat}\th^\mhat \th^\nhat
  =-\Bigl(\frac{Z_I}{Z}\Bigr)_{,\,\hat1}
  \left[(\th^{\1hat})^2+(\th^{\2hat})^2+(\th^{\3hat})^2-(\th^{\4hat})^2\right]
  +2 F^{\4hat}_{I\,\ahat\1hat} \th^\ahat \th^{\4hat}
  +2\th^{\1hat} d\Bigl(\frac{Z_I}{Z}\Bigr),
\end{equation}
with expressions for $(h_{I\2hat})_{\mhat\nhat}$ and
$(h_{I\3hat})_{\mhat\nhat}$ obtained by cyclic permutation of $\1hat$,
$\2hat$, and $\3hat$.


\subsection{Equivalence}
\label{App: Equivalence}

With the identification 
\begin{equation}
  \d x^{I\ahat} = \th^\ahat{}_\b\,\d x^{I\b},
\end{equation}
the metric deformations of the last two sections agree, provided the
latter is supplemented by a diffeomorphism
\begin{equation}
 x^m\mapsto x'^m = x^m -\d N^m.
\end{equation}
It is convenient to lower the index on $\d N^m$ and describe the
1-form $\d N = \d N_mdx^m$, which appears in the metric transformation
\begin{equation}\label{eq:Bdiffeomorphism}
  ds^2 \mapsto ds'^2 = ds^2 - \bigl(\nabla_m \d N_n + \nabla_n \d N_m\bigr)
  dx^m dx^n.
\end{equation}
In the remainder of this section, we prove that
\begin{equation}\label{eq:CompSoln}
  \d N = -\d x^{I\ahat}\Bigl(\frac{Z_I}{Z}\Bigr)\th_\ahat,
\end{equation}
where $\th_\ahat = \d_{\ahat\bhat}\th^\bhat = \th^\ahat$, or
equivalently, in the coordinate basis,
\begin{equation}
  \d N^\a = -\d x^{I\a}\Bigl(\frac{Z_I}{Z}\Bigr),\quad \d N^4 = 0.
\end{equation}

For simplicity, consider the case that $\d x^I_\a$ is nonzero only for
$\a=1$.  The generalization is straightforward.  The difference
between the metric deformations of the previous two sections is
\begin{multline}\label{eq:diffdef}
  \d(ds^2) - \d x^{I\a}(h_{I\a})_{\mhat\nhat}\th^\mhat\th^\nhat =
  -\d x^{I\1hat}
  \left[\frac{Z_I}{Z}\frac{\pd_{\1hat} Z}{Z}
    \left[(\th^{\1hat})^2+(\th^{\2hat})^2+(\th^{\3hat})^2-(\th^4)^2\right]
  \right.\\
  \left.-2\frac{Z_I}{Z}F^{\4hat}_{I\,\ahat\1hat} \th^\ahat \th^{\4hat} 
    + 2\pd_{\ahat}\Bigl(\frac{Z_I}{Z}\Bigr)\th^\ahat\th^1\right].
\end{multline}
To relate this to a diffeomorphism~\eqref{eq:Bdiffeomorphism}, we need
an explicit expression for the covariant derivative operator.


\paragraph{Maurer-Cartan equations.}

It is convenient to work in the basis $\th^\ahat,\th^{\4hat}$ and
deduce the connection from the first Maurer-Cartan equations
\begin{equation}\label{eq:MC1}
  d\th^\a = \o^\ahat{}_\bhat\w \th^\bhat + \o^\a{}_{\4hat}\w\th^{\4hat},
  \quad
  d\th^{\4hat} = \o^{\4hat}{}_\bhat\w \th^\bhat,
\end{equation}
where $\o_{\mhat\nhat}=\o_\mhat{}^{\phat}\d_{\phat\nhat}$ is
antisymmetric.  Then,
\begin{equation}
  \nabla_\mhat B_\nhat = \pd_\mhat B_\nhat - (\o_\mhat)^\phat{}_\nhat B_\phat.
\end{equation}


\paragraph{Connection 1-form.}

The coframe was given in Eq.~\eqref{eq:coframehat}.  Taking the
exterior derivative gives
\begin{equation}
  \begin{split}
    d\th^\ahat &= \tfrac12 d(\log Z)\w \th^\ahat,\\
    d\th^{\4hat} &= -\tfrac12 d(\log Z)\w \th^{\4hat} + F^{\4hat},
  \end{split}
\end{equation}
where $F^{\4hat} = \D^{1/2}Z^{-1/2}F$ and $F=dA$.  In components, we
write $F^{\4hat} = \tfrac12 F^{\4hat}_{\a\b} dx^\a\w dx^\b = \tfrac12
F^{\4hat}_{\ahat\bhat}\th^\ahat\w \th^\bhat$.

Comparing to the Maurer Cartan equation~\eqref{eq:MC1}, we have
\begin{align*}
  &\o^\ahat{}_\bhat\w\th^\bhat + \o^\ahat{}_{\4hat}\w\th^{\4hat}
  = \tfrac12 \pd_\bhat\log Z \th^\ahat\w\th^\bhat,\\
  &\o^{\4hat}{}_{\bhat}\w\th^\bhat 
  = -\tfrac12\pd_\bhat\log Z \th^{\4hat} -\tfrac12
  F^{\4hat}_{\ahat\bhat}\th^\ahat.
\end{align*}
From the antisymmetry of $\o_{\ahat\bhat}$ we deduce that
\begin{equation}
  \begin{split}
    \o^\ahat{}_\bhat &= \tfrac12\left(\pd_\bhat\log Z\th^\ahat 
      -\pd^\ahat\log Z\th_\bhat\right)
    +\tfrac12 F_{\4hat\,bhat}{}^\ahat\th^{\4hat},\\
    \o^{\4hat}{}_{\bhat} &= -\tfrac12\pd_{\bhat}\log Z\th^{\4hat}
    -\tfrac12 F^{\4hat}_{\ahat\bhat}\th^\ahat\\
    \o^\bhat{}_{\4hat} &=  \tfrac12\pd^\bhat\log Z\th_{\4hat}
    + \tfrac12 F_{\4hat\,\ahat}{}^\bhat\th^\ahat.
  \end{split}
\end{equation}


\paragraph{Diffeomorphism.}

Therefore,
\begin{align*}
  2\nabla_{(\ahat} \d N_{\bhat)}
  &= 2\pd_{(\ahat} \d N_{\bhat)}
  -\left[(\o_\ahat)^\ghat{}_\bhat + (\o_\bhat)^\ghat{}_\ahat\right]\d N_\ghat
  -\left[(\o_\ahat)^{\4hat}{}_\bhat + (\o_\bhat)^{\4hat}{}_\ahat\right]\d N_{\4hat}\\
  &= 2\pd_{(\ahat} \d N_{\bhat)} 
  -\tfrac12\Bigl[\pd_\bhat\log Z\d^\ghat_\ahat + \pd_\ahat\log Z\d^\ghat_\bhat
    -2\pd^\ghat\log Z\d_{\ahat\bhat}\Bigr]\d N_\ghat,\\
  2\nabla_{(\ahat} \d N_{\4hat)}
  &= 2\pd_{(\ahat} \d N_{\4hat)}
  -\left[(\o_\ahat)^\ghat{}_{\4hat} + (\o_{\4hat})^{\ghat}{}_\ahat\right]\d N_\ghat\\
  &= 2\pd_{(\ahat} \d N_{\4hat)}
  - F_{\4hat\,\ahat}{}^\ghat \d N_\ghat\\
  2\nabla_{(\4hat} \d N_{\4hat)}
  &= 2\pd_{\4hat} \d N_{\4hat}
  - 2(\o_{\4hat})^\ghat{}_{\4hat} \d N_\ghat\\
  &= 2\pd_{\4hat} \d N_{\4hat} - \bigl(\pd^\ghat\log Z\bigr) \d N_\ghat.
\end{align*}
Writing
\begin{equation*}
  \d N = -\d x^{I\1hat}\Bigl(\frac{Z_I}{Z}\Bigl)\th_{\1hat},
\end{equation*}
and remembering that $\d x^{I\1hat} = \th^{\1hat}{}_\b\d a^{I\b}$ with
$\th^{\hat1}{}_\a$ proportional to $Z^{1/2}$ (so that
$\pd_\ahat\th^{\1hat}{}_\b = \tfrac12 (\pd_\ahat\log Z)
\th^{\1hat}{}_\b$), we find
\begin{align*}
  \nabla_{\1hat} \d N_{\1hat} + \nabla_{\1hat} \d N_{\1hat} 
  & = -\d x^{I\1hat}\left[
  2\pd_{\1hat}\Bigl(\frac{Z_I}{Z}\Bigr) + \frac{Z_I}{Z}\pd_{\1hat}\log Z\right],\\
  \nabla_{\2hat} \d N_{\2hat} + \nabla_{\2hat} \d N_{\2hat} 
  &= -\d x^I_1\frac{Z_I}{Z}\pd_{\1hat}\log Z,\\
  \nabla_{\3hat} \d N_{\3hat} + \nabla_{\3hat} \d N_{\3hat} 
  &= -\d x^{I\1hat}\frac{Z_I}{Z}\pd_{\1hat}\log Z,\\
  \nabla_{\4hat} \d N_{\4hat} + \nabla_{\4hat} \d N_{\4hat} 
  &= \d x^{I\1hat}\frac{Z_I}{Z}\pd_{\1hat}\log Z,\\
  \nabla_{\1hat} \d N_{\2hat} + \nabla_{\2hat} \d N_{\1hat} 
  &= -\d x^{I\1hat}\pd_{\2hat}\Bigl(\frac{Z_I}{Z}\Bigr),\\
  \nabla_{\1hat} \d N_{\3hat} + \nabla_{\3hat} \d N_{\1hat} 
  &= -\d x^{I\1hat}\pd_{\3hat}\Bigl(\frac{Z_I}{Z}\Bigr),\\
  \nabla_{\1hat} \d N_{\4hat} + \nabla_{\4hat} \d N_{\1hat} &= 0,\\
  \nabla_{\2hat} \d N_{\3hat} + \nabla_{\3hat} \d N_{\2hat} &= 0,\\
  \nabla_{\2hat} \d N_{\4hat} + \nabla_{\4hat} \d N_{\2hat} 
  &= \d x^{I\1hat}F^4_{\2hat\1hat}\Bigl(\frac{Z_I}{Z}\Bigr),\\
  \nabla_{\3hat} \d N_{\4hat} + \nabla_{\4hat} \d N_{\3hat} 
  &= \d x^{I\1hat}F^4_{\3hat\1hat}\Bigl(\frac{Z_I}{Z}\Bigr).
\end{align*}
Therefore, $-\bigl(\nabla_m\d N_n + \nabla_n\d N_m\bigr)dx^m dx^n$
indeed agrees with the difference between the two metric
deformations~\eqref{eq:diffdef}, as desired.  The analogous results
for $\d x^{I\2hat}$ and $\d a^{I\3hat}$ are obtained by cyclic
permutation of $\1hat$, $\2hat$, and $\3hat$.  Thus, we obtain
Eq.~\eqref{eq:CompSoln} for general $\d x^{I\ahat}$.


\subsection{$\int\o_I\w\o_J$ and related integrals}
\label{app:Integrals}

In this Appendix we evaluate $\int\o_I\w\o_J$ for the anti-selfdual
harmonic 2-forms $\o_I$ in the Gibbons-Hawking multicenter
metric~\eqref{eq:GHmetric3rdform}, as well as a few related integrals.
Let
\begin{equation}
  \CI_{IJ} = - \int_{X^4} \o_I\w\o_J.
\end{equation}
From Eq.~\eqref{eq:L2forms}, we have
\begin{equation}\label{eq:Iintegral}
  \CI_{IJ} = 2\int_{\IR^3} d^3x\,Z G^{\a\b}\Bigl(\frac{Z_I}{Z}\Bigr)_{,\,\a} 
  \Bigl(\frac{Z_J}{Z}\Bigr)_{,\,\b},
\end{equation}
where $Z = 1 + \sum_I Z_I$ and
\begin{equation}
  -\nabla^2 Z_I = \d^3(\bx-\bx^I).
\end{equation}
If Eq.~\eqref{eq:Iintegral} is integrated by parts, we find a
cancellation between two Laplacians (using $Z_I/Z = 1$ at
$\bx=\bx^I$), leaving
\begin{equation}
  \CI_{IJ} = \CJ_{IJ},
\end{equation}
where
\begin{equation}\label{eq:Jdef}
  \CJ_{IJ}\equiv -2\int_{\IR^3} d^3x\, \frac{Z_I}{Z}G^{\a\b}
  (\pd_\a Z) \pd_\b\Bigl(\frac{Z_J}{Z}\Bigr).
\end{equation}
At this point we can apply the following trick:
\begin{enumerate}
\item Observe that $\CJ_{IJ}$ must be symmetric in $IJ$ as a
  consequence of the equality $\CI_{IJ} = \CJ_{IJ}$ and the definition
  of $\CI_{IJ}$.  This is not obvious from the definition of $\CJ_{IJ}$
  in Eq.~\eqref{eq:Jdef}.
\item Integrate $\CJ_{IJ}$ by parts and use the $\nabla^2 Z$
  expression together with the fact that $Z_I/Z = \delta_{IJ}$ at
  $\bx=\bx_J$.  This relates $\CJ_{IJ}$ to $\CJ_{JI}$ plus a term
  proportional to $\delta_{IJ}$.
\end{enumerate}
Steps 1~and~2 together allow us to solve for $\CI_{IJ}$. The
result is~\cite{Ruback:1986ag}
\begin{equation}
  \CI_{IJ} = \delta_{IJ}.
\end{equation}
In a similar manner, the following integrals can be evaluated:
\begin{align}
  \CK_{IJ} &= 2\int_{\IR^3} d^3x \frac1{Z}G^{\a\b}(\pd_\a Z_I)(\pd_\b Z_J) = 3\d_{IJ},\\
  \CL_{IJ} &= 2\int_{\IR^3} d^3x \frac{Z_IZ_J}{Z^3}G^{\a\b}(\pd_\a Z)(\pd_\b Z) = 0,\\
  \CM_{IJ} &= -2\int_{\IR^3} d^3x \frac{Z_I}{Z^2}G^{\a\b}(\pd_\a Z)(\pd_\b Z_J) = -\d_{IJ}.
\end{align}
In particular, for $F_I = dA_I = \star dZ_I$, where $\star$ is defined
in the 3D metric $G_{\a\b}$, we have
\begin{equation}
    \int_{X^4}d^4x\frac1Z (F_I)_{\a\g}(F_I)^{\a\g} 
    = 2\int_{X^4} F_I\w \star F_J
    = \CK_{IJ} = 3\d_{IJ}.
\end{equation}
A closely related integral is
\begin{equation*}
   \int_{X^4}d^4x\frac1Z (F_I)_{\a\g}(F_I)_\b{}^\g,
\end{equation*}
which, by symmetry, is equal to $\frac13 G_{\a\b}$ times the previous
one:
\begin{equation}\label{eq:Fintegral}
   \int_{X^4}d^4x\frac1Z (F_I)_{\a\g}(F_I)_\b{}^\g = G_{\a\b}\,\d_{IJ}.
\end{equation}
This integral (or rather its analog with $F_I$ replaced by
$F_I-F_{I'}$ and $\int_{X^4}$ replaced by $\half\int_{T^4}$) appears
in Sec.~\ref{sec:Method2}.


\end{document}